\documentclass[a4paper,twoside,12pt]{article}

\usepackage{psfig}
\usepackage{amsfonts}

\newtheorem{theorem}{Theorem}
\newtheorem{proposition}{Proposition}
\newtheorem{corollary}{Corollary}

\newtheorem{definition}{Definition}
\newtheorem{conjecture}{Conjecture}

\def\halmos{\hbox{\vrule height0.31cm width0.01cm\vbox{\hrule height
 0.01cm width0.3cm \vskip0.29cm \hrule height 0.01cm width0.3cm}\vrule
 height0.31cm width 0.01cm}}

\newcommand{\ZZ}{\mathbb{Z}}

\newcommand{\CC}{\mathbb{C}}

\newcommand{\MA}{\mathcal{A}}
\newcommand{\MB}{\mathcal{B}}

\newcommand{\lker}{{\rm LKer}}
\newcommand{\rker}{{\rm RKer}}

\newcommand{\spn}[1]{\mbox{$\langle\,#1\,\rangle$}}
\newcommand{\sptp}{\,\tilde{\otimes}\,}

\begin{document}
\begin{flushright}
ITFA-2002-13\\
EMPG-02-08\\
HWM-02-16\\
hep-th/0205114
\end{flushright}
\vspace{0.0cm}
\begin{center}
\baselineskip 24 pt
{\Large \bf Hopf symmetry breaking and confinement  } \\
 {\Large \bf in (2+1)-dimensional gauge theory }

\baselineskip 18pt
\parskip 7pt

\vspace{0.3cm}
{\large F.~A.~Bais$^1$,\,
 B.~J.~Schroers$^2$ and     
J.~K.~Slingerland$^1$ }\\

\vspace{0.4cm}

$^1$Institute for Theoretical Physics, University of Amsterdam, \\
       Valckenierstraat 65, 1018 XE Amsterdam, The Netherlands.\\
$^2$ Department of Mathematics, Heriot-Watt University,\\
Edinburgh EH14 4AS, United Kingdom  \\
{\footnotesize
E-mail: {\tt bais@science.uva.nl, bernd@ma.hw.ac.uk, slinger@science.uva.nl}
}

\vspace{0.5cm}

{\large May 2002}

\end{center}

\begin{abstract}

\noindent
Gauge theories in  2+1 dimensions whose gauge symmetry
is spontaneously broken to a finite group enjoy a quantum group  symmetry
which includes the  residual gauge symmetry. This symmetry
provides a framework in which   fundamental excitations (electric charges) 
and  topological excitations (magnetic fluxes) 
can be treated on equal footing. In order to study symmetry
breaking by both electric and magnetic condensates
we develop a theory of symmetry breaking which is applicable to 
models whose  symmetry is described by a quantum group (quasitriangular
Hopf algebra). Using this general framework we
investigate the symmetry breaking and confinement phenomena which
occur in (2+1)-dimensional gauge theories. Confinement of particles
is linked to the formation of string-like defects. Symmetry
breaking by an electric condensate leads to magnetic confinement
and vice-versa. We illustrate the general formalism with examples 
where the symmetry is broken by electric, magnetic and dyonic condensates.

\end{abstract}

%\vspace{0.5cm}
%\centerline{MSC 17B37, 81R50, 81S10, 83C45}

\section{Introduction}
One of the roads towards an understanding of confinement starts with
the proposal of 't Hooft and Mandelstam \cite{hooftconf,mandelconf} to
think of it in terms of the breaking of a dual or magnetic symmetry by
a condensate of magnetic monopoles.  While this idea has been very
fruitful, it has not yet led to a rigorous proof of confinement. One
reason for this is that the supposed magnetic symmetry is not manifest
in the usual formulation of gauge theory.  It is therefore difficult
to study its breaking in detail. Other approaches also try to link the
phenomenon of electric confinement to condensate physics in the
magnetic sector, but there seems to be no general consensus as to
which magnetic excitations should be the ones to condense. 

In this paper, we will study the Higgs and confinement transitions in
a class of theories where both the electric and the magnetic symmetry
are manifest and where we have a clear picture of the possible
magnetic excitations. The theories in question are (2+1)-dimensional
gauge theories in which the gauge group $G$ is broken down to a
discrete group $H$. The full electric-magnetic symmetry in such
``discrete gauge theories'' is described by a quantum group or
quasitriangular Hopf algebra: the quantum double $D(H)$ of the
discrete unbroken gauge group. If $H$ is Abelian, then $D(H)$ is just
the group algebra of the group $\tilde{H}\times H$, so that we have
the electric group $H$ and a dual magnetic group $\tilde{H}\cong
H$. However, if $H$ is non-Abelian, then $D(H)$ is not a group algebra
and so the total symmetry is not described by a group. As a
consequence, the discussion of $D(H)$-symmetry breaking cannot proceed
in the usual way when $H$ is non-Abelian; we need to generalize the
concepts involved in the discussion of symmetry breaking so that we
can study symmetry breaking not only for symmetries described by
groups, but also for symmetries described by quantum groups. A large
part of this paper (sections \ref{breaksec} and \ref{confsec}) is thus
devoted to setting up a formalism for the description of symmetry
breaking and confinement in theories with a quantum group
symmetry. The procedure that emerges has two steps: the original
symmetry algebra is first broken down to a residual Hopf algebra and
this algebra is subsequently projected onto a Hopf algebra which
classifies the unconfined excitations. This formalism can in principle
be applied to any planar system with a quantum group symmetry and
could thus have applications in physics ranging from
(2+1)-dimensional quantum gravity to the quantum Hall effect.

In the rest of the paper, we apply our formalism to discrete gauge
theories.  The results on discrete gauge theories that we have
obtained include descriptions of
\begin{itemize}
\item
electric gauge symmetry breaking and the corresponding
magnetic confinement,
\item
symmetry breaking by manifestly gauge invariant magnetic
condensates and the ensuing electric confinement,
\item
symmetry breaking and confinement by various other types
of condensates, such as non-gauge invariant magnetic condensates and
dyonic condensates.
\end{itemize}
All these results will be illustrated with explicitly worked
examples. These will include a complete treatment of the discrete
gauge theories whose gauge group is an odd dihedral group.

The detailed setup of the paper is as follows: In section
\ref{settingsec}, we give a quick introduction to discrete gauge
theories. We describe the fundamental and topological excitations and
the topological interactions between these excitations, which play an
important role in the confinement discussion. At the end of this
section, we also give a non-technical preview of the effects of
symmetry breaking by a condensate of fundamental, electrically charged
particles. In section \ref{defsec}, we review some basic notions from
the theory of quantum groups and establish notation. In section
\ref{dhsec}, we describe the quantum groups which reproduce the
spectrum and interactions of discrete gauge theories: the quantum
doubles of finite groups. Some examples of quantum doubles which we
use throughout the paper are introduced in section \ref{examsec}. In
section \ref{breaksec}, we develop a general method for the study of
spontaneous symmetry breaking in systems with a quantum group symmetry
and in section \ref{confsec}, we give a general discussion of the
confinement phenomena that accompany this symmetry breaking. Section
\ref{stabsec} gives some motivation for the choice of the specific
condensates we study in the rest of the paper. In section
\ref{elcondsec}, we discuss the phases of discrete gauge theories that
occur when the electric symmetry is broken by condensation of an
electrically charged particle that does not carry magnetic flux. In
section \ref{magcondsec}, we discuss the phases that occur when the
magnetic symmetry is broken by a gauge invariant magnetic
condensate. In section \ref{purecondsec}, we discuss the simultaneous
breaking of the electric and magnetic symmetries as a consequence of
the condensation of a pure, non gauge invariant, magnetic flux. In
section \ref{dyonsec}, we present some results on dyonic
condensates. Finally, in section \ref{conclusec}, we give a brief
summary and an outlook.

\section{Discrete gauge theories: physical setting}
\label{settingsec}

We use the term discrete gauge theory for a (2+1)-dimensional
Yang-Mills-Higgs theory in which the Higgs field has broken the
(continuous) gauge group $G$ down to a finite group $H$.  Such
theories are discussed in detail in
\cite{fluxmetam,dgt1,dgt2,dgt3,dgt4,dgt5}. As a consequence of the
symmetry breaking these theories contain topological defects which are
labeled by elements of $\pi_1(G/H)$. By the exact homotopy sequence,
this corresponds to $\pi_0(H)=H$ when $\pi_1(G)$ is trivial. It
follows that, when $G$ is simply connected, the defects are
characterized by elements of the unbroken group $H$\footnote{One may
argue that even in a theory where $G$ is not simply connected, the set
of {\em stable} fluxes will still be labeled by elements of $H$, due
to the presence of (Dirac) magnetic monopoles in the three-dimensional
theory that underlies the two-dimensional theory we are considering
here \cite{dgt5}}.  The element of $H$ that characterizes a defect may
be identified as the value of a Wilson loop integral around the
defect. In analogy with electromagnetism, we will call the value of
this loop integral the ``magnetic flux'' through the loop and we will
call the defects magnetic fluxes. In this setting, fluxes are thus
group valued. It is clear that the values of the Wilson loop integrals
are indeed elements of the unbroken group $H$, since, if they were
not, parallel transport around the closed loop would not leave the
Higgs field's expectation value invariant.

The action of the unbroken group $H$ on fluxes is given by
conjugation: a flux $g\in H$ is sent to $hgh^{-1}$ by the element $h$
of $H$. This transformation rule is just the transformation rule for
the Wilson loop integral. As a consequence, the fluxes are organized
into gauge multiplets, one for each conjugacy class of $H$. Thus, the
distinct types of flux-carrying particle are labeled by the conjugacy
classes $A,B,\ldots$ of $H$, while a particle of, say, type $A$, has
an internal Hilbert space of dimension equal to the number of elements
of the conjugacy class $A$.

Apart from the topological fluxes, we also allow fundamental charged
particles which are labeled by the irreducible representations of the
unbroken group $H$. The internal Hilbert space of particle that
carries the irrep $\alpha$ of $H$ is just the module $V_\alpha$ of
$\alpha$ and the action of the gauge group on this space is just the
action given by the matrices of the irrep $\alpha$. In addition to
charges and fluxes, one has dyons: particles which carry both flux and
charge. The charges of dyons with flux $g \in H$ are characterised by
the irreducible representations of the centraliser associated with the
conjugacy class of $g$ in $H$. In other words, the charge of a dyon is
characterised by a representation of the subgroup of the gauge group
that leaves the flux of the dyon invariant. The set of electric
charges available to a dyon thus depends on the flux of the dyon,
indicating that there must be a non-trivial interplay between the
electric and magnetic symmetries in this theory.

Now that we have given the natural set of quantum numbers labeling
the different sectors (or particle charges) in this theory, let us
turn to the interactions. Because the unbroken group is discrete, the
gauge fields of the theory are massive, with mass proportional to the
length $v$ of the vacuum expectation value of the Higgs field. As a
consequence, the electric and magnetic gauge interactions are screened
with a screening length inversely proportional to $v$. We are
interested in the low energy or long distance limit of these theories,
or equivalently in the limit in which the expectation value of the
Higgs field becomes large. In this limit, the theory becomes
topological; the only interacions between the particles that survive
are ultra-short range interactions, that may be described by fusion
rules, and non-local Aharonov-Bohm interactions (the Aharonov-Bohm
effect is not screened by the Higgs effect
\cite{krausswil,preskrauss,dgt3,dgt5}). These Aharonov-Bohm
interactions may be described by the action of a (coloured) braid
group on the multi-particle states involved and hence we will refer to
it as ``braiding''

The fusion rules for charges are given by the tensor product
decomposition of $H$-irreps. The fusion product of two fluxes is found
by concatenation of the associated Wilson loops, which leads to the
conclusion that the fusion product of fluxes $g_1,g_2\in H$ is the
flux labeled by $g_1 g_2 \in H$. The effect of braiding a charge
$\alpha$ around a flux $g\in H$ is given by the action of $g$ on
$V_{\alpha}$. If $\alpha$ is a one-dimensional irrep of $H$, then this
is the usual Aharonov-Bohm phase factor, but if $\alpha$ is
higher-dimensional, then the action of $g$ on $V_{\alpha}$ will be
described by the matrix $\alpha(g)$, which will not necessarily reduce
to a phase factor. It follows that, if the unbroken group is
non-Abelian, discrete gauge theories allow for non-Abelian braiding
between charges and fluxes; if a charge is first taken around a flux
$g_1$ and then around a flux $g_2$, then the effect on the wave
function of the charge may be different than if it is taken first
around $g_2$ and then around $g_1$, simply because one may have
$\alpha(g_1)\alpha(g_2)\neq\alpha(g_2)\alpha(g_1)$. The braiding
between fluxes may be found by contour manipulation. If a  particle
with flux $g_1$ is taken around a particle with flux $g_2$, then its
flux will change from $g_1$ to $g_2 g_1 g_{2}^{-1}$, i.e. braiding
between fluxes is given by conjugation.  Braiding and fusion of dyons
will be described in section \ref{dhsec}.

Much of the rest of this paper is devoted to the development of a
mathematical formalism which will help us describe the phases that
result when the symmetry of a discrete gauge theory is broken by the
formation of a condensate. However, in the special case where the
condensate is purely electric, one may already get a fairly accurate
picture of what happens using only the information in this
section. The reason for this is that the formation of an electric
condensate can be viewed as a simple modification of the Higgs
condensate which broke $G$ down to $H$ in the first place. After this
modification, the new residual gauge group will be the subgroup $N$ of
$H$ which leaves the new condensate invariant. The spectrum of free
excitations should thus consist of fluxes labeled by conjugacy
classes of $N$, charges labeled by irreps of $N$ and dyons with flux
and centralizer charge. In short, everything should be as it was
before, except that the role of $H$ has been taken over by $N$. Any
fluxes $h\in H\backslash N$ that were present when the new condensate
formed, will pull a string. That is, their presence makes it
impossible for the new Higgs condensate to be single valued and hence
causes the expectation value of the Higgs field to develop a line-like
discontinuity. The energy associated with this discontinuity will grow
linearly with its length and as a consequence, the fluxes $h$ outside
$N$ will be confined in ``hadrons'' whose overall flux does lie in
$N$. The line discontinuities themselves can be viewed as domain walls
between regions with different values $v_1,v_2$ of the Higgs
expectation value. They may thus be characterized by an element $h\in
H$ such that $hv_1=v_2$, but this characterization is not unique,
since $hnv_1$ will also equal $v_2$ for any $n\in N$. Therefore, the
strings (or walls) should be characterised by a coset $hN$ in $H/N$,
or more precisely by a gauge orbit of such cosets. This picture of
what happens when a purely electric condensate is formed is an
important part of the intuition that will be used in the rest of the paper
and it is a non-trivial test of the formalism we will develop that it
must repoduce this picture.

\section{Hopf symmetry in (2+1) dimensions} 
\label{defsec}

The spectrum, fusion and braiding that we have sketched in section
\ref{settingsec}, and also the fusion and braiding of dyons, may be
derived in a purely algebraic way, using a quantum group symmetry that
is present in the theory. The quantum group in question is the quantum
double $D(H)$ of the unbroken group $H$. A description of this quantum
group and its role in discrete gauge theory appears in section
\ref{dhsec}, followed by some concrete examples in section
\ref{examsec}. The aim of the present section is to quickly remind the
reader of some aspects of Hopf algebra theory and its application to
the description of many particle systems that are relevant to us. We
also establish notation and collect some formulae for reference. For
much more information on Hopf algebras and quantum groups, one may
consult for example \cite{mont, abe, sweedler, chapress, majid}.

A Hopf algebra is an associative algebra $\mathcal{A}$ with
multiplication $\mu$ and unit $1$, that has extra structures called
counit, antipode and coproduct. These defining structures of a Hopf
algebra guarantee that the spectrum of irreducible representations of
a Hopf algebra has properties that mimic those of the particle
spectrum of a physical theory.

The {\em coproduct} $\Delta$ is an algebra map from $\mathcal{A}$ to
$\mathcal{A}\otimes \mathcal{A}$ with the following property, called
coassociativity:
\begin{equation}
({\rm id}\otimes\Delta)\Delta=(\Delta\otimes {\rm id})\Delta.
\end{equation} 
Here, ${\rm id}$ is the identity map on $\mathcal{A}.$
Given two representations $\pi^{1},\pi^{2}$ of the
quantum group, the coproduct makes it possible to define a tensor 
product representation $\pi^1\otimes\pi^2$ by the formula
\begin{equation}
\pi^{1}\otimes\pi^{2}:~ x \rightarrow
(\pi^{1}\otimes\pi^{2})(\Delta(x)). 
\end{equation} 
The coassociativity of $\Delta$ ensures that the tensor product of
representations is associative. If we have a theory where the
different particles carry irreducible representations of
$\mathcal{A}$, then $\Delta$ provides us with a way of describing the
action of $\mathcal{A}$ on multi-particle states. The decomposition of
the tensor product representations defined by means of $\Delta$ will
describe the fusion rules of the theory.

The {\em counit} $\epsilon$ of $\mathcal{A}$ is an algebra map from
$\mathcal{A}$ to $\CC$, or equivalently, a one-dimensional
representation of $\mathcal{A}$, which satisfies
\begin{equation}
(\epsilon\otimes {\rm id})\Delta=({\rm id}\otimes\epsilon)\Delta={\rm id}.
\end{equation}
It follows that $\epsilon\otimes\pi \cong \pi\otimes\epsilon \cong
\pi$ for any representation $\pi$ of the quantum group. Thus, if we
assume that the vacuum (or an $\mathcal{A}$-neutral particle)
transforms in the representation $\epsilon$ of $\mathcal{A}$, then we
get the fusion properties that one would expect. In other words,
$\epsilon$ provides us with a vacuum representation of the algebra.

The {\rm antipode} of $\mathcal{A}$ is a linear algebra
antihomomorphism $S:\mathcal{A}\rightarrow\mathcal{A}$ which satisfies
\begin{equation}
\label{antipode}
\mu(S\otimes {\rm id})\Delta(a)=\mu({\rm id}\otimes
S)\Delta(a)=\epsilon(a) 1
\end{equation}
for each $a \in \mathcal{A}$. If we are given a representation $\pi$
of $\mathcal{A}$, then the antipode makes it possible to define the
representation $\bar{\pi}$ conjugate to $\pi$ by the formula
\begin{equation}
\bar{\pi}(a)=(\pi(S(a)))^{\rm t},
\end{equation}
where the $t$ denotes matrix transposition. The properties
(\ref{antipode}) then ensure that the tensor product representations
$\pi\otimes\bar{\pi}$ and $\bar{\pi}\otimes\pi$ will contain the
trivial representation $\epsilon$ in their decomposition. Thus, a
particle which carries the representation $\pi$ and its antiparticle,
which carries the representation $\bar{\pi}$ may indeed annihilate.

We now briefly recall the notion of a Hopf subalgebra and of the dual
of a Hopf algebra, as they will turn out to be important in our
discussion of Hopf symmetry breaking.
\begin{definition}
 A {\em Hopf subalgebra} of a Hopf algebra $\mathcal{A}$ is a
subalgebra $\mathcal{B}$ of $\mathcal{A}$ which satisfies
\begin{equation}
\label{subhopf}
1\in \mathcal{B},~~~S(\mathcal{B})\subset \mathcal{B},~~~
\Delta(\mathcal{B})\subset \mathcal{B}\otimes\mathcal{B}.
\end{equation}
This implies that $\mathcal{B}$ is itself a Hopf algebra, with ``the
same'' structures as $\mathcal{A}$.
\end{definition}
\begin{definition}
For any finite dimensional Hopf algebra $\mathcal{A}$, the {\em dual
Hopf algebra} is the vector space $\mathcal{A}^*$ of linear
functionals from $\mathcal{A}$ to $\CC$ with Hopf algebra structures
$1^{*}, \mu^{*}, \Delta^{*}, \epsilon^{*}$ and $S^{*}$ defined by:
\begin{equation} 
\begin{array}{rlrl}
1^* :& a\mapsto \epsilon(a)&
\mu^*(f_1,f_2):& a \mapsto f_1\otimes f_2\circ\Delta(a) \\
\epsilon^*:& f \mapsto f(1)& 
\Delta^*(f):& (a_1,a_2) \mapsto f(\mu(a_1,a_2))\\
S^*(f):& a \mapsto f(S(a))&~&~
\end{array}
\end{equation}
Here, $a,a_1,a_2$ are arbitrary elements of $\mathcal{A}$ and
$f,f_1,f_2$ are arbitrary elements of $\mathcal{A}^*$.
\end{definition}
In this paper, we will deal only with finite-dimensional semisimple
Hopf algebras. We call Hopf algebra $\mathcal{A}$ semisimple if all
$\mathcal{A}$-modules are fully reducible. A simple and important
example of a finite dimensional semisimple Hopf algebra is the group
algebra $\CC H$ of a finite group $H$. Its comultiplication, antipode
and counit are given on the basis of group elements $h\in H$ by the
formulae
\begin{equation}
\label{grouphopf}
\Delta(h)=h\otimes h ~~~ S(h)=h^{-1} ~~~ \epsilon(h)=1. 
\end{equation}
Clearly, we may similarly associate a Hopf algebra to any group. Thus,
if we can describe the excitation spectrum and fusion properties of a
physical system by means of a group then we may also describe them by
means of a Hopf algebra. On the other hand, the converse is not
necessarily true; there are physical systems whose symmetry algebra is
a Hopf algebra, but not a group algebra. In fact, most systems which
can be described by means of two-dimensional conformal field theory
have this property (for reviews, see for instance
\cite{ags90,gras,fuchs}). Examples in (2+1) dimensions are the
discrete gauge theories we will treat in this paper, but also
(2+1)-dimensional gravity \cite{bamu,bamusch} and certain fractional
quantum Hall systems \cite{moread,sliba}.

In two spatial dimensions, the exchanges of a system of $n$ particles
are governed by the braid group $B_n$. There is a special class of
Hopf algebras, called quasitriangular Hopf algebras or quantum groups,
whose extra structure makes it possible to include a description of
the braiding between particles in the Hopf algebraic framework. The
extra feature which makes a Hopf algebra $\mathcal{A}$ quasitriangular
is an invertible element $R \in\mathcal{A}\otimes \mathcal{A}$ which
is called the {\rm universal $R$-matrix} and which satisfies the
following equations
\begin{eqnarray}
\label{rprop1}
\Delta^{\rm op}R&=&R\Delta\\
\label{rprop2}
(\Delta\otimes 1)R&=&R_{13}R_{23} \\
\label{rprop3}
(1\otimes \Delta)R&=&R_{13}R_{12}.
\end{eqnarray} 
Here, $\Delta^{\rm op}$ is the comultiplication, followed by an exchange
of the tensor factors in $\mathcal{A}\otimes\mathcal{A}$ and
$R_{ij}$ is an abbreviation for the action of $R$ on the factors $i$
and $j$ of $\mathcal{A}^{\otimes 3},$ so for example
$R_{12}=R\otimes 1.$ 

The $R$-matrix is used to describe braiding in the following way. If
we have a system of $n$ identical particles that transform in the
representation $\pi$ of $\mathcal{A}$, then the corresponding
(internal) Hilbert space is the $n$-fold tensor product
$(V_{\pi})^{\otimes n}$, where $V_{\pi}$ is the module of $\pi$. We can
now exchange adjacent particles numbered $i$ and $i+1$ by letting
$R_{i,i+1}$ act on the state of the system and then flipping the
$i^{\rm th}$ and $(i+1)^{th}$ tensor factors. We will call the map
which flips the factors of a tensor product $\sigma$. 

The defining properties of the $R$-matrix are chosen so that they make
sure that exchanging particles by means of $\sigma R$ makes physical
sense.  The properties (\ref{rprop2}) and (\ref{rprop3}) make sure
that braiding of two particles around a third one and then fusing them
together gives the same result as fusing the two particles first and
then braiding the result around the third one. The property
(\ref{rprop1}) ensures that the exchanges commute with the action of
the quantum group. Using either (\ref{rprop1}) and (\ref{rprop2}) or
(\ref{rprop1}) and (\ref{rprop3}), one may also prove that in any
representation, we have
\begin{equation}
(\sigma R \otimes 1)(1\otimes\sigma R)
(\sigma R\otimes 1) =
(1\otimes\sigma R)(\sigma R \otimes1)
(1\otimes\sigma R).
\end{equation}
It follows that, for a system of $n$ identical particles that carry a
representation of the quantum group, the exchanges of adjacent
particles, as performed using $\sigma R,$ satisfy the relations of the
braid group. Hence, since $R$ is invertible, they generate a
representation of this group. Since the exchanges commute with the
action of the quantum group $\mathcal{A}$, it follows that the system
carries a representation of $\mathcal{A}\times B_{N}.$

Sometimes, a description of the spin of the particles in a two
dimensional theory can also be incorporated into the Hopf algebraic
description of the system. This is the case when the quasitriangular
Hopf algebra that describes the system has an invertible central
element $c$ that satisfies the equations
\begin{eqnarray}
c^2=uS(u), ~~~S(c)=c,~~~\epsilon(c)=1 \nonumber\\
\Delta(c)=(R_{21}R_{12})^{-1}(c\otimes c),
\end{eqnarray}
where $u=\mu(S\otimes {\rm id})(R_{21})$. The element $c$ is called
the {\em ribbon element} and a quasitriangular Hopf algebra that has a
ribbon element is called a {\em ribbon Hopf algebra.} The action
of the ribbon element on the physical Hilbert space is interpreted as
the action of a rotation of the system over $2\pi$. In particular, the
action of $c$ on an irrep of $\mathcal{A}$ describes the effect of
rotating the particle that carries this irrep. Because $c$ is central,
the action of $c$ on an irrep may always be described by a scalar
factor, which is called the spin factor of the irrep and of the
corresponding particle. The equations that $c$ has to satisfy make
sure that the spin of the vacuum is trivial, that the spin of
a particle and its antiparticle are equal and that rotating a system
of two particles over an angle of $2\pi$ may be accomplished both by
acting with $c$ on the two-particle system (making use of $\Delta$)
and through braiding the two particles around each other and then
rotating them separately.

\section{The quantum double of a finite group}
\label{dhsec}

\subsection{The double and its dual}

The ribbon Hopf algebra that describes the fusion and braiding of the
discrete gauge theory with unbroken group $H$ is the quantum double
$D(H)$ of $H$. As a vector space, $D(H)$ is $F(H)\otimes \CC H$, the
tensor product of the group algebra $\CC H$ of $H$ and its dual, the
space $F(H)$ of functions on $H$. Since $H$ is finite, we may identify
this vector space with $F(H\times H)$, the space of functions on
$H\times H$, and we may write elements of the double as such
functions. On the double, we have the usual structures of a
Hopf algebra: a multiplication $\bullet$, identity 1,
comultiplication $\Delta$, counit $\epsilon$ and antipode $S$:
\begin{equation} 
\label{algebra}
\begin{array}{rcl}
1(x,y)&:=&\delta_e(y)\\
(f_1\bullet f_2)(x,y)&:=&\int_H f_1(x,z)\,f_2(z^{-1}xz,z^{-1}y)\,dz \\
\epsilon(f)&:=&\int_H f(e,y)\,dy\\
(\Delta f)(x_1,y_1;x_2,y_2)&:=&f(x_1x_2,y_1)\,\delta_{y_1}(y_2) \\
(S f)(x,y)&:=&f(y^{-1}x^{-1}y,y^{-1}).
\end{array}
\end{equation}
Here, the integrals over $H$ are a convenient notation for the sum
over all elements of $H$. We see that $D(H)$ is generated as an
algebra by the elements $1\otimes g ~(g \in H)$ and $\delta_g\otimes e
~(g\in H)$. The elements $1\otimes g$ together form the gauge group
$H$, while the elements $\delta_g\otimes e$ are a basis of $F(H)$ and
can be interpreted as projections on the set of states with flux $g$
in the theory. Both multiplication and comultiplication of the double
are consistent with this interpretation.  The universal $R$-matrix of
$D(H)$ is given by the formula
\begin{equation}
\label{rmatdh}
R(x_1,y_1;x_2,y_2) = \delta_e(y_1)\delta_e(x_1y_2^{-1}) 
\end{equation}
and the ribbon element $c$ is given by
\begin{equation}
c(x,y) = \bullet\circ(S\otimes{\rm id})(R_{21}) =  \delta_e(xy).
\end{equation}
The dual $D(H)^*$ of $D(H)$ is $\CC H\otimes F(H)$ as a vector
space. This space may again be identified with $F(H\times H)$, so that
we may realize both the structures of $D(H)$ and those of $D(H)^*$ on
this space.  The multiplication $\star$, unit $1^*$, comultiplication
$\Delta^*$, counit $\epsilon^*$ and antipode $S^*$ of $D(H)^{*}$ are
given by
\begin{equation}
\label{coalgebra}
\begin{array}{rcl}
1^* (x,y)&:=&\delta_e(x)\\
(f_1\star f_2)(x,y)&:=&\int_H f_1(z,y)\,f_2(z^{-1}x,y)\,dz \\
\epsilon^*(f)&:=&\int_H f(x,e)\,dx \\
(\Delta^* f)(x_1,y_1;x_2,y_2)&:=&
f(x_1,y_1y_2)\,\delta_{x_2}(y_1^{-1}x_1y_1)\\
(S^*  f)(x,y)&:=&f(y^{-1}x^{-1}y,y^{-1}).
\end{array}
\end{equation}
$D(H)$ and $D(H)^*$ have a canonical Hermitian inner product, given by the
same formula for both $D(H)$ and $D(H)^*$:
\begin{equation}
\label{inprod}
(f_1,f_2)=\int_H\int_H f_1(x,y)\,\overline{f_2(x,y)}\,dx\,dy.
\end{equation} 
The matrix elements of the irreps of both 
$D(H)$ and $D(H)^*$ are orthogonal with respect to this inner product.
This follows from the theory of Woronowicz \cite{woronowicz}
for compact matrix quantum groups, which holds both for $D(H)$ and
$D(H)^*$ \cite{ftart}, but it may also be proved directly.

\subsection{Irreducible representations}
\label{irrepsec}

The irreducible representations $\pi^A_\alpha$ of $D(H)$ have been
classified in \cite{dpr}, using the fact that the double is a based
ring in the sense of Lusztig \cite{lusztig}. An alternative way to
classify the irreps of quantum doubles of groups makes use of the fact
that $D(H)$ is a transformation group algebra \cite{kobamu}. Since we
will be making rather extensive use of transformation group algebras
in the sequel, we will follow this path. We follow the notation and
conventions of \cite{kobamu}. First, we give a simplified definition
of a transformation group algebra, adjusted to our needs, which
involve only finite groups:
\begin{definition}
\label{tragrodef}
Let $H$ be a finite group acting on a finite set $X$. Then
$F(X\times H)$ is called a transformation group algebra if it is
equipped with the multiplication $\bullet$ given by
\begin{equation}
F_1\bullet F_2 (x,y)=\int_{H}F_1(x,z)F_2(z^{-1}x,z^{-1}y) dz. 
\end{equation}
\end{definition}
For a more general definition and references, see \cite{packer}.  When
we take $X=H$ and the action of $H$ given by conjugation, then we
regain the algebra structure of quantum double $D(H)$. There is a
general theorem which classifies the irreducible representations of
all transformation group algebras as defined above, but before we give
this, we must first define the Hilbert spaces that the representations
will act upon. Let $N$ be a subgroup of $H$, let $\alpha$ be a unitary
representation of $N$, and let $V_{\alpha}$ be its module, then we
define
\begin{equation}
\label{irrepmod}
F_{\alpha}(H,V_{\alpha}):=\{\phi:H\rightarrow V_{\alpha}|\phi(xn)=
\pi_\alpha(n^{-1})\phi(x),
\forall x\in H,\forall n\in N\}.
\end{equation}
The irreps of our transformation group algebras are then described by
the following theorem, which is a simple consequence of theorem 3.9 in
\cite{komu}
\begin{theorem}
\label{irrepth}
Let $F(X\times H)$ be a transformation group algebra and let
$\{\mathcal{O}_{A}\}$ be the collection of $H$-orbits in $X$ ($A$
takes values in some index set). For each $A$, choose some $\xi_A \in
\mathcal{O}_{A}$ and let $N_A$ be the stabilizer of $\xi_A$ in
$H$. Then, for each pair $(\mathcal{O}_{A},\alpha)$ of an orbit
$\mathcal{O}_{A}$ and an irrep $\alpha$ of the stabilizer $N_A$ of
this orbit, we have an irreducible unitary representation
$\tau^{A}_{\alpha}$ of $F(X\times H)$ on $F_{\alpha}(H,V_{\alpha})$
given by
\begin{equation}
(\tau^{A}_{\alpha}(F)\phi)(x):=\int_H F(x\xi_A,z)\phi(z^{-1}x) dz.
\end{equation}
Moreover, all unitary irreducible representations of $F(X\times H)$
are of this form and irreps $\tau^{A}_{\alpha}$ and $\tau^{B}_{\beta}$
are equivalent only if $\mathcal{O}_A=\mathcal{O}_B$ and $\alpha\cong\beta$.
\end{theorem}
In the case of $D(H)$, the orbits $\mathcal{O}_{A}$ are just the
conjugacy classes of $H$. The irreps $\Pi^{A}_{\alpha}$ of $D(H)$ are
thus labeled by pairs $(A,\alpha)$ of a conjugacy class label $A$ and
an irrep $\alpha$ of the centralizer $N_A$ of a specified element
$g_A\in C_A$. We see that the spectrum of irreps of $D(H)$ is in one
to one correspondence with the spectrum of excitations of the discrete
gauge theory that we described in section \ref{settingsec}. In
particular, the pure (uncharged) magnetic fluxes correspond to the
$\Pi^{A}_{1}$, where $1$ denotes the trivial representation of $H$,
and the pure charges correspond to the $\Pi^{e}_{\alpha}$.  We will
call the element $g_A\in A$ the {\em preferred element} of $A$. Any
choice of preferred element yields the same isomorphy class of
representations of $D(H)$.  The carrier space of $\Pi^A_{\alpha}$ is
just the space $F_{\alpha}(H,V_{\alpha})$ defined above and for
brevity we will denote it by $V^{A}_{\alpha}$ The dimension
$d^{A}_{\alpha}$ of $V^{A}_{\alpha}$ is the product of the number of
elements of the conjugacy class $A$ and the dimension $d_{\alpha}$ of
$\alpha$. To see this, note that the functions in
$V^{A}_{\alpha}$ are completely determined once their value on one
element of each $N_{A}$-coset is chosen. Now the number of
$N_{A}$-cosets is just $|A|$, the number of elements of $A$, which
shows that $d^{A}_{\alpha}=|A|d_{\alpha}$. In fact, there is a
canonical correspondence between cosets of $N_{A}$ and elements of
$A$: the coset $hN_{A}$ corresponds to the element $h g_A
h^{-1}$. Thus, a state with pure flux $h g_A h^{-1}$ will be
represented by a wave function with support on $h N_A$.

The action of an element $F\in D(H)$ on $V^{A}_{\alpha}$ is given by
the formula in the theorem above, which in the case of $D(H)$ becomes
\begin{equation}
\label{dblaction}
\left(\pi^A_{\alpha}(F)\phi\right)(x):=\int_H\,dz\,F(x g_A x^{-1},z)
\, \phi(z^{-1}x).
\end{equation} 
{}From this formula, it is easy to see that the action of the gauge
group elements $1\otimes g$ in the purely electric representation
$\Pi^{e}_{\alpha}$ is indeed isomorphic to the action of the gauge
group in the representation $\alpha$. Also, the action of the gauge
group on magnetic fluxes is given by conjugation, which can be seen as
follows: the state with flux $hg_A h^{-1}$ is represented by the
function $\delta_{hN_{A}} \in V^{A}_{1}$, i.e. by the characteristic
function of the coset $hN_{A}$. The action of the element $1\otimes g
\in D(H)$ sends this function to the function $ghN_A$, which in turn
corresponds to the flux $g(hg_{A}h^{-1})g^{-1}$.

The {\em spin} of a particle that transforms in the irrep
$\Pi^{A}_{\alpha}$ is given by the action of the ribbon element $c$. We have
\begin{equation}
\left(\pi^A_{\alpha}(c)\phi\right)(x):=\alpha(g_{A}^{-1})\phi(x).
\end{equation} 
Since the element $g_{A}^{-1}$ is central in $N_A$, the matrix
$\alpha(g_{A}^{-1})$ is a constant multiple of the unit matrix: we
have $\alpha(g_{A}^{-1})=s^{A}_{\alpha} I$, where $s^{A}_{\alpha} \in
\CC$ is a root of unity which we call the {\em spin factor} of
$\Pi^{A}_{\alpha}$. Clearly, we have $s^{A}_{\alpha}={\rm
\frac{1}{d_{\alpha}}Tr(\alpha(g_{A}^{-1}))}$, where $d_{\alpha}$ is
the dimension of the representation $\alpha$ of $N_A$. A consistent
description of the {\em braiding} for arbitrary representations of
$D(H)$ is given by the $R$-matrix (\ref{rmatdh}).

\subsection{Matrix elements and Characters}

The fusion rules for representations of $D(H)$ and of more general
transformation group algebras may be calculated by means of a
character formalism. The character of a representation $\pi$ of a
transformation group algebra $\mathcal{T}=C_c(X\times H)$ is a linear
functional $\chi:\mathcal{T}\rightarrow \CC$ defined as follows:
$\chi(t)$ is the trace of the matrix $\pi(t)$. Clearly, the character
of the representation $\pi$ depends only on the isomorphism class of
$\pi$. Let us calculate the character of an irrep $\tau^{A}_{\alpha}$
of $\mathcal{T}$ from the formula in theorem \ref{irrepth}. The first
thing we need is a basis for the vector space
$F_{\alpha}(G,V_{\alpha})$. To get this, we choose a basis
$e^{\alpha}_{i}$ for $V_{\alpha}$ and a representative for each left
coset of $N_A$. For a given coset of $N_A$, all the elements of this
coset send the preferred element $\xi_A$ of the orbit
$\mathcal{O}_{A}$ to the same element $\zeta$ of this orbit. Moreover,
each element $\zeta\in \mathcal{O}_A$ uniquely determines a
coset. Therefore, we call the representative elements of the cosets
$x_{\zeta}$ and these $x_{\zeta}$ are just arbitrarily chosen elements
of $H$ with the property that $x_{\zeta}\xi_{A}=\zeta$. A basis for
$F_{\alpha}(G,V_{\alpha})$ is now given by the functions
$\phi^{i}_{\zeta}$ defined by
\begin{equation}
\label{traforepbas}
\phi^{i}_{\zeta}(y)=1_{x_{\zeta}N_{A}}(y)\alpha(y^{-1}x_{\zeta})e^{\alpha}_{i}.
\end{equation} 
$\phi^{i}_{\zeta}$ is the unique element of $F_{\alpha}(G,V_{\alpha})$
which takes the value $e^{\alpha}_{i}$ at $x_{\zeta}$ and one may
easily verify that these elements do indeed form a basis for
$F_{\alpha}(G,V_{\alpha})$. The matrix elements of 
$\tau^{A}_{\alpha}$ in this basis are given by
\begin{equation}
\label{trafomatelts}
\tau^{A}_{\alpha}(F)^{i,j}_{\zeta,\eta}=
\int_{N_A}F(x_{\eta}\xi_{A},x_{\eta}n x_{\zeta}^{-1})\alpha_{i,j}(n)\,dn,
\end{equation} 
where the $\alpha_{i,j}$ are the matrix elements of $\alpha$ with
respect to the basis of $e^{\alpha}_{i}$.
As a consequence, the character $\chi^{A}_{\alpha}$ of
$\tau^{A}_{\alpha}$ is given by
\begin{equation}
\label{trfc}
\chi^{A}_{\alpha}(F)=
\int_{\mathcal{O}_A} d\zeta \int_{N_A} dn \,
F(x_{\zeta}\xi_{A},x_{\zeta}n x_{\zeta}^{-1})\chi_{\alpha}(n),
\end{equation} 
where $\chi_{\alpha}$ denotes the character of $\alpha$.  We can
remove the arbitrarily chosen elements $x_{\zeta}$ from this formula
by adding an integration over $N_A$, thus changing the integration
over $\mathcal{O}_A$ into an integration over $H$:
\begin{equation}
\label{trafochar}
\chi^{A}_{\alpha}(F)=
\int_H dz \int_{N_A} dn \,F(z\xi_{A},znz^{-1})\chi_{\alpha}(n).
\end{equation} 
When $\mathcal{T}=D(H)$, this reduces to the formula given in
\cite{kobamu}. Clearly, the characters are fully determined by their
values on a basis for $\mathcal{T}$. When $X$ is finite, we can take the basis
of delta functions $\delta_{\eta}\delta_{h} ~(\eta\in X,h\in H)$ and we
can take the characters to be elements of $F(X\times H)$ by writing
$\chi^{A}_{\alpha}(\eta,h):=\chi^{A}_{\alpha}(\delta_{\eta}\delta_{h})$.
We have
\begin{equation}
\label{trafochar2}
\chi^{A}_{\alpha}(\eta,h)=
1_{N_{\eta}}(h)1_{\mathcal{O}_A}(\eta)\chi_{\alpha}(x_{\eta}^{-1}h x_{\eta}).
\end{equation} 
When $\mathcal{T}=D(H)$, this gives the formula for the characters in
\cite{dpr}.  

We may define an inner product $\langle\cdot,\cdot\rangle$ on the
space of functions $X\times H$ by the formula
\begin{equation}
\label{trafoinprod}
\langle \chi_1,\chi_2\rangle=\int_X \,d\eta\,\int_H\,dh\,
\chi_1(\eta,h)\overline{\chi_2(\eta,h)}. 
\end{equation}
One may check that the characters are orthogonal with respect to this
inner product:
\begin{equation}
\label{charorth}
\left \langle \chi^{A}_{\alpha},\chi^{B}_{\beta}\, 
\right\rangle = |H|\delta_{A,B}\delta_{\alpha,\beta}. 
\end{equation}
When $\mathcal{T}=D(H)$, the inner product defined here is just the
invariant inner product (\ref{inprod}) on $D(H)^*$ and the
orthogonality of the characters with respect to this inner product
follows from Woronowicz's general theory. The decomposition of a
tensor product of irreps of $D(H)$ may be found by calculating the
inner products of the characters of the irreps with the character of
the tensor product. In this way, the fusion properties of pure fluxes
and charges that we have described in section \ref{settingsec} are
reproduced and we may also calculate the fusion rules for dyons.

\section{Examples of quantum doubles}
\label{examsec}

We briefly present examples of quantum doubles of finite groups. These
will be the standard examples in our discussion of symmetry breaking
in the remainder of the paper.

\subsection{$D(H)$ for Abelian $H$}
\label{abdoubsec}

The quantum double of an Abelian group $H$ is isomorphic to the group
algebra of $H\times H$ as a Hopf algebra. One way to see this is the
following: First recall that any finite Abelian group is isomorphic to
some $\ZZ_{k_1}\times\ldots\times \ZZ_{k_n}$, where $k_i|k_j$ for
$i<j$. Then recall that, by Pontryagin duality, the character group of
an Abelian group $H$ is isomorphic to $H$. We may thus denote the
elements of $H$ by $n$-tuples $(m_1,\ldots,m_n)$, with $0\le m_i\le
k_i$ and we may also label the characters $\chi_{m_1,\ldots,m_n}$ of
$H$ with such $n$-tuples in such a way that the map $(m_1,\ldots,m_n)
\rightarrow \chi_{m_1,\ldots,m_n}$ is an isomorphism of groups. The
canonical way to do this labeling is such that
$\chi_{m_1,\ldots,m_n}$ is the character given by
\begin{equation}
\label{abchar}
\chi_{m_1,\ldots,m_n}(l_1,\ldots,l_n) = \exp(2\pi i (\frac{m_1
l_1}{k_1}+\dots+\frac{m_n l_n}{k_n})).
\end{equation} 
The characters are linearly independent functions on $H$ and thus
$D(H)$ is spanned by the elements $\chi\otimes \delta_h$, where $\chi$
is a character of $H$ and $h$ is an element of $H$. But one calculates
easily that
\begin{eqnarray}
(\chi_1\otimes\delta_{h_1})\bullet (\chi_2\otimes\delta_{h_2}) &=&
(\chi_1\chi_2\otimes \delta_{h_1 h_2}) \nonumber \\
1_{D(H)} &=& 1\otimes \delta_e \equiv e_{H\times H} \nonumber \\
\Delta(\chi\otimes\delta_h) &=& (\chi\otimes\delta_h)\otimes
(\chi\otimes\delta_h) \nonumber \\
S(\chi\otimes\delta_h) &=& 
(\bar{\chi}\otimes\delta_{h^{-1}}) = (\chi\otimes\delta_h)^{-1} \nonumber \\
\epsilon (\chi\otimes \delta_h) &=& 1,
\end{eqnarray}
so that, comparing to (\ref{grouphopf}), we see that we indeed have
$D(H)\cong \CC (H\times H)$ as a Hopf algebra.

As a consequence, the irreducible representations of $D(H)$ are just
the tensor products $\chi_1\otimes\chi_2$ that may be formed from two
irreps $\chi_1,\chi_2$ of $H$. These correspond to the irreps
$\Pi^{A}_{\alpha}$ that we described in section \ref{irrepsec} in the
following way. When $H$ is Abelian, all conjugacy classes of $A$
consist of just one element, so that the label $A$ may be identified
with the element $g_A$ of $H$. This element may in turn be identified
with a character $\chi_A$ using the isomorphism between $H$ and its
character group that we indicated above. Also, we have $N_A=H$ and so
$\alpha$ is already a character of $H$. One may now check easily that
the irrep $\Pi^{A}_{\alpha}$ of $D(H)$ corresponds to the irrep
$\chi_{A}\otimes \alpha$ of $H\otimes H$. The tensor product of two
$D(H)$-irreps is just the usual tensor product of $H\times H$
representations. The only difference with the usual representation
theory of $H\times H$ is that the representations now have non-trivial
spin-factors and non-trivial braiding, given by the ribbon element and
the $R$-matrix of $D(H)$ respectively. We have
\begin{eqnarray}
s^{A}_{\alpha}=\alpha(g_{A}^{-1}) \nonumber \\
\Pi^{A}_{\alpha}\otimes\Pi^{B}_{\beta}(R)=\beta(g_A). 
\end{eqnarray}
As one may read off using (\ref{abchar}), these are just the usual
phase factors one expects for Abelian dyons, involving products of
charge and flux quantum numbers.

\subsection{$D(D_{2m+1})$}

Perhaps the simplest non-Abelian groups are the dihedral groups $D_n$
which describe the symmetries of the regular $n$-gons. Among these is
the smallest non-Abelian group: the dihedral group $D_3$, which is
isomorphic to the symmetric group $S_3$. $D_n$ has $2n$ elements: the
unit, $n-1$ non-trivial rotations and $n$ reflections. It can be
presented on two generators as follows:
\begin{equation}
D_n=\{s,r| \, s^2=r^n=1,sr=r^{n-1} s\}.
\end{equation}
The $2n$ elements may all be written in the form $r^k$ or $sr^k$ (with
$e=r^0$). The powers of $r$ are the rotations, the elements that
involve $s$ are the reflections. 

In this paper, we will deal exclusively with the odd dihedral groups
$D_{2m+1}$. $D_{2m+1}$ has $m+2$ conjugacy classes, which we will
label by their preferred  elements and which we will denote by their
preferred elements in square brackets (e.g.$[r]$) if confusion
between class and element might arise. The classes are
\begin{eqnarray}
[e] &=& \{e\} \nonumber \\
{}[r^k] &=& \{r^k,r^{-k}\} ~~~(0<k\le m) \nonumber \\
{}[s] &=& \{sr^k| \, 0\le k \le 2m+1\}.
\end{eqnarray}
The centralizers of these classes are given by

\begin{equation}
N_e=D_{2m+1},~~~~ N_{r^k}=\spn{r}\cong\ZZ_{2m+1},~~~~ N_s=<s> \cong \ZZ_2.
\end{equation}
$D_{2m+1}$ has two one dimensional representations: the trivial
representation, which we will denote $J_0$ and a representation $J_1$
given by $J_1(r)=1,~J_1(s)=-1$. The remaining irreps of $D_{2m+1}$ are all
two dimensional and faithful. We will call them
$\alpha_1,\ldots,\alpha_m$ and they may be given by
\begin{equation}
\label{doddmats}
\alpha_k(r)=\left(\begin{array}{cc}
\cos(\frac{2k\pi}{2m+1})& -\sin(\frac{2k\pi}{2m+1}) \\
\sin(\frac{2k\pi}{2m+1})& \cos(\frac{2k\pi}{2m+1})
\end{array}\right)~~~~~
\alpha(s)=\left(\begin{array}{cc}
-1& 0\\
 0& 1
\end{array}\right)
\end{equation}
The character table for $D_{2m+1}$ can now be read off; it is given in
table \ref{doddchartab}
\begin{table}[h,t,b]
\[
\begin{array}{|c|c|c|c|} \hline
       ~&[e]&[r^k]&[s] \\ \hline
       J_0&  1&     1&  1 \\
       J_1&  1&     1& -1 \\
\alpha_j&  2&q^{jk}+q^{-jk}&  0 \\ \hline 
\end{array} 
\]
\caption{\footnotesize character table for $D_{2m+1}$. We have defined
$q=e^{2\pi i/(2m+1)}$. }
\label{doddchartab}
\end{table}
The representations of the $\ZZ_{2m+1}$ and $\ZZ_2$ centralizers will
be denoted $\beta_0,\beta_1,\ldots,\beta_{2m}$ and
$\gamma_0,\gamma_1\equiv \gamma$ respectively. They are as given in
the previous section. The representations of $D(D_{2m+1})$ will thus
be labeled $\Pi^{e}_{J_0}, \Pi^{e}_{J_1}, \Pi^{e}_{\alpha_{k}},
\Pi^{r^{k}}_{\beta_{l}},\Pi^{s}_{1}$ and $\Pi^{s}_{\gamma}$. All in
all this yields $2(m^2+m+2)$ representations. The dimensions
$d^{A}_{\alpha}$ and spin factors $s^{A}_{\alpha}$ of these irreps are
given in table \ref{ddoddspintab}:
\begin{table}[h,t,b]
\[
\begin{array}{|c|c|c|c|c|c|c|} \hline
~&\Pi^{e}_{J_0}&\Pi^{e}_{J_1}&\Pi^{e}_{\alpha_{j}}&
\Pi^{r^{k}}_{\beta_{l}}&\Pi^{s}_{1}&\Pi^{s}_{\gamma}  \\ \hline
d^{A}_{\alpha}&1&1&2&2&2m+1&2m+1 \\
s^{A}_{\alpha}&1&1&1&q^{-kl}&1&-1 \\ \hline 
\end{array}
\]
\caption{\footnotesize dimensions and spin factors for the irreps of
$D(D_{2m+1})$}
\label{ddoddspintab}
\end{table}
The fusion rules of the irreps of $D(D_{2m+1})$ may be determined by
means of the characters (\ref{trafochar2}) and the orthogonality
relations (\ref{charorth}). They have been given explicitly in
\cite{thesismark}. One may also show that tensor products of multiple
$D(D_{2m+1})$-irreps can carry non-Abelian representations of the
braid group.

%\vspace*{\fill}
%\pagebreak

\section{Symmetry Breaking}
\label{breaksec}

\subsection{Hopf symmetry breaking}

Consider the situation where a condensate has formed, carrying the
representation $\Pi^{A}_{\alpha}$ of $D(H)$. The ground state or
``vacuum'' of the theory is then a background of identical particles,
all in the same state $\phi\in V^{A}_{\alpha}$. We model this situation
with a tensor product state $\phi\otimes \phi\otimes \ldots\otimes\phi$.
This state breaks the $D(H)$-symmetry of the theory and we want to
find out what the residual symmetry algebra of the system after this
breaking is. Now if the original symmetry were described by a group,
then finding the residual symmetry would in principle be
straightforward; we would find out which of the group elements leave
the condensate state $\phi$ invariant, i.e. we would find the
stabilizer of $\phi$, and this stabilizer would be the residual
symmetry. If the original symmetry is described by a Hopf algebra,
then we cannot use this recipe, for several reasons. First of all, we
cannot expect to find a subalgebra of the Hopf algebra which leaves
$\phi$ invariant in the usual sense of the word; any such subalgebra
would have to contain the element $0$ which would obviously send
$\phi$ to $0$. Hence, we need a new definition of
invariance. Fortunately, there is a natural definition, namely the
following (cf. \cite{chapress})

\begin{definition}
\label{invardef}
Let $\mathcal{A}$ be a Hopf algebra with counit $\epsilon$, let
$a\in\mathcal{A}$ and let $\phi$ be a vector in some
$\mathcal{A}$-module. Then we say that $\phi$ is left invariant by $a$ if
the action of $a$ on $\phi$ is given by $a \cdot \phi =
\epsilon(a)\phi$. 
\end{definition}
This definition is natural, since it just says that the vector $\phi$
transforms under $a\in\mathcal{A}$ in the same way that the vacuum
would. Also, if the Hopf algebra $\mathcal{A}$ is a group algebra,
then we see that this definition of invariance reduces to the usual
one on the group elements. Nevertheless, when we apply the above
definition of invariance to a group algebra, then we see that the
subalgebra which leaves a vector $\phi$ invariant is {\em not} the
group algebra of the stabilizer of $\phi$. In fact, it is a much
larger algebra, which is not a Hopf algebra. On the other hand, the
maximal Hopf subalgebra of the group algebra which leaves $\phi$
invariant (with the above definition of invariance), is exactly the
group algebra of the stabilizer of $\phi$. This follows easily from
the fact that the Hopf subalgebras of a group algebra $\CC G$ are
exactly the group algebras of the subgroups of $G$ (this is well known
to Hopf algebra theorists, but we also give an explanation of why it
is so in section {\ref{doubbreaksec}}). This suggests that we should
define the residual symmetry algebra after breaking as follows:

\begin{definition}
\label{residdef}
Suppose we have a theory with Hopf symmetry $\MA$.  If a condensate of
particles in the state $\phi$ forms in this theory, then the residual
symmetry algebra is the maximal Hopf subalgebra of $\MA$ that leaves
$\phi$ invariant. We will call this algebra the {\em Hopf stabilizer}
of $\phi$
\end{definition}
A maximal Hopf subalgebra with a certain property is defined as a Hopf
subalgebra with this property which is not a subalgebra of a larger
Hopf subalgebra with his property. The maximal Hopf subalgebra in the
above definition is unique, since, if we have two different Hopf
subalgebras which leave the same vector invariant, then the subalgebra
generated by these two is itself a Hopf subalgebra which leaves this
vector invariant and which contains the original two Hopf
subalgebras. The above definition reduces to the usual definition in
the case of group algebras and it has the further advantage that the
residual symmetry algebra will always be a Hopf algebra\footnote{In
fact, it is also semisimple since any Hopf subalgebra of a finite
dimensional semisimple Hopf algebra is itself semisimple. This is
proved in \cite{mont}, using the Nichols-Zoeller theorem
\cite{nichzoel}.}. The spectrum of the residual algebra will thus have
the desirable properties that we discussed in section \ref{defsec};
there will be a natural description of many-particle states, there
will be a trivial or vacuum representation and given an irrep of the
algebra that labels a ``particle'' (an excitation over the
condensate), there will also be an irrep (possibly the same) that
labels the ``antiparticle''.  The fact that the residual symmetry
algebra is a Hopf algebra also makes sure that the invariance of
$\phi$ implies the invariance of all the states
$\phi\otimes\phi\otimes \ldots \otimes \phi$. This follows easily from
the fact that $(\epsilon\otimes\epsilon)\circ \Delta =\epsilon$.
Thus, we might have taken the condensate to be a superposition of
states with different numbers of particles (still all in the state
$\phi$) and this would have yielded exactly the same residual algebra.

\subsection{Hopf subalgebras and Hopf quotients}
\label{techsec}

In view of the definition \ref{residdef} of the residual symmetry
algebra after the formation of a condensate, it is useful to find out
all we can about Hopf subalgebras of the quantum double $D(H)$, or
more generally, about Hopf subalgebras of finite dimensional
semisimple Hopf algebras. In the present section, we give a
characterization of the Hopf subalgebras of such Hopf algebras, which
will provide us with a way of finding all these Hopf subalgebras and
in particular the residual symmetry algebras of definition
\ref{residdef} in a systematic way. Along with the results on Hopf
subalgebras, we also prove some results on Hopf quotients or quotient
Hopf algebras which will be useful in our discussion of confinement
further on. The main theorems in this section are closely related to
results in \cite{nikshych} and \cite{nichrich}. We include elementary
proofs here in order to make the paper more self-contained. We write
the results in a form which is useful for our needs, rather than
maximally general or compact. A lot of background for this section can
be found in \cite{mont}.

We define a Hopf quotient as follows
\begin{definition}
Let $\MA$ and $\MB$ be Hopf algebras. If we have a surjective Hopf map
$\Gamma:\MA \rightarrow \MB$, then we call $\MB$ a {\em Hopf quotient}
of $\MA$.
\end{definition}
The Hopf algebra $\MB$ is in fact isomorphic to the quotient of $\MA$
by the kernel of the map $\Gamma$, explaining the terminology. 
Our first step in characterizing Hopf subalgebras and Hopf quotients
is to relate them to each other, using the following proposition
\begin{proposition}
Let $\MA$ and $\MB$ be finite dimensional Hopf algebras and let
$\Gamma:\MA\rightarrow \MB$ be a Hopf map. Then the dual map
$\Gamma^*:\MB^*\rightarrow \MA^*$ is also a Hopf map. Moreover, if
$\Gamma$ is injective then $\Gamma^*$ is surjective and if $\Gamma$ is
surjective then $\Gamma^*$ is injective. Finally, if we identify $\MA$
and $\MA^{**}$ and $\MB$ and $\MB^{**}$ in the canonical way, then we
have $\Gamma^{**}=\Gamma$.
\end{proposition}
{\bf Proof:} The proof that $\Gamma^*$ is a Hopf map is straightforward
calculation. As an example, we show that $\Gamma^*\otimes\Gamma^* \circ
\Delta_{\MB^*}=\Delta_{\MA^*}\circ\Gamma^*$. For any $f\in \MB^*$, we
have
\begin{equation}
\Gamma^*\otimes\Gamma^* \circ \Delta_{\MB^*}(f)=
f\circ\mu_{\MB}\circ \Gamma\otimes\Gamma= 
f \circ \Gamma \circ \mu_{\MA}=
\Delta_{\MA^*}\circ\Gamma^* (f).
\end{equation} 
We used the fact that $\Gamma$ is a Hopf map in the second equality. The
other properties that make $\Gamma^*$ into a Hopf map can be verified
analogously. The proof that $\Gamma^{**}=\Gamma$ is also
easy. We identify the element $a \in \MA$ with the functional $E_a
\in \MA^{**}$ that is evaluation in $A$, i.e. $E_a:f\rightarrow
f(a)$. Similarly, we identify $b\in \MB$ with $E_b \in \MB^{**}$. One
may check that these identifications are Hopf isomorphisms. The
action of $\Gamma^{**}$ on $a$ is calculated as follows: 
\begin{equation}
\Gamma^{**}(a)(f)=\Gamma^{**}(E_a)(f)=E_a\circ(\Gamma^*)(f)=E_a(f\circ\Gamma)
=E_{\Gamma(a)}(f)=\Gamma(a)(f)
\end{equation} 
Hence, we see that indeed $\Gamma^{**}=\Gamma$.  The statements about
injectivity or surjectivity of $\Gamma^*$ are basic properties of the
dual map. $\halmos$

\noindent From this proposition, we have the following corollaries
\begin{corollary}
\label{subcorr}
Let $\MA$ be a finite dimensional Hopf algebra and let $\MB$ be a Hopf
subalgebra of $\MA$. Then $\MB^*$ is a Hopf quotient of $\MA^*$.
The corresponding surjective Hopf map is restriction to $\MB$, which
is the dual map of the embedding of $\MB$ in $\MA$
\end{corollary}
\begin{corollary}
\label{quotcorr}
Let $\MA$ be a finite dimensional Hopf algebra and let $\MB$ be a Hopf
quotient of $\MA$, with corresponding surjective Hopf map $\Gamma$. Then
$\MA^*$ has a Hopf subalgebra isomorphic to $\MB^*$, namely the image
of the embedding $\Gamma^*$.
\end{corollary}
The next proposition shows that if $\MB$ is a Hopf quotient of $\MA$,
then the set of representations of $\MB$ naturally corresponds to a
subring of the representation ring of $\MA$.
\begin{proposition}
\label{quotmatprop}
Let $\MA$ be a Hopf algebra and let $\MB$ be a Hopf quotient of
$\MA$. Denote the associated surjective Hopf map from $\MA$ to $\MB$
by $\Gamma$. Then the representations of $\MB$ are in one-to-one
correspondence with the representations of $\MA$ that factor over
$\Gamma$. Also, this correspondence preserves irreducibility. It
follows that, if $\MA$ is semisimple, then so is $\MB$. The
correspondence map between representations also commutes with taking
conjugates and tensor products of representations. As a consequence,
the tensor product of irreps of $\MA$ that factor over $\Gamma$ will
decompose in the same way as the tensor product of the corresponding
irreps of $\MB$.
\end{proposition}
{\bf Proof:} Let $\rho$ be a representation of $\MB$. Then
$\rho\circ\Gamma$ is a representation of $\MA$, since $\Gamma$ is a Hopf
map. Moreover, if $\rho$ is irreducible then so is $\rho\circ\Gamma$,
since $\Gamma$ is surjective. On the other hand, let $\tau:\MA\rightarrow
M_{n\times n}$ be a representation of $\MA$ which factors over $\Gamma$,
that is, $\tau=\rho\circ\Gamma$ for some map $\rho:\MB\rightarrow
M_{n\times n}$. Then $\rho$ is a representation of $\MB$, since $\Gamma$
is surjective and irreducibility of $\tau$ implies that $\rho$ is
irreducible. Also, $\tau$ uniquely determines $\rho$ and vice
versa. Hence, the representations of $\MB$ are in one-to-one
correspondence with the representations of $\MA$ which factor over
$\Gamma$ and irreducibility is preserved in this correspondence.
Semisimplicity of $\MA$ is equivalent to the property that all
$\MA$-modules decompose into irreducibles. This holds in particular
for all $\MA$-modules in which the action of $\MA$ factors over $\Gamma$,
and hence also for all $\MB$-modules, implying that $\MB$ is
semisimple. The remaining statements follow easily from the fact that
$\Gamma$ is a Hopf map. If $\tau=\rho\circ\Gamma$ then
$\overline{\tau}=\overline{\rho}\circ\Gamma$. This can be seen by looking
at the matrix elements $\overline{\tau}_{i,j}$ of $\overline{\tau}$:
\begin{equation}
\overline{\tau}_{i,j}= (\tau_{j,i}\circ
S_{\MA})=\rho_{j,i}\circ\Gamma\circ S_{\MA}=\rho_{j,i}\circ
S_{\MB}\circ\Gamma = \overline{\rho}_{i,j}\otimes \Gamma.
\end{equation}
Here, we have used $\Gamma\circ S_{\MA}=S_{\MB}\circ\Gamma$.
For the tensor product of irreps $\tau^{a}=\rho\circ\Gamma$ and
$\tau^{b}=\rho^{b}\circ\Gamma$, we have
\begin{equation}
\tau^{a}\otimes\tau^{b}\circ\Delta_{\MA}=
\rho^{a}\otimes\rho^{b}\circ\Gamma\otimes\Gamma\circ\Delta_{\MA}=
\rho^{a}\otimes\rho^{b}\circ\Delta_{\MB}\circ\Gamma,
\end{equation}
where we used that
$\Gamma\otimes\Gamma\circ\Delta_{\MA}=\Delta_{\MB}\circ\Gamma$. We see that
$\tau^{a}\otimes\tau^{b}\circ\Delta_{\MA}$ and
$\rho^{a}\otimes\rho^{b}\circ\Delta_{\MB}$ act on the same module by
the same matrices (since $\Gamma$ is surjective). Hence the decomposition
of tensor product representations of $\MA$ will be the same as the
decomposition of tensor product representations of
$\MB$. $\halmos$

\noindent Before the next proposition, we need another definition
\begin{definition}
We call a set $X$ of irreps of a Hopf algebra $\MA$ {\rm closed under
tensor products and conjugation} if $\tau\in X\Rightarrow
\overline{\tau}\in X$ and if $\tau^{a},\tau^{b} \in X$ implies that
all the irreps in the decomposition of the tensor product of
$\tau^{a}$ and $\tau^{b}$ are contained in $X$.
\end{definition}
Note that, in the previous proposition, the set of irreps of $\MA$
that factor over $\Gamma$ is an example of a set of irreps of $\MA$ that
close under tensor products and conjugation. Also note that a set of
irreps that closes under tensor products and conjugation will always
contain the counit.  We now prove a basic fact about sets of
irreducibles of $\MA$ that close under conjugation and tensor
products:
\begin{proposition}
Let $\MA$ be a semisimple Hopf algebra and let $X$ be a set of irreps
of $\MA$ that closes under conjugation and tensor products. Then the
linear space $V_X$ spanned by the matrix elements of the
representations in $X$ is a Hopf subalgebra of $\MA^*$
\end{proposition}
{\bf Proof:} First, let us take the product of two matrix elements. We
have $\mu_{\MA^*}(\tau^{a}_{i,j}\tau^{b}_{k,l})=
(\tau^{a}\otimes\tau^{b}\circ\Delta_{\MA})_{(i,j),(k,l)}$. In other
words, the product of matrix elements of $\tau^{a}$ and $\tau^{b}$ in
$\MA^*$ is a matrix element of the tensor product representation
$\tau^{a}\otimes\tau^{b}\circ\Delta_{\MA}$. Since $\MA$ is semisimple,
this tensor product may be decomposed into irreps and the matrix
elements of the tensor product representation are linear combinations
of the matrix elements of the irreps this decomposition. But since
these irreps are contained in $X$, it follows that $V_X$ is closed
under multiplication. Clearly, $V_X$ also contains
$1_{\MA^*}=\epsilon_{A}$, so $V_X$ is a unital subalgebra of $\MA^*$.
We also have $S(V_X)\subset V_X$, since
\begin{equation}
S_{\MA^*}(\tau_{i,j})=\tau_{i,j}\circ S_{\MA}=\overline{\tau}_{j,i}
\end{equation}
and $\tau\in X\Rightarrow \overline{\tau}\in X$.
For the comultiplication of a matrix element of any representation of
$\MA$, we have
\begin{equation}
\Delta_{\MA^*}(\tau_{i,j})=\sum_{k}\tau_{i,k}\otimes \tau_{k,j}
\end{equation}
and hence we have $\Delta_{\MA^*}(V_X) \subset V_X\otimes V_X$. $\halmos$

\noindent Now we arrive at one of our main goals, which is a partial
converse of the previous proposition:
\begin{theorem}
\label{shth}
Let $\MA$ be a finite dimensional semisimple Hopf algebra over the
complex numbers. Let $\MB$ be a Hopf subalgebra of
$\MA=\MA^{**}$. Then $\MB$ is spanned by the matrix elements of a set
of irreps of $\MA^*$ which closes under conjugation and tensor
products.
\end{theorem} 
{\bf Proof:} Let $\iota$ be the inclusion of $\MB$ into $\MA$. Then
$\MB^*$ is a Hopf quotient of $\MA^*$ with the associated Hopf map
given by $\iota^*$ (cf. corollary \ref{subcorr}). Because $\MA$ is
semisimple and defined over the complex numbers, its dual $\MA^*$ is
also semisimple (see \cite{larred88} and also \cite{mont}). Hence,
using proposition \ref{quotmatprop}, $\MB^*$ is also semisimple. But
then it follows that the matrix elements of the irreps of $\MB^{*}$
span $\MB^{**}=\MB$. On the other hand, we know from proposition
\ref{quotmatprop} that the irreps of $\MB^*$ are identified (through
$\iota^{**}=\iota$) with a set of irreps of $\MA^*$ which closes under
conjugation and tensor products. $\halmos$

\noindent This theorem is the characterization of Hopf subalgebras
that we will be using in our discussion of $D(H)$-symmetry breaking in
section \ref{doubbreaksec}. More generally, it can simplify the
problem of finding all the Hopf subalgebras of a finite dimensional
semisimple Hopf algebra $\MA$ enormously. If the irreps of
$\mathcal{A}^*$ and the decompositions of their tensor products are
known, then finding all sets of irreps of $\MA^*$ that close under
tensor products is a process that can be carried out easily on a
computer.  Finally, we prove a similar statement about Hopf quotients:
\begin{theorem}
Let $\MA$ be a finite dimensional semisimple Hopf algebra over the
complex numbers. Then any Hopf quotient of $\MA=\MA^{**}$ is
isomorphic to a quotient obtained by restriction to a Hopf subalgebra
of $\MA^*$ generated by matrix elements of a set of irreps of $\MA$
which closes under conjugation and tensor products.
\end{theorem}
{\bf Proof:} Let $\MB$ be a Hopf quotient of $\MA$ and let $\Gamma$ be
the associated projection. Then $\Gamma^*(\MB^*)\cong\MB^*$ is a Hopf
subalgebra of $\MA^*$ (cf. corollary \ref{quotcorr}) and
$\MB=\MB^{**}$ is isomorphic to the quotient of $\MA=\MA^{**}$
obtained by restriction to $\Gamma^*(\MB^*)$. Since $\MA$ is semisimple,
$\MA^*$ is also semisimple. Thus, we can now apply the previous theorem to
the pair $(\MA^*,\Gamma^*(\MB^*))$ and it follows that $\Gamma^*(\MB^*)$ is
the desired Hopf subalgebra. $\halmos$

\subsection{Hopf subalgebras of quantum doubles}
\label{doubbreaksec}

In section \ref{techsec}, we showed that the Hopf subalgebras of a
finite dimensional semisimple Hopf algebra (such as $D(H)$) are in
one-to-one correspondence with sets of irreps of the dual Hopf algebra
that close under tensor products and conjugation. Therefore, we now
construct the representations of the dual algebra $D(H)^*$. From
(\ref{coalgebra}), we see that, as an algebra (but not as a Hopf
algebra), $D(H)^{*}$ is isomorphic to $\CC H \otimes F(H)$. As a
consequence, the irreducible representations of $D(H)^{*}$ are tensor
products of irreps of $\CC H$ and irreps of $F(H)$. The irreps of $\CC
H$ just correspond to the irreps of $H$ and we will denote them
$\rho_i$. The irreps of $F(H)$ are all one dimensional and are labeled
by the elements of $H$. We have an irrep $E_g$ for each $g \in H$,
given by
\begin{equation}
\label{fhirreps}
E_g(f)=f(g).
\end{equation}
We can thus label each representation of $D(H)^*$ by a pair
$(\rho_i,g)$. Tensor products of the irreps of $D(H)^*$ may be formed
by means of $\Delta^*$. Although this coproduct is not the same as the
usual coproduct for $\CC H \otimes F(H)$, the decomposition of tensor
products into irreps is not affected by this (the Clebsch-Gordan
coefficients for the decomposition are affected). Thus we have
\begin{equation}
\rho_i\otimes \rho_j = \bigoplus_{k} N_{ij}^{k}\rho_k ~\Longrightarrow ~
(\rho_i,g)\otimes (\rho_j,h) = \bigoplus_{k} N_{ij}^{k}(\rho_k,gh)
\end{equation}
where the $N^{ij}_{k}$ are the usual multiplicities in the
decomposition of tensor products of $H$-irreps. Also, we have
\begin{equation}
\overline{(\rho_i,g)}=(\bar{\rho_{i}},g^{-1}). 
\end{equation}
{}From these formulae, we see that any set $X$ of irreps of $D(H)^*$
that closes under tensor products and conjugation is associated to a
set of irreps of $H$ and a set of irreps of $F(H)$ with the same
property. These sets just consist of the irreps that may occur as a
factor of one of the irreps in $X$. Consequently, for any Hopf
subalgebra $\mathcal{B}$ of $D(H)$, there are minimal Hopf subalgebras
$\mathcal{C}$ of $F(H)$ and $\mathcal{D}$ of $\CC H$ such that
$\mathcal{B} \subset \mathcal{C}\otimes\mathcal{D}\subset D(H)$. In
the other direction, we see that, for any pair of Hopf subalgebras
$\mathcal{C}\subset F(H)$ and $\mathcal{D}\subset\CC H$, the vector
space $\mathcal{C}\otimes\mathcal{D}$ is a Hopf subalgebra of
$D(H)$. Note that this Hopf subalgebra is usually not isomorphic to
$\mathcal{C}\otimes\mathcal{D}$ as a Hopf algebra. Also, not all Hopf
subalgebras of $D(H)$ are of this form.  However, the ones that are
will be quite important in the sequel. Therefore, we now find all the
Hopf subalgebras of the group algebra $\CC H$ and of the function
algebra $F(H)$.  This also gives us two simple examples of the use of
theorem \ref{shth}.
\begin{proposition}
The Hopf subalgebras of a group algebra $\CC H$ are the group
algebras of the subgroups of $H$.
\end{proposition}

\noindent {\bf Proof:} The Hopf algebra dual to $\CC H$ is the vector
space $F(H)$ of functions on $H$, with Hopf algebra structure given by
\begin{eqnarray} 
1^* : g \mapsto 1, ~~~
&\mu^*(f_1,f_2): g \mapsto f_1(g) f_2(g),~~~
&\Delta^*(f): (g_1,g_2) \mapsto f(g_1 g_2),\nonumber \\
~&\epsilon^*: f \mapsto f(e),~~~ 
&S^*(f): g \mapsto f(g^{-1}),
\end{eqnarray}
where $g, g_1, g_2$ are arbitrary elements of the group $H$. Note
that, in the formula for the comultiplication, we have identified
$F(H)\otimes F(H)$ with $F(H\times H)$ in the usual way. The
irreducible representations $E_g$ of $F(H)$ were given in
(\ref{fhirreps}). One checks easily that the tensor product of two if
these irreps, as defined using $\Delta^*$, is given by
\begin{equation}
E_g\otimes E_h=E_{gh}.
\end{equation}
Also, we have $\bar{\pi}_{g}=E_{g^{-1}}$. Hence, the sets of irreps of
$(\CC H)^*$ that close under conjugation and tensor products correspond
exactly to the subgroups of $H$. The proposition follows. $\halmos$
\begin{proposition}
A Hopf subalgebra of an algebra $F(H)$ of functions on a group $H$
is isomorphic to the algebra $F(H/K)$ of functions on the quotient of
$H$ by some normal subgroup $K$.
\end{proposition}

\noindent {\bf Proof:} Let $\mathcal{C}$ be a Hopf subalgebra of
$F(H)$ and let us denote the irreps of $\CC H$ whose matrix elements
span $\mathcal{C}$ by $\rho_i$. Then the intersection of the kernels
of the $\rho_i$ is a normal subgroup $K$ of $H$ and any function in
$\mathcal{C}$ will be constant on the cosets of $K$. We can also say
that $\mathcal{C}$ really consists of functions on the quotient group
$H/K$. Now let us show the opposite inclusion. If we form the direct
sum $\oplus_i \rho_i$ of all the representations $\rho_i$, then this
representation of $H$ will have exactly $K$ as its kernel and hence it
can be identified with a faithful representation of $H/K$. Now it is a
theorem in the theory of finite groups that the tensor powers of any
faithful representation of a group contain all irreducible
representations of this group (see for instance \cite{jalie}). Hence
all irreps of $H/K$ are contained in the tensor powers of $\oplus_i
\rho_i$ and hence the matrix elements of these irreps are contained in
$\mathcal{C}$. But since the matrix elements of the irreps of $H/K$
span $F(H/K)$, it follows that $\mathcal{C}\cong F(H/K)$. $\halmos$

Thus we see that for any Hopf subalgebra $\mathcal{B}$ of $D(H)$,
there is a maximal normal subgroup $K$ of $H$ and a minimal subgroup
$N$ of $H$ such that $\mathcal{B}$ is in fact a Hopf subalgebra of
$F(H/K)\otimes \CC N$. Also, every subalgebra of $D(H)$ of the form
$F(H/K)\otimes \CC N$ is automatically a Hopf subalgebra. These
particular Hopf subalgebras are in fact also transformation group
algebras, with the group $N$ acting on $H/K$ by conjugation. This will
be very useful later on, since it will allow us to apply the
representation theory of transformation group algebras that we
described in section \ref{dhsec}.

Now let us turn to the problem of finding the Hopf subalgebra of
$D(H)$ which leaves a given condensate vector $\phi$ invariant.
\begin{proposition}
\label{tprop}
The Hopf stabiliser $\mathcal{T}\subset D(H)$ of a given vector
$\phi\in V^{A}_{\alpha}$ is spanned by the matrix elements of those
irreps $(\rho,g)$ of $D(H)^*$ for which
\begin{equation}
\label{inveq}
\forall x: \phi(g x)=\frac{\chi_{\rho}(g_A)}{d_{\rho}}\phi(x).
\end{equation}  
Here, $\chi_{\rho}$ denotes the character of the irrep $\rho$ of $H$
and $d_{\rho}$ denotes its dimension. This equation for $(\rho,g)$ can
only be satisfied if $\frac{\chi_{\rho}(g_A)}{d_{\rho}}$ is a root of
unity. $\mathcal{T}$ is a transformation group algebra of the
form $F(H/K)\otimes \CC N$ if and only if this root of unity equals
$1$ for all $(\rho,g)$ which satisfy (\ref{inveq})
\end{proposition}

\noindent {\bf Proof:} $\mathcal{T}$ is by definition the
maximal Hopf subalgebra of $D(H)$ which leaves $\phi$
invariant. Therefore it is spanned by the matrix elements of a set of
irreps $(\rho_{i},g)$ of $D(H)^*$ which closes under conjugation and
tensor products (cf. theorem \ref{shth}). The requirement that the
matrix elements of the irrep $(\rho_{i},g)$ leave $\phi$ invariant is
just
\begin{equation}
\label{invareq}
(\rho_i)_{ab}(xg_{A}x^{-1})\phi(g^{-1}x)=\delta_{ab}\phi(x).
\end{equation}  
If we take the trace of the left and right hand side of this equation,
we obtain
\begin{equation}
\chi_{\rho_i}(g_{A})\phi(g^{-1}x)=d_{\rho_i}\phi(x).
\end{equation}  
Here, we have used the invariance of $\chi_{\rho_i}$ under conjugation
to remove the conjugation with $x$. This equation is equivalent
to (\ref{inveq}), so any solution to (\ref{invareq}) satisfies
(\ref{inveq}). The converse is also true. From (\ref{inveq}), we see
that $\phi$ has to be an eigenvector of the action of $g^{-1}$ with
eigenvalue $\frac{\chi_{\rho_i}(g_A)}{d_{\rho_i}}$. Since $g$ has
finite order, this implies that
$\frac{\chi_{\rho_i}(g_A)}{d_{\rho_i}}$ is a root of unity. This means
that $\rho_i(g_A)$ must be $\frac{\chi_{\rho_i}(g_A)}{d_{\rho_i}}$
times the unit matrix, since $\chi_{\rho_i}(g_A)$ is the sum of the
eigenvalues of $\rho_i(g_A)$, which are all roots of unity ($\rho_i$
is unitary). But if this holds, then $\rho_i(xg_Ax^{-1})$ is also
$\frac{\chi_{\rho_i}(g_A)}{d_{\rho_i}}$ times the unit matrix and
hence (\ref{invareq}) is satisfied. 

Using that the coproduct of $D(H)^{*}$ corresponds to the product of
$D(H)$ and also that $\Pi^{A}_{\alpha}$ and $\epsilon$ are algebra
homomorphisms, one may easily show that the set of irreps $(\rho_i,g)$
whose matrix elements solve (\ref{invareq}) (or \ref{inveq}) closes
under tensor products. It also clearly closes under
conjugation. Therefore, $\mathcal{T}$ is spanned by the matrix
elements of those irreps.

If, for all irreps $(\rho_i,g)$ whose matrix elements span
$\mathcal{T}$, we have $\frac{\chi_{\rho}(g_A)}{d_{\rho}}=1$, then
all the $\rho_{i}$ are paired up with the same set of elements
$g$ of $H$, namely those elements whose action leaves $\phi$
invariant. These elements form a subgroup $N_{\phi}$ of
$H$. IN this situation, $\mathcal{T}$ is the transformation group algebra
$F(H/K)\otimes N_{\phi}\subset D(H)$, where $K$ is the intersection of
the kernels of the $\rho_{i}$. If one of the roots of unity
$\frac{\chi_{\rho_{i}}(g_A)}{d_{\rho}}$ does not equal $1$, then the
representation $(\rho_{i},e)$ of $D(H)^*$ does not occur in
$\mathcal{T}$, but $(\rho_i,g)$ does, for some $g\neq e$. Hence
$\mathcal{T}$ cannot be a transformation group algebra of the form
$F(H/K)\otimes N$ in this case. $\halmos$

\section{Confinement}
\label{confsec}

\subsection{Confinement and Hopf quotients}
\label{genconfsec}

As we have seen, the formation of a condensate of particles in the
state $\phi$ breaks the Hopf symmetry $\MA$ of a theory down to
the Hopf stabilizer $\mathcal{T}\subset\MA$ of $\phi$. The particles
in the effective theory which has the condensate as its ground state
will thus carry irreducible representations of $\mathcal{T}$. However,
not all the particles in the effective theory will occur as free
particles; some will be confined. The intuition behind this is simple:
if a particle in the effective theory has non-trivial monodromy with
the condensate particles, then it will ``draw a string'' in the
condensate. That is, the condensate's order parameter has to have a
(half)line discontinuity as a consequence of the non-trivial parallel
transport around the location of the particle. This line discontinuity
corresponds physically to a domain wall and will cost a fixed amount
of energy per unit of length\footnote{Note that we have not specified
the Hamiltonian in our model, but we assume here that, behind the
scenes, there is a ``Higgs potential'' which causes the symmetry
breaking condensation. Such a potential will make strings cost
an amount of energy that increases linearly with their length.} and
hence it may not extend to infinity. As a consequence, single
particles that have non-trivial braiding with the condensate cannot
occur. On the other hand, configurations such as a particle and its
antiparticle connected by a (short) finite length string may occur and
we may compare these to the mesons of QCD. Similarly, one may have
baryon-like excitations, which are bound states of three or more
elementary excitations which do not match in pairs. Thus, we expect
all the irreps of the Hopf stabilizer of the condensate to occur as
particles in the broken theory, but some of them will occur as free
particles, while others will occur only as constituents of mesonic or
baryonic excitations.

There are some requirements which should hold for the set of
representations of $\mathcal{T}$ that do not get confined. Clearly,
this set should contain the vacuum representation or counit of
$T$. Also, it should be closed under tensor products and charge
conjugation; we would not want two non-confined particles to fuse to a
confined particle, and, given that a particle is not confined, we
would like the same to hold for its charge-conjugate. From these
conditions on the non-confined representations it follows, using the
results of section \ref{techsec}, that the matrix elements of the
representations of the non-confined irreps of $\mathcal{T}$ span a
Hopf-subalgebra of $\mathcal{T}^*$. We will call this subalgebra
$\mathcal{U}^*$. Again using the results in section \ref{techsec}, it
follows that the dual $\mathcal{U}$ of $\mathcal{U}^*$ will be a Hopf
algebra whose irreducible representations are exactly the
representations of $\mathcal{T}$ which are not confined (and whose
matrix elements span $\mathcal{U}^*$). The dual map of the embedding
of $\mathcal{U}^*$ into $\mathcal{T}^*$ is a Hopf map from
$\mathcal{T}$ onto $\mathcal{U}$ and therefore $\mathcal{U}$ is a Hopf
quotient of $\mathcal{T}$. This Hopf quotient $\mathcal{U}$ may be
seen as the symmetry which classifies the non-confined excitations of
the system. A schematic picture of the main symmetry algebras defined
in this paper and their relations may be found on page
\pageref{finalpic}.

To determine which irreps of $\mathcal{T}$ correspond to free
particles and which are confined, we need to have a notion of braiding
between a representation $\pi$ of the original Hopf algebra and a
representation $\rho$ of $\mathcal{T}$. Clearly, the braiding should
be derived from the $R$-matrix of $\mathcal{A}$. Let us write this as
$R=\sum_{k}R^{1}_{k}\otimes R^{2}_{k}$. Unfortunately, we cannot just
define the matrix for an exchange of a $\rho$ and a $\pi$ as
$\sigma\circ(\rho\otimes\pi)(R)$, since the $R^{1}_{k}$ are not
usually elements of $\mathcal{T}$. However, we can take the exchange
matrix to be $\sigma\circ((\rho\circ P)\otimes\pi)(R)$, where $P$ is
the orthogonal projection of $\mathcal{A}$ onto $\mathcal{T}$.  We
also define the braid matrix for the product $\pi\otimes\rho$ as
$\sigma\circ(\pi\otimes(\rho\circ P))(R)$. A representation $\rho$ of
$\mathcal{T}$ should now correspond to a free particle excitation if
these braiding matrices have trivial action on the product of the
condensate vector with an arbitrary vector in the module of
$\rho$. That is, for a non-confined representation $\rho$, we would
like to demand
\begin{eqnarray}
\label{braidcond}
\sum_{k}\rho(P(R^{1}_{k}))\otimes \pi(R^{2}_{k})\phi&=&
\rho(1)\otimes \phi \nonumber \\
\sum_{k}\pi(R^{1}_{k})\phi \otimes\rho(P(R^{2}_{k}))&=& 
\phi\otimes\rho(1).
\end{eqnarray}
This gives us a requirement on every matrix element of each of the
non-confined representations $\rho$. These matrix elements are of
course elements of $\mathcal{T}^*$ and we may in fact write down a
corresponding requirement for arbitrary elements of
$\mathcal{T}^*$. To do this, we first define a left and a right action
of $\mathcal{T}^*$ on the module $V_{\pi}$ of the representation $\pi$
of $\mathcal{A}$. We take
\begin{eqnarray}
\label{actatt}
f\cdot v&:=& \sum_{k}f(P(R^{1}_{k}))\pi(R^{2}_{k}) v \nonumber \\
v\cdot f&:=& \sum_{k}f(P(R^{2}_{k}))\pi(R^{1}_{k}) v,
\end{eqnarray}
where $f\in \mathcal{T}^*$ and $v\in V_{\pi}$.
\begin{proposition}
Let $\star$ denote the multiplication on $\mathcal{T}^*$. Then
\begin{eqnarray}
(f_1 \star f_2)\cdot v&=&f_1\cdot(f_2\cdot v) \nonumber\\
v\cdot(f_1\star f_2)&=&(v\cdot f_1)\cdot f_2.
\end{eqnarray}
\end{proposition}

\noindent{Proof:}  We have
\begin{eqnarray}
(f_1\star f_2)\cdot v &=& 
(f_1 \otimes f_2 \otimes \pi) \circ
(\Delta\otimes {\rm id})\circ (P\otimes {\rm id})(R) v \nonumber \\
~&=& (f_1 \otimes f_2 \otimes \pi) \circ (P\otimes P\otimes {\rm id})\circ
(\Delta\otimes {\rm id})(R) v \nonumber \\ 
~&=& (f_1 \otimes f_2 \otimes \pi) \circ (P\otimes P\otimes {\rm id})
(R_{13}R_{23}) v \nonumber \\ 
~&=& f_1(P(R^{1}_{l}))
f_2(P(R^{1}_{k}))\pi(R^{2}_{l}R^{2}_{k}) 
\nonumber \\
~&=& f_1(P(R^{1}_{l}))\pi(R^{2}_{l})
f_2(P(R^{1}_{k}))\pi(R^{2}_{k}) v 
\nonumber \\ 
~&=& f_1\cdot(f_2 \cdot v).
\end{eqnarray}
In the fourth equality, we used that $\pi$ is a representation of
$\mathcal{A}$.  In the third equality, we used $(\Delta\otimes {\rm
id})(R)=R_{13}R_{23}$. In the second equality, we used the fact that
the orthogonal projection $P$ commutes with the comultiplication, that
is
\begin{equation}
(P\otimes P)\circ\Delta = \Delta\circ P.
\end{equation}
One may see that this holds by evaluating both sides on the basis of
$\mathcal{A}$ that is given by the matrix elements of the irreps of
$\mathcal{A}^*$. Using this property of $P$ and $({\rm
id}\otimes\Delta)(R)=R_{13}R_{12}$, one may similarly prove that
$v\cdot (f_1\star f_2)=(v\cdot f_1)\cdot f_2$. $\halmos$

\noindent The requirements (\ref{braidcond}) on the matrix elements of $\rho$
may now be generalized to
\begin{equation}
\label{ustarcond}
f\cdot \phi = \phi\cdot f=f(1)\phi=\epsilon^{*}(f)\phi.
\end{equation}
Hence, the requirement that the representation $\rho$ has trivial
braiding with the condensate becomes the requirement that the left and
right action of the matrix elements of $\rho$, as defined above, leave
the condensate invariant (in the sense of definition
\ref{invardef}). Thus, we may say that passing from the unbroken
symmetry $\mathcal{T}$ to the unconfined symmetry $\mathcal{U}$ is
equivalent to breaking the dual $\mathcal{T}^*$ down to
$\mathcal{U}^*$ by the condensate $\phi$. If we take this point of
view, then the fact that we are talking about braiding is hidden in
the definition of the actions above.
 
Unfortunately, it turns out that (\ref{ustarcond}) does not always
have solutions. In particular, the counit $\epsilon_{\mathcal{T}}$ of
$\mathcal{T}$ does not always solve (\ref{ustarcond}) (or
(\ref{braidcond})). This is linked to the fact that the ``action'' we
defined above preserves the multiplication, but not necessarily the
unit $\epsilon_{\mathcal{T}}$ of $\mathcal{T}^*$.

Therefore, to ensure that $\mathcal{U}^*$ contains at least the
``vacuum representation'' $\epsilon_{\mathcal{T}}$ of $\mathcal{T}^*$,
we change the condition (\ref{ustarcond}) to
\begin{eqnarray}
\label{modustarcond}
f\cdot \phi &=& f(1) \epsilon\cdot \phi \nonumber \\
\phi\cdot f &=& f(1) \phi\cdot \epsilon.
\end{eqnarray}
In other words, we no longer demand that the elements of
$\mathcal{U}^*$ leave the condensate invariant, but instead, we ask
that they act on the condensate in the same way as $\epsilon$. To put
it yet another way, we say that the representations of $\mathcal{T}$
that are not confined are those representations that have the same
braiding with the condensate as the vacuum representation. Note that,
since $\epsilon_\mathcal{T}=1_{\mathcal{T}^*}$, we may also write the
above condition as
\begin{eqnarray}
f\cdot (\epsilon \cdot \phi) &=& 
\epsilon^{*}(f) (\epsilon\cdot \phi) \nonumber \\
(\phi\cdot \epsilon) \cdot f &=& 
\epsilon^{*}(f) (\phi\cdot \epsilon).
\end{eqnarray}
We may thus still see confinement as a dual symmetry breaking, but now
$\mathcal{T}^*$ is not broken to $\mathcal{U}^*$ by the original
condensate $\phi$, but by the vectors $\epsilon\cdot\phi$ and
$\phi\cdot\epsilon$. Clearly, when $\epsilon\cdot\phi=\phi$ and
$\phi\cdot\epsilon=\phi$, the new condition on elements of
$\mathcal{U}^*$ reduces to (\ref{ustarcond}).

We have now defined an algebra $\mathcal{U}$ (through its dual
$\mathcal{U}^*$) whose representations should classify the
non-confined excitations over the condensate. There should be an
action of the braid group on the Hilbert space for a number of such
excitations. Therefore, we should like $\mathcal{U}$ to be
quasitriangular with an $R$-matrix related to the $R$-matrix of the
original Hopf algebra $\mathcal{A}$ (for example $(P\otimes
P)(R_{\mathcal{A}})$). As we will see in the examples, the conditions
(\ref{modustarcond}) are often enough to ensure that $\mathcal{U}$ has
such a quasitriangular structure. Nevertheless, it does not always
seem to the case (this will be made somewhat clearer in section
\ref{purecondsec}). Therefore we expect that the requirements
(\ref{modustarcond}) will in general have to be supplemented by some
extra condition and the non-confined algebra could then be smaller
than the algebra $\mathcal{U}$ defined here.

\subsection{Domain walls and Hopf kernels}
\label{wallsec}

In the previous section, we spoke of confined particles pulling
strings in the condensate. These were line discontinuities in the
condensate's wave function, induced by the non-trivial parallel
transport around the confined particle. Evidently, the internal state
of the condensate particles on one side of such a string will differ
from that on the other side. Therefore, we may also view these strings
as domain walls between regions with different condensates which
exhibit the same symmetry breaking pattern\footnote{One may in
principle also have domain walls between condensates with different
symmetry breaking patterns, but this requires the parameters which
govern the system (the symmetry breaking potential) to vary as one
crosses the walls, breaking translational symmetry at the level of the
Lagrangian or Hamiltonian}. We would like to classify such
walls. Clearly, a wall is uniquely determined by the confined
particles on which it may end, or in other words, by a representation
of the residual algebra $\mathcal{T}$ that does not correspond to a
representation of its non-confined quotient $\mathcal{U}$. However,
there may be several irreps of $\mathcal{T}$ that cause the same
parallel transport in the condensate and these will all pull the same
string (or wall). In fact, let $\rho$ be an irrep of $\mathcal{T}$ and
let $\tau$ be a non-confined irrep of $\mathcal{T}$, then any irrep of
$\mathcal{T}$ in the tensor product representation
$(\rho\otimes\tau)\circ\Delta$ will pull the same string as $\rho$,
since the non-confined irrep $\tau$ has trivial braiding with the
condensate. In short, we may say that walls are unaffected by fusion
with non-confined particles.

In view of the above, we expect that the wall that corresponds to a
$\mathcal{T}$-representation $\rho$ is already determined by the
restriction of $\rho$ to a subalgebra $\mathcal{W}$ of
$\mathcal{T}$. This subalgebra should be such that, if $\tau$ is a
non-confined irrep of $\mathcal{T}$, then the restriction to
$\mathcal{W}$ of the tensor product representation
$(\rho\otimes\tau)\circ\Delta$ should be isomorphic to a direct sum of
copies of the restriction of $\rho$ to $\mathcal{W}$ (the number of
copies being the dimension of $\tau$). Now it turns out that such a
$\mathcal{W}\subset \mathcal{T}$ exists, and in fact, there are two logical
options. Denote the Hopf map from the residual symmetry algebra
$\mathcal{T}$ onto the non-confined algebra $\mathcal{U}$ by
$\Gamma$. The {\em left Hopf kernel} $\lker(\gamma)$ of $\Gamma$ is
then the subset of $\mathcal{T}$ defined as
\begin{equation}
\lker(\Gamma):=\{t\in \mathcal{T}\,|\,(\Gamma\otimes{\rm
id})\circ\Delta(t)=1_{\mathcal{U}}\otimes t\}
\end{equation}
and similarly, the {\em right Hopf kernel} of $\Gamma$ is defined as 
\begin{equation}
\rker(\Gamma):=\{t\in \mathcal{T}\,|\,({\rm
id}\otimes\Gamma)\circ\Delta(t)=t\otimes 1_{\mathcal{U}}\}.
\end{equation}
These are our two candidates for $\mathcal{W}$. One may check that the
left Hopf kernel is a right coideal subalgebra (that is,
$\lker(\Gamma)$ is a subalgebra and $\Delta(\lker(\Gamma))\subset
\mathcal{T}\otimes \lker(\Gamma)$) and similarly that the right Hopf
kernel is a left coideal subalgebra.  Moreover, one has
$S(\rker(\Gamma))=\lker(\Gamma)$ and
$S(\lker(\Gamma))=\rker(\Gamma)$. Thus, $\lker(\Gamma)$ is a Hopf
subalgebra of $\mathcal{T}$ exactly if
$\lker(\Gamma)=\rker(\Gamma)$. In the examples we will meet, this will
not usually be the case. Note that, even if
$\lker(\Gamma)\neq\rker(\Gamma)$, the representations of
$\lker(\Gamma)$ and $\rker(\Gamma)$ are in one to one correspondence:
the representation $\rho$ of $\lker(\Gamma)$ corresponds to the
representation $\bar{\rho}:=\rho^{t}\circ S$ of $\rker(\Gamma)$. If
$\lker(\Gamma)$ and $\rker(\Gamma)$ are semisimple algebras (which we
will assume), then this isomorphism of representations induces an
isomorphism of algebras and so $\lker(\Gamma) \cong\rker(\Gamma)$. In
other words: our candidates for $\mathcal{W}$ are isomorphic and it
does not really matter which one we take.
\begin{conjecture}
The wall corresponding to the $\mathcal{T}$-irrep $\Omega^{B}_{\beta}$ is
charcterized by the restriction of $\Omega^{B}_{\beta}$ to either
$\lker(\Gamma)$ or $\rker(\Gamma)$.
\end{conjecture}
We provide the following evidence for this conjecture

\vspace{1mm}
\noindent{\bf 1.} If $\rho$ is a representation of $\lker(\Gamma)$ and
$\tau$ is a representation of $\mathcal{T}$, then we define the tensor
product as $(\tau\otimes\rho)\circ\Delta$. This is well-defined, since
$\lker(\Gamma)$ is a right coideal of $\mathcal{T}$. It is also
clearly a representation of $\lker(\Gamma)$. Now suppose that $\tau$
corresponds to a representation $\tilde{\tau}$ of $\mathcal{U}$, that
is, $\tau=\tilde{\tau}\circ\Gamma$. Then, the defining property of
$\lker(\Gamma)$ guarantees that we have
\begin{equation}
(\tau\otimes\rho)\circ(\Gamma\otimes{\rm id})\circ\Delta(t) =
\tau(1)\otimes\rho(t)
\end{equation}
for all $t\in \lker(\Gamma)$. In other words, $\lker(\Gamma)$ is
indeed defined in such a way that its representations are not affected
by fusion with representations of the non-confined algebra
$\mathcal{U}$, just as walls are not affected by fusion with
non-confined particles. Clearly, we may also define the tensor product
$(\bar{\rho}\otimes\tau)\circ\Delta$ of an $\rker(\Gamma)$
representation $\bar{\rho}$ with a representation $\tau$ of
$\mathcal{T}$ and again, the fusion will be trivial if $\tau$
corresponds to a representation of $\mathcal{U}$.

\vspace{1mm}
\noindent {\bf 2.} Every representation of $\mathcal{T}$ corresponds to a representation of
$\lker(\Gamma)$ by restriction. In particular, if $\rho$ is an irrep
of $\mathcal{T}$ which factors over $\Gamma$, that is $\rho=\tau\circ\Gamma$,
with $\tau$ a representation of $\mathcal{U}$, then we have, for all $t\in
\lker(\Gamma)$:
\begin{eqnarray}
\rho(t)&=&\tau\circ\Gamma(t) \nonumber \\
~&=&
\tau\circ\Gamma\circ ({\rm id}\otimes \epsilon)\circ\Delta
(t)\nonumber \\
~&=&
\tau\circ ({\rm id}\otimes \epsilon)\circ 
(\Gamma\otimes{\rm id}) \circ\Delta (t)= \epsilon(t)\tau(1).
\end{eqnarray}
Thus, the non-confined irreps of $\mathcal{T}$ all correspond to the trivial
representation of $\lker(\Gamma)$. This result is consistent with the
fact that the non-confined representations of $\mathcal{T}$ do not pull
strings. Again, a similar result holds for $\rker(\Gamma)$.

\vspace{1mm}
\noindent {\bf 3.}  If $B$ is a finite dimensional Hopf algebra, $C$ a
Hopf quotient of $B$ and $A$ the corresponding left Hopf kernel, then
it is known by a theorem of Schneider (theorem 2.2 in
\cite{schneider}, see also \cite{mont} for background) that $B$ is
isomorphic to a crossed product of $A$ and $C$ as an algebra and also
as a left $A$-module and as a right $C$-comodule. Such crossed
products are defined as follows:
\begin{definition} Let $C$ be a Hopf algebra, let $A$ be an algebra
and let $\sigma:C\otimes C\rightarrow A$ be a convolution-invertible
linear map. Also, suppose we have a linear map
from $B\otimes A$ to $A$, which we write as $b\otimes a\mapsto b\cdot
a$. We require that $A$ is a twisted $B$-module, that is, $1\cdot a=a$
for all $a$ and 
\begin{equation}
c \cdot(d\cdot a)=\sum \sigma(c_1,d_1)(c_2 d_2 \cdot a)\sigma^{-1}(c_3,d_3)
\end{equation}
Here we use Sweedler notation for the coproduct.
We also require  that $\sigma$ is a cocycle, that is
\begin{eqnarray}
\sigma(c,1)=\sigma(1,c)=\epsilon(c)1 \nonumber \\
\sum \left(c_1)\cdot \sigma(d_1,e_1)\right)\sigma(c_2,d_2e_2)=
\sum \sigma(c_1,d_1)\sigma(c_2 d_2,e)
\end{eqnarray}
and that $C$ measures $A$:

\begin{equation}
c\cdot 1= \epsilon(c) 1,~~~ c \cdot(ab)=\sum (c_1\cdot a)(c_2 \cdot b).
\end{equation}
Now the {\em crossed product algebra} $A\#_{\sigma}B$ is the vector space
$A\otimes B$ with the product given by
\begin{equation}
(a\otimes c)(b\otimes d)=\sum a (c_1 \cdot b)\sigma(c_2,d_1)\otimes
c_3 d_2.
\end{equation}
\end{definition}
These crossed products were introduced in \cite{bcm86,dt86}. An
accessible treatment may be found in \cite{mont} or \cite{majid}. When
it is given that $C$ measures $A$, one may show that the conditions
that involve $\sigma$ are equivalent to the associativity of the
product of $A\#_{\sigma}C$. Some elementary properties of the crossed
product are
\begin{itemize}
\item
$A$ is embedded into $A\#_{\sigma}C$ through $a\mapsto a\otimes 1$,
that is, we have \\ $(a\otimes 1)(b\otimes 1)=(ab\otimes 1)$.
\item
The map $j:C\rightarrow A\#_{\sigma}C$ given by $c\mapsto 1\otimes c$
is clearly a $C$-comodule morphism when $A\#_{\sigma}C$ and $C$ are
given the comodule structures ${\rm id}_{A}\otimes\Delta_{C}$ and
$\Delta_{C}$ respectively, but it is usually not an algebra morphism;
we have $(1\otimes c)(1\otimes d)=\sum (\sigma(c_1,d_1)\otimes c_2 d_2)$.
\item
When $\sigma$ is trivial, that is
$\sigma(c,d)=\epsilon(c)\epsilon(d)$, the cross product is just the
ordinary smash product; we have 
$(a\otimes b)(c\otimes d)=\sum a (c_1 \cdot b)\otimes
c_2 d_2$. In this case, $j$ is an algebra morphism.
\end{itemize}
Thus, we see that our residual algebra $\mathcal{T}$ is isomorphic to
the cross product $\lker(\Gamma)\#_{\sigma}\mathcal{U}$ for some
cocycle $\sigma$. This lends support to the idea that
$\mathcal{T}$-excitations are characterized by a wall, corresponding
to a representation of $\lker(\Gamma)$ and by further quantum numbers,
which can be associated to the non-confined algebra $\mathcal{U}$. If
the cross product was just the tensor product $\lker(\Gamma)\otimes
\mathcal{U}$, then these ``non confined quantum numbers'' would be
labels of $\mathcal{U}$-representations, but here, we cannot expect
this, because the actions of $\lker(\Gamma)$ and $\mathcal{U}$ on a
$\mathcal{T}$-module will not commute. In fact, $\mathcal{U}$ is
typically not even a subalgebra of $\mathcal{T}$. Therefore, finding
the quantum numbers associated to $\mathcal{U}$ for the general case
is a non-trivial task, which we postpone to future work\footnote{Note
that much more than we have written here is known when $\lker(\Gamma)$
is a Hopf algebra. For such results, see for instance
\cite{maj90,andev95,andrus96}}.

\subsection{Confinement for transformation group algebras}
\label{trafoconf}

Suppose the $D(H)$ symmetry of a discrete gauge theory has been broken
by a condensate $\phi\in V^{A}_{\alpha}$ and the residual symmetry
algebra $\mathcal{T}$ is a transformation group algebra of the kind
referred to in section \ref{doubbreaksec}. The explicit definition is
\begin{equation}
\label{Ttgafrm}
\mathcal{T}=\left\{ F\in D(H)|F(xk,y)=F(x,y)1_{N}(y)~ (\forall k \in
K)\right\},
\end{equation}
where $N$ is a subgroup of $H$ and $K$ is a normal subgroup of
$H$. Such algebras will arise frequently in our examples.  Here, we
investigate which representations of such a $\mathcal{T}$ are confined
and which are not. In particular, we will find that there is a set of
non-confined representations of $\mathcal{T}$ such that the irreps in
this set are in one to one correspondence with those of $D(N/(N\cap
K))$.

First, we find some properties of the condensate vector $\phi$. If
$\mathcal{T}$ is of the form given above then the invariance of $\phi$
under elements of the form $1\otimes n$ (with $n\in N$) and $f\otimes
e$ implies that we have
\begin{eqnarray}
\label{phiprops}
(\forall n\in N):&~& \phi(nx)=\phi(x) \nonumber \\
(\forall (f\otimes e) \in \mathcal{T}):&~&f(x g_A x^{-1})\phi(x)=f(e)\phi(x).
\end{eqnarray}
The second equation and $\phi\neq 0$ imply that there is an $x \in H$
such that $f(xg_A x^{-1})=f(e)$ for all $f$ that are constant on
$K$-cosets. As a consequence, we have $xg_A x^{-1}\in K$ and hence,
since $K$ is normal in $H$, we have $A\subset K$ and in particular
$g_A \in K$. 

Now let us write down an explicit formula for the orthogonal
projection $P$ of $D(H)$ onto $\mathcal{T}$:
\begin{equation}
P(F)(x,y)=\frac{1}{|K|}\int_{K}{\rm d}k\, F(xk,y)1_{N}(y).
\end{equation}
In the following, we will sometimes omit the characteristic function
of $N$ and just keep in mind that the projected function has support
in $N$. With the above formula, we can find $(P\otimes {\rm
id}) (R)$ and $({\rm id}\otimes P)(R)$ from the formula (\ref{rmatdh})
for the $R$-matrix of $D(H)$:
\begin{eqnarray}
(P\otimes {\rm id}) (R) (x_1,y_1,x_2,y_2) &=& 
\frac{1}{|K|}\int_{K}{\rm d}k\,\delta_{e}(x_1 k (y_2)^{-1})\delta_{e}(y_1) 
\nonumber \\
({\rm id}\otimes P)(R)(x_1,y_1,x_2,y_2) &=& 
\delta_{e}(x_1 (y_2)^{-1})\delta_{e}(y_1) 1_{N}(y_2).
\end{eqnarray}
Using these formulae, we can write down the left and right actions of
$\mathcal{T}^*$ on the condensate vector, as defined in equation
(\ref{actatt}). They are given by
\begin{eqnarray}
(\tau \cdot \phi) (x) &=& \int_{H}{\rm d}z\, (\tau(F_L))(z)\phi(z^{-1}x) 
\nonumber \\
(\phi\cdot \tau) (x) &=& (\tau(F_R))\phi(x), 
\end{eqnarray}
where $\tau$ is an arbitrary element of $\mathcal{T}^*$ and we have defined
\begin{eqnarray}
F_L(a,b\,;z)&=& \frac{1}{|K|}\int_{K}{\rm d}k\,
\delta_{e}(a k z^{-1})\delta_{e}(b)
\nonumber \\
F_R(a,b\,;x,z) &=& \delta_{e}(x g_A x^{-1} b^{-1}) 1_{N}(b).
\end{eqnarray}
$F_R$ and $F_L$ should be read as functions of $a$ and $b$ with
parameters $x$ and $z$.  We want to find the maximal Hopf subalgebra
of $\mathcal{T}^*$ for which the condition (\ref{modustarcond})
holds. This will be spanned (as a linear space) by the matrix elements
of a set of representations of $\mathcal{T}$. Since $\mathcal{T}$ is
isomorphic to a transformation group algebra, we know its
representations (see section \ref{irrepsec}). They are labeled by an
orbit $B$ of the action of $N$ on $H/K$ and by an irreducible
representation $\beta$ of the stabilizer $N_{B}\subset N$ of this
orbit. The matrix elements of the representation labeled by $B$ and
$\beta$ in the basis of formula (\ref{traforepbas}) can be read off
from formula (\ref{trafomatelts}), which in this case becomes
\begin{equation}
\label{helpmatelts}
\tau^{B}_{\beta}(F)^{i,j}_{\zeta,\eta}=
\int_{N_B}F(x_{\eta}g_{B}x^{-1}_{\eta},x_{\eta}n x_{\zeta}^{-1})
\beta_{i,j}(n)\,{\rm d}n.
\end{equation} 
In this formula, we have $x_{\eta},x_{\zeta}\in N$ as in
(\ref{trafomatelts}), while $g_B$ is an arbitrary element of the
$K$-coset $\xi_B$ that features in (\ref{trafomatelts}). Note that it
does not matter which element of this coset we take, since the
function $F$ in the integrand is constant on $K$-cosets in its left
argument. 
\begin{proposition}
\label{calcprop}
The requirements (\ref{modustarcond}) which determine which of the
irreps $\tau^{B}_{\beta}$ of $\mathcal{T}$ are not confined, reduce to
\begin{eqnarray}
\label{tgacond1}
(\forall \eta\in \mathcal{O}_{B}):~
\frac{1}{|K|}\int_{K}{\rm d}k\,\phi(k x_{\eta}g_{B}^{-1} x^{-1}_{\eta}x)
=\frac{1}{|K|}\int_{K}{\rm d}k\, \phi(kx) \\
\label{tgacond2}
(g_A \not\in N) \vee \left( \forall x \in {\rm supp}(\phi),\,\forall \eta
\in \mathcal{O}_{B}:~ \beta(x_{\eta}^{-1} x g_{A}x^{-1}x_{\eta})=I\right).
\end{eqnarray}
Here, $\mathcal{O}_{B}$ is the orbit of $\xi_{B}$ in $H/K$ and
$I$ is the unit matrix in the module of $\beta$.
\end{proposition}
{\bf Proof:} Substituting (\ref{helpmatelts}) into to the formulae for
the actions above, we find for the left action
\begin{eqnarray}
((\tau^{B}_{\beta})^{i,j}_{\zeta,\eta} \cdot \phi) (x) &=& 
\frac{1}{|K|}\int_{H}{\rm d}z\,\int_{N_B}{\rm d}n\, 
\int_{K}{\rm d}k\,\delta_{e}(x_{\eta}g_{B}x^{-1}_{\eta} k z^{-1})
\delta_{e}(x_{\eta}n x_{\zeta}^{-1})\beta_{i,j}(n)\phi(z^{-1}x) 
\nonumber \\
~&=&
\frac{1}{|K|}\int_{N_B}{\rm d}n\, 
\int_{K}{\rm d}k\,\delta_{e}(x_{\eta}n x_{\zeta}^{-1})
\beta_{i,j}(n)\phi(k^{-1} x_{\eta}g_{B}^{-1} x^{-1}_{\eta} x) 
\nonumber \\
~&=&
1_{N_B}(x_{\eta}^{-1} x_{\zeta})\beta_{i,j}(x_{\eta}^{-1} x_{\zeta})
\frac{1}{|K|}\int_{K}{\rm d}k\,\phi(k x_{\eta}g_{B}^{-1} x^{-1}_{\eta} x)
\end{eqnarray} 
and similarly, for the right action
\begin{eqnarray}
(\phi\cdot (\tau^{B}_{\beta})^{i,j}_{\zeta,\eta}) (x) &=& 
\int_{N_B}{\rm d}n\,\delta_{e}(xg_{A}x^{-1} x_{\zeta} n^{-1} x_{\eta}^{-1})
\beta_{i,j}(n)\phi(x) 
\nonumber \\
~&=&
1_{N_B}(x_{\eta}^{-1} x g_{A}x^{-1} x_{\zeta})
\beta_{i,j}(x_{\eta}^{-1} x g_{A}x^{-1} x_{\zeta})\phi(x).
\end{eqnarray} 
As a special case, we can find the left and right action of the
counit $\epsilon \in \mathcal{T}^*$, which corresponds to the one-dimensional
representation $\tau^{[e]}_{1}$. We have
\begin{eqnarray}
(\epsilon\cdot \phi)(x) &=& \frac{1}{|K|}\int_{K}{\rm d}k\, \phi(kx) 
\nonumber \\
(\phi\cdot \epsilon)(x) &=&  1_N(g_A) \phi(x).
\end{eqnarray} 
The final ingredient we need in order to write down the
requirements (\ref{modustarcond}) for the matrix elements, is the
value of $(\tau^{B}_{\beta})^{i,j}_{\zeta,\eta}$ in
$1_\mathcal{T}=1_{D(H)}$. This is given by
\begin{equation}
(\tau^{B}_{\beta})^{i,j}_{\zeta,\eta}(1_{D(H)})=
1_{N_{B}}(x_{\eta}^{-1}x_{\zeta})\beta_{ij}(x_{\eta}^{-1}x_{\zeta}).
\end{equation}
Thus, the conditions (\ref{modustarcond}) that the matrix elements of
$\tau^{B}_{\beta}$ have to fulfill, in order for $\tau^{B}_{\beta}$
not to be confined, become
\begin{equation}
\label{vrlcond1}
1_{N_B}(x_{\eta}^{-1} x_{\zeta})\beta_{i,j}(x_{\eta}^{-1} x_{\zeta})
\frac{1}{|K|}\int_{K}{\rm d}k\,\phi(k x_{\eta}g_{B}^{-1} x^{-1}_{\eta}x)
=
1_{N_{B}}(x_{\eta}^{-1}x_{\zeta})\beta_{ij}(x_{\eta}^{-1}x_{\zeta})
\frac{1}{|K|}\int_{K}{\rm d}k\, \phi(kx) 
\end{equation}
and
\begin{equation}
\label{vrlcond2}
1_{N_B}(x_{\eta}^{-1} x g_{A}x^{-1} x_{\zeta})
\beta_{i,j}(x_{\eta}^{-1} x g_{A}x^{-1} x_{\zeta})\phi(x) 
=
1_N(g_A)1_{N_{B}}(x_{\eta}^{-1}x_{\zeta})
\beta_{ij}(x_{\eta}^{-1}x_{\zeta})\phi(x).
\end{equation}
In the special case where $\eta=\zeta$ and $i=j$, the condition
(\ref{vrlcond1}) reduces to (\ref{tgacond1}). On the other hand, if
(\ref{tgacond1}) holds, then (\ref{vrlcond1}) will also hold for
general $(\eta,\zeta,i,j)$ and hence (\ref{tgacond1}) is equivalent to
(\ref{vrlcond1}). The condition (\ref{vrlcond2}) is trivially
satisfied when $g_A$ is not contained in $N$ (this is the first
alternative in (\ref{tgacond2})). When $g_A$ is an element of $N$ (and
hence of $N\cap K$, using (\ref{phiprops})), it may also be
simplified; in the special case where $\eta=\zeta$, (\ref{vrlcond2})
reduces to
\begin{equation}
1_{N_B}(x_{\eta}^{-1} x g_{A}x^{-1}x_{\eta})
\beta_{i,j}(x_{\eta}^{-1}x g_{A}x^{-1}x_{\eta})\phi(x) 
=\delta_{ij}\phi(x).
\end{equation}
Now since $g_A\in N\cap K$, it follows that $x_{\eta}^{-1} x
g_{A}x^{-1}x_{\eta}\in N\cap K$. But $N\cap K$ acts trivially on $H/K$
and hence $N\cap K \subset N_B$ for any $B$. Hence the condition above
reduces to the second alternative in (\ref{tgacond2}). In the other
direction, it is not difficult to see that (\ref{vrlcond2}) will be
satisfied for general $(\eta,\zeta,i,j)$if (\ref{tgacond2}) is
satisfied. Thus, we see that (\ref{vrlcond2}) is equivalent to
(\ref{tgacond2}). $\halmos$

\noindent Proposition \ref{calcprop} indicates how far we can go
towards the general solution of (\ref{modustarcond}) without
specifying the condensate vector $\phi$. The following proposition
describes a set of solutions that is present for any $\phi$, but
that is not always the full set of solutions.
\begin{proposition}
\label{uprop}
Independently of the condensate vector $\phi$, there is a set of
unconfined irreps of $\mathcal{T}$ which closes under conjugation and
tensor products. The corresponding Hopf quotient of $\mathcal{T}$ is
isomorphic to the quantum double of the group $N/(N\cap K)$. This
quantum double may be realized naturally on the space of functions on
$N\times N$ which are constant on $(N\cap K)$-cosets in both
arguments. The Hopf surjection $\Gamma:\mathcal{T}\rightarrow
D(N/(N\cap K))$ is then given by
\begin{equation}
\Gamma(f)(x,y)=\int_{N\cap K}f(x,yk) dk.
\end{equation}
\end{proposition}
{\bf Proof:} First, we find our set of irreps. Note that the left hand
side of (\ref{tgacond1}) is $\frac{1}{|K|}$ times the sum of the
values of $\phi$ over the $K$-coset
$x_{\eta}g_{B}^{-1}x_{\eta}^{-1}xK$. Similarly, the right hand side
involves a sum over $xK$. Using the fact that $\phi(nx)=\phi(x)$ for
any $n\in N$ (cf. (\ref{phiprops})), we see that (\ref{tgacond1}) will
be satisfied if $x_{\eta}g_{B}^{-1}x_{\eta}^{-1}xK=nxK$ for some $n\in
N$, or equivalently if there is an $n\in N$ such that $g_B
K=nK$. Furthermore, (\ref{tgacond2}) is clearly satisfied for all
$\beta$ that are trivial on $N\cap K\subset N_B$. Thus, the irreps
$\tau^{B}_{\beta}$ of $\mathcal{T}$ for which $g_B K=nK$ and
$\beta|_{N\cap K}=1$ are never confined.

Second, we show that these irreps are in one-to one correspondence with
the irreps of $D(N/(N\cap K))$. To see this, first note that, for
$n_{1},n_{2}\in N$, we have
\begin{equation}
n_{1}K=n_{2}K \Leftrightarrow 
n_{1}(K\cap N)=n_{2}(K\cap N).
\end{equation}
In fact, let $\bar{N}$ denote the subgroup of $H/K$ which consists of
the classes $nK$ with $n\in N$, then this correspondence is an
isomorphism between $\bar{N}$ and $N/(N\cap K)$. It follows that the
$N$-orbits in $H/K$ whose elements lie in $\bar{N}$ are in one to one
correspondence with the conjugacy classes of $N/(K\cap N)$. Now fix an
arbitrary such orbit $B\subset \bar{N}$. The irreps $\beta$ of the
stabilizer $N_B\subset N$ of this orbit which are trivial on $K\cap N$
are in one to one correspondence with the irreps of $N_B/(K\cap
N)$. But $N_B/(K\cap N)$ is exactly the centralizer of the conjugacy
class of $N/(K\cap N)$ that corresponds to $B$.  Hence the
non-confined irreps of $T_A$ are labeled by a conjugacy class of
$N/(K\cap N)$ and an irrep of the centralizer of this class in
$N/(K\cap N)$. But this means that they are in one to one
correspondence with the irreps of $D(N/(K\cap N))$.

Now let us have a closer look at the map $\Gamma:\mathcal{T}\rightarrow
D(N/(N\cap K))$. For convenience, we will realize $D(N/(N\cap K))$ on
the space of functions on $N\times N$ which are constant on $(N\cap
K)$-cosets in both arguments. The isomorphism with the usual
formulation in terms of functions on $N/(N\cap K)$ is taken as
follows. Let $\bar{f}\in F(N/(N\cap K)\times N/(N\cap K))$ then
$\bar{f}$ corresponds to the function $f \in F(N \times N)$ given by
$f(x,y)=\bar{f}(x(N\cap K),y(N\cap K))$. The demand that this
identification is an isomorphism fixes the Hopf algebra structure on
$F(N\times N)$. For example, the product of two functions on $N\times
N$ may now be written as
\begin{equation}
f_1 \bullet f_2 (x,y)= \frac{1}{|N\cap K|}\int_{N}
f_1(x,z)f_2(z^{-1}xz,z^{-1}y) dz. 
\end{equation}
It is straightforward to prove that $\Gamma$, as defined above, is
indeed a Hopf homomorphism. For example, to see that $\Gamma$
preserves the product, we write
\begin{eqnarray}
\Gamma(f_1)\bullet\Gamma(f_2)(x,y)&=&
\frac{1}{|N\cap K|}\int_{N} dz \int_{N\cap K} dk_1\int_{N\cap K} dk_2\, 
f_1(x,zk_1)f_2(z^{-1}xz,z^{-1}yk_2) \nonumber \\
~&=&
\frac{1}{|N\cap K|}\int_{N} dz \int_{N\cap K} dk_1\int_{N\cap K} dk_2\, 
f_1(x,z)f_2(z^{-1}xz,k_1 z^{-1}yk_2) \nonumber \\
~&=&
\int_{N} dz \int_{N\cap K} dk_2\, f_1(x,z)f_2(z^{-1}xz,z^{-1}yk_2)=
\Gamma(f_1\bullet f_2)(x,y).
\end{eqnarray}
In going from the first to the second line, we used the invariance of
the integral over $N$ and the fact that $f_2$ is constant on
$K$-cosets in its left argument. In going from the second to the third
line, we used the invariance of the $k_2$-integral to remove the
$k_1$-dependence from the integrand and subsequently removed the
$k_1$-integral. The rest of the proof that $\Gamma$ is a Hopf algebra
morphism is similar. It is also easy to see that $\Gamma$ is
surjective.

To complete the proof of the proposition, we need to show that the set
of unconfined irreps we have found consists precisely of those irreps
of $\mathcal{T}$ that factor over $\Gamma$. Thus, let
$\Pi^{\tilde{B}}_{\tilde{\beta}}$ be an irrep of $D(N/N\cap K)$ and
let $\chi^{\tilde{B}}_{\tilde{\beta}}$ be its character, as given in
(\ref{trfc}). Then we have 
\begin{equation}
\chi^{\tilde{B}}_{\tilde{\beta}}(\Gamma(f))= \int_{\tilde{B}} d\zeta
\int_{N_{\tilde{B}}} dn \int_{N\cap K} dk\, f(x_{\zeta}
g_{\tilde{B}}x_{\zeta}^{-1},x_{\zeta}n
x_{\zeta}^{-1}k)\chi_{\tilde{\beta}}(n).
\end{equation} 
Here, we have abused notation slightly: in stead of elements of
$N/(N\cap K)$ one should read representatives of these elements in $N$
where appropriate. It should be clear that the choice of
representatives does not affect the result. We may now change the sum
over the conjugacy class $\tilde{B} \subset N/(N\cap K)$ into a sum
over the corresponding $N$-orbit $B\subset \bar{N}\subset
H/K$. Similarly, we may change the sums over $N_{\tilde{B}}$ and
$N\cap K$ into one sum over $N_{B} \subset N$. This yields
\begin{equation}
\chi^{\tilde{B}}_{\tilde{\beta}}(\Gamma(f))=
\int_{B} d\zeta \int_{N_{B}} dn \,
f(x_{\zeta} g_{B}x_{\zeta}^{-1},x_{\zeta}n x_{\zeta}^{-1})\chi_{\beta}(n),
\end{equation} 
where $\beta$ is the irrep of $N_{B}$ that corresponds to the irrep
$\tilde{\beta}$ of $N_{\tilde{B}}$ (of course, $\beta$ is trivial on
$N\cap K$). The expression above is just the value on $f$ of the
character of the irrep $\tau^{B}_{\beta}$ of
$\mathcal{T}$. $\tau^{B}_{\beta}$ indeed belongs to our set of
unconfined irreps and from the one to one correspondence between
irreps of $D(N/(K\cap N))$ and irreps in our unconfined set, we see
that must get all irreps in the unconfined set in this way. Thus, the
set of unconfined irreps of $\mathcal{T}$ that we have found
corresponds precisely to the set of irreps of $\mathcal{T}$ that
factor over $\Gamma$ and the proposition follows. $\halmos$

\noindent Note that the $R$-matrix and ribbon-element of $D(N/(K\cap N))$
provide the set of non-confined irreps that we have found above with a
well defined braiding and spin. It is not clear that we will have such
properties for the full set of solutions to (\ref{tgacond1}) and
(\ref{tgacond2}). Therefore, we expect that the physically relevant
set of solutions to these equations is the one given in the
proposition above.This issue will not be very important in the sequel,
since the set of solutions in the proposition is actually complete in
all our examples.
\begin{proposition}
\label{kerprop}
The left and right Hopf kernels of $\Gamma$ are given by
\begin{eqnarray}
\label{tgalker}
\lker(\Gamma)=\{f \in \mathcal{T}|(\forall x_1\in N):\, f(x_1 x_2,y)=f(x_2,y)  
\wedge (\forall y \not\in N\cap K):\,f(x,y)=0 \}
\\
\label{tgarker}
\rker(\Gamma)=\{f \in \mathcal{T}|(\forall x_2\in N):\, f(x_1 x_2,y)=f(x_1,y)  
\wedge (\forall y \not\in N\cap K):\,f(x,y)=0 \}
\end{eqnarray}
\end{proposition}
{\bf Proof:}
We have 
\begin{eqnarray}
\Gamma\otimes{\rm id}(\Delta(f))(x_1,y_1,x_2,y_2)&=&
\int_{N\cap K}f(x_1 x_2,y_1 k)\delta_{e}(y_1 k y_2^{-1}) \nonumber \\
(1_{D(N/(N\cap K))}\otimes f)(x_1,y_1,x_2,y_2)&=&
1_{N\cap K}(y_1)f(x_2,y_2)
\end{eqnarray}
and the left Hopf kernel of $\Gamma$ consists of those functions $f$ for
which the right hand sides of these equations are equal:
\begin{equation}
\label{kernelreq}
\int_{N\cap K}f(x_1 x_2,y_1 k)\delta_{e}(y_1 k y_2^{-1})=
1_{N\cap K}(y_1)f(x_2,y_2).
\end{equation}
Now if we take $y_1=y_2$, then this requirement reduces to
\begin{equation}
f(x_1 x_2,y_1)=1_{N\cap K}(y_1)f(x_2,y_1),
\end{equation}
{}from which we see that $f(x,y)$ equals zero for all $x \in H$ when $y$
is not an element of $N\cap K$, while for $y \in N\cap K$, we have
$f(x_1 x_2,y)=f(x_2,y)$ for all $x_1 \in N$. On the other hand, all
$f$ which satisfy these requirements automatically satisfy
(\ref{kernelreq}). One may see this by noting that both the left hand
side and the right hand side of (\ref{kernelreq}) can be non-zero only
if both $y_1$ and $y_2$ are elements of $K\cap N$, in which case left
hand side and right hand side are equal. The formula for
$\lker(\Gamma)$ now follows. The proof of the expression for
$\rker(\Gamma)$ is similar and we leave it to the reader. $\halmos$

\noindent 
If we once again let $\bar{N}$ be the subgroup of $H/K$ that consists
of the cosets $nK$ of the elements of $N$, then we see that we have
the following 
\begin{corollary}
\label{kercorr}
As algebras:
\begin{eqnarray}
\lker(\Gamma)\cong F(\bar{N}\backslash (H/K))\otimes\CC(N\cap K) \nonumber \\
\rker(\Gamma)\cong F((H/K)/\bar{N})\otimes\CC(N\cap K)
\end{eqnarray}
\end{corollary}
{\bf Proof:} To see that the isomorphisms are algebra isomorphisms,
note that the elements of $1\otimes \CC(K\cap N)$ commute with those
of $F(\bar{N}\backslash (H/K))\otimes 1$ and $F((H/K)/\bar{N})\otimes
1$. This is because the elements of $F(H/K)\otimes 1$ already
commuted with those of $1\otimes\CC(K\cap N)$ in $\mathcal{T}$. $\halmos$

\noindent As a consequence of this corollary, each irreducible
representation of the left kernel is a product of an irrep of
$F(\bar{N}\backslash(H/K))$ and an irrep of $N\cap K$. The irreps of
$F(\bar{N}\backslash (H/K))$ are of course labeled by the elements of
$\bar{N}\backslash (H/K)$ and hence each irrep of $\lker(\Gamma)$ is
labeled by an element of $\bar{N}\backslash (H/K)$ and an irrep of
$N\cap K$. Similarly, each irrep of $\rker(\Gamma)$ is labeled by an
element of $(H/K)/\bar{N}$ and an irrep of $N\cap K$.

\section{Requirements on condensates}
\label{stabsec}

Before we turn to the study of explicit examples of symmetry breaking
and confinement, let us first motivate the choices of
condensate vectors that we will use in our examples.

Up to now we have assumed that one may form a condensate of any kind
of particle in the theory, in any internal state $\phi$. However, if
we want to have true Bose condensates, then we should demand that the
state $\phi$ has trivial self-braiding and also trivial
spin factor\footnote{In some applications, it could be more useful to think
of our condensate as a background of particles in the same {\em
internal} state, but not necessarily with the same external quantum
numbers. Then the restrictions we give here are not
necessary. Examples of ``condensates'' of particles with a non-trivial
spin factor would be the fractional quantum Hall ground states
proposed in \cite{haldane, halperin}}.
In other words:
\begin{itemize}
\item
The condensate must have trivial spin factor, i.e $\alpha(g_A)=I$.
\item
The condensate must have trivial self-braiding, i.e. 
\begin{equation} 
\sigma \circ \Pi^{A}_{\alpha}\otimes \Pi^{A}_{\alpha}(R) \phi\otimes\phi
=\phi\otimes\phi. 
\end{equation}
\end{itemize}
The examples that we will treat in the rest of this paper will all
have trivial spin and self-braiding.  The rest of this section is
devoted to finding out which kinds of electric, magnetic and dyonic
condensates will satisfy these requirements.

\vspace{1mm} For any purely {\em electric} condensates $\phi\in
V^{e}_{\alpha}$ (see section \ref{elcondsec}), the requirements are
both trivially satisfied.

\vspace{1mm}
A vector $\phi$ in a purely {\em magnetic} $D(H)$-module $V^{A}_{1}$ will
automatically have trivial spin, but may have non-trivial
self-braiding.  Nevertheless, there will always be at least two gauge
orbits of magnetic states with trivial self-braiding for every class
$A$ which has more than a single element. The first of these orbits
contains all the states with pure fluxes $hg_A h^{-1}$, which have
wave functions $1_{hN_A}$. We will study the corresponding condensates
in sections \ref{purecondsec}). The second orbit, which
will be studied in section \ref{magcondsec}, consists of the single
gauge invariant state which is the superposition of all these pure
fluxes. Its wave function is the function that sends all elements of
$H$ to $1$. Of course if $A$ has only a single element, then these
orbits coincide. Note that, when the orbits are different, they will
also have different symmetry breaking patterns. In particular, the
gauge singlet will leave the electric group unbroken, while the states
in the other orbit will not. To see that the states in the two orbits
we have mentioned do indeed have trivial self-braiding and to see if
there are more states with this property, we write down the
expression for the self-braiding of an arbitrary $f \in V^{A}_{1}$. We
have
\begin{eqnarray}
f\otimes f &:& (g,h)\mapsto f(g)f(h) \nonumber \\
\sigma \circ R\, (f\otimes f) &:&  (g,h)\mapsto f(g)f(gg^{-1}_{A}g^{-1}h)
\end{eqnarray}
and hence $f$ has trivial self-braiding precisely when
\begin{equation}
f(g)f(h)=f(g)f(gg^{-1}_{A}g^{-1}h) ~~~(\forall g,h\in H).
\end{equation}
One may readily check that the states we have already mentioned are
always solutions to this equation. Depending on $H$ and $A$, there may
also be extra solutions. For example, if $A\subset N_A$, then all $f$
are allowed, since in that case $hN_A=gg^{-1}_{A}g^{-1}hN_A$.

\vspace{1mm} {\em Dyons} (see section \ref{dyonsec})can have
non-trivial spin, but dyons with trivial spin also exist for many
groups $H$. In fact, given a magnetic flux $A$, there will be dyons
with flux $A\neq [e]$ and trivial spin factor precisely when $g_A$ is
contained in a proper normal subgroup of its centralizer $N_A$. For
Abelian $H$, this just means that the cyclic group generated by $g_A$
must be a proper subgroup of $H$. For non-Abelian $H$, one may note
that $g_A$ is contained in the center of $N_A$, which is a proper
normal subgroup if $N_A$ is still non-Abelian. If $N_A$ is Abelian,
then we have the requirement that the cyclic group generated by $g_A$
must be a proper subgroup of $N_A$. When $H$ is Abelian, the
requirement of trivial self-braiding is equivalent to that of trivial
spin and hence all the spinless dyons we have found may be
condensed. When $H$ is non-Abelian, this is not the case and the
requirement of trivial self-braiding then restricts the possibilities
further. In particular, using the ribbon property of $D(H)$, it gives
the necessary condition that two of the condensed dyons should be able
to fuse into a particle with trivial spin. In spite of this
restriction, there are still many non-Abelian groups $H$ which allow
for dyonic states with trivial self-braiding. One may for example show
that they occur for any non-Abelian $H$ with a non trivial center.

\section{Electric condensates}
\label{elcondsec}

\subsection{Symmetry breaking} 
\label{elbreaksec}

In this section, we study symmetry breaking by an electric condensate
$\phi\in \Pi^{e}_{\alpha}$. The first thing to do is to find the
residual symmetry algebra, which is the Hopf stabilizer of
$\phi$. This means finding all representations $(\rho,g)$ of $D(H)^*$
which solve equation (\ref{inveq}) in the special case where the flux
$A$ is trivial. In this case, we see immediately that all $\rho$ are
allowed. The requirement on $g$ is just that $\phi(g x)=\phi(x)$, or
equivalently, $\phi(g^{-1}x)=\phi(x)$, for all $x\in H$. Using the
invariance property of $\phi$, this reduces to
$\alpha(x^{-1}gx)\phi(x)=\phi(x)$ and using the invariance
property once again, we see that this reduces to the single
requirement
\begin{equation}
\alpha(g)\phi(e)=\phi(e).
\end{equation}
Thus, if we define $v:=\phi(e)\in V_{\alpha}$, then $g$ has to be an
element of the stabilizer $N_{v}$ of $v$. Since $\rho$ was
unrestricted, it follows that the residual symmetry algebra is the
Hopf subalgebra $\mathcal{T}_v(H)$ of the double which is
$F(H)\otimes\CC N_v$ as a vector space, or in terms of functions on
$H\times H$:
\begin{equation}
\label{tvdef}
\mathcal{T}_v(H) := \left\{ F\in D(H)|{\rm supp} (F) \subset H\times
N_v \right\}.
\end{equation}
$\mathcal{T}_v(H)$ is a transformation group algebra, with $N_v$ acting on $H$
by conjugation.  Hence we may immediately write down all its
irreducible representations, using theorem \ref{irrepth}. They are
labeled by an $N_v$-orbit $\mathcal{O}$ in $H$ and by a
representation $\tau$ of the stabilizer $N_{\mathcal{O}}$ of of a
chosen element $g_{\mathcal{O}} \in \mathcal{O}$ in
$N_A$. We will denote them $\Omega^{\mathcal{O}}_{\tau}$ The Hilbert
space on which $\Omega^{\mathcal{O}}_{\tau}$ acts is the space
$F_{\tau}(N_v,V_{\tau})$ defined in (\ref{irrepmod}). We will call it
$V^{\mathcal{O}}_{\tau}$ for short. The action of
$\Omega^{\mathcal{O}}_{\tau}$ on this space is given by the formula in
theorem \ref{irrepth}, which in this case becomes
\begin{equation}
\label{Tvaction}
\left(\Omega^{\mathcal{O}}_{\tau}(F)\phi\right)(x):=\int_{N_v}\,dz\,
F(x g_{\mathcal{O}} x^{-1},z)\, \phi(z^{-1}x).
\end{equation} 
The characters $\psi^{\mathcal{O}}_{\tau}$ of these representations
are given by formula \ref{trafochar} or equivalently by formula
(\ref{trafochar2}). We have 
\begin{equation}
\psi^{\mathcal{O}}_{\tau}(\eta,h)=
1_{N_{\eta}}(h)1_{\mathcal{O}}(\eta)\psi_{\tau}(x_{\eta}^{-1}h x_{\eta}).
\end{equation} 
Using these characters and the inner product (\ref{trafoinprod}), one
may calculate the fusion rules for $\mathcal{T}_v(H)$-irreps.

Clearly, any representation of $D(H)$ also gives a representation of
$\mathcal{T}_v(H)$ by restriction. When we consider the irreps of $D(H)$ as
$\mathcal{T}_v(H)$-representations in this way, they will usually no longer be
irreducible. Their decomposition into $\mathcal{T}_v(H)$-irreps may be
calculated by taking the inner product (\ref{trafoinprod}) of their
character with the characters $\psi^{\mathcal{O}}_{\tau}$. The
character of the $D(H)$-irrep $\Pi^{B}_{\beta}$, seen as a
$\mathcal{T}_v(H)$-irrep, is just the restriction of the original character
$\chi^{B}_{\beta}$; we have 
\begin{equation}
\chi^{B}_{\beta}(\eta,h)=
1_{N_{B}}(h)1_{B}(\eta)\chi_{\alpha}(x_{\eta}^{-1}h x_{\eta}).
\end{equation} 
{}From this formula and the formula for $\psi^{\mathcal{O}}_{\tau}$, we
see immediately that the irreps $\Omega^{\mathcal{O}}_{\tau}$ of $\mathcal{T}_v$
which constitute $\Pi^{B}_{\beta}$ will all have $\mathcal{O}\subset
B$. Also we see that a purely magnetic $D(H)$-irrep $\Pi^{B}_{1}$ will
decompose into the purely magnetic $\mathcal{T}_v(H)$-irreps
$\Omega^{\mathcal{O}}_{1}$ with $\mathcal{O}\subset B$. A purely electric
irrep $\Pi^{e}_{\beta}$ of $D(H)$ will decompose into the purely
electric irreps $\Omega^{e}_{\tau}$ of $\mathcal{T}_v(H)$ which are such, that the
irrep $\tau$ of $N_A\subset H$ is contained in the decomposition of
the irrep $\beta$ of $H$.

\subsection{Confinement}

Let us now determine which representations of the residual algebra
$\mathcal{T}_{v}(H)$ of the previous section will be confined and
which will not. The non-confined representations have to satisfy the
conditions (\ref{modustarcond}). Since $\mathcal{T}_{v}(H)$ is
isomorphic to a transformation group algebra, we may apply the results
of section \ref{trafoconf} (with $N=N_v$ and $K=\{e\}$) to simplify
these to the conditions (\ref{tgacond1}) and (\ref{tgacond2}). From
section \ref{trafoconf}, proposition \ref{uprop}, we know that these
equations are solved at least by those $\Omega^{\mathcal{O}}_{\tau}$
for which $g_\mathcal{O} K=n K$ for some $n\in N_v$ and $\tau$ is
trivial on $K$. Since we have $K=\{e\}$ here, this reduces to just the
requirement that $g_{\mathcal{O}} \in N_{v}$. We have also shown that
this set of irreps closes under conjugation and tensor products and
that they are in fact the irreps of a quotient $\mathcal{U}_{v}(H)$ of
$\mathcal{T}_{v}(H)$ that is isomorphic to $D(N_v/(N_v\cap K))$, which
is here just $D(N_v)$.

It turns out that the irreps $\Omega^{\mathcal{O}}_{\tau}$ with
$g_{\mathcal{O}}\in N_v$ are actually all the irreps that meet the
requirements (\ref{tgacond1}) and (\ref{tgacond2}). Let us check this.
In the case at hand, (\ref{tgacond2}) is always satisfied, since $g_A$
is the unit element of $H$. Thus, we are left with condition
(\ref{tgacond1}). Since the $K=\{e\}$, this reduces to
\begin{equation}
\phi(x_{\eta}g_{\mathcal{O}}^{-1} x^{-1}_{\eta}x)=\phi(x) .
\end{equation}
Using the invariance property of $\phi$, this becomes
$\alpha(x^{-1}x_{\eta}g_{\mathcal{O}}x^{-1}_{\eta}x)\phi(x)=\phi(x)$.
Multiplying with $\alpha(x)$ from the left and using the invariance of
$\phi$ once more, we see
$\alpha(x_{\eta}g_{\mathcal{O}}x^{-1}_{\eta})\phi(e)$ must equal
$\phi(e)$. If we now recall that $v=\phi(e)$ and that the $x_{\eta}$
are elements of $N_v$, then we see that we are left with the
requirement that $g_{\mathcal{O}}$ should be an element of
$N_v$. Thus, the class of solutions that we had already is indeed
complete and the non-confined algebra is just the quantum double of
$N_v$.

The fact that the non-confined algebra $\mathcal{U}$ is the quantum
double $D(N_v)$ of the stabilizer $N_v$ of the condensate vector comes
as no surprise; the original $D(H)$-theory was obtained from a gauge
theory with a continuous gauge group $G$ by breaking this group down
to $H$ through condensation of an electric excitation. All we have
done by condensing one of the electric particles of the $D(H)$-theory
is to modify the electric condensate of the $G$-theory in such a way
that the residual gauge group is now $N_v$ rather than $H$. We
referred to this replacement of $H$ with $N_v$ already at the end of
section \ref{settingsec} and it is encouraging to see that our
formalism for symmetry breaking and confinement in quantum groups
produces the result we anticipated there. 

The result that the non confined irreps of
$\Omega^{\mathcal{O}}_{\tau}$ of $\mathcal{T}_v$ are exactly those for
which $g_{\mathcal{O}} \in N_v$ is also in accordance with our
intuitive treatment in section \ref{settingsec}; the
$\Omega^{\mathcal{O}}_{\tau}$ whose ``flux'' $g_{\mathcal{O}}$ acts
trivially on the condensate are not confined, because they will have
trivial braiding with the condensate. The remaining
$\Omega^{\mathcal{O}}_{\tau}$ will be confined, because they pull
strings in the condensate.  In fact, all the expectations we voiced in
section \ref{settingsec} come true and are now under precise
mathematical control. ``Hadronic'' excitations with overall flux in
$N_v$ can be classified by means of the fusion rules of
$\mathcal{T}_v$, which can be obtained using the inner product
(\ref{trafoinprod}) on the space of characters.  Also, the theory of
section \ref{wallsec} implies that the classification of strings or
domain walls does indeed involve the elements of $H/N_v$, as we now
show.

In section \ref{wallsec} we asserted that
the string associated with an irrep $\Omega^{\mathcal{O}}_{\tau}$ may
be characterized by the restriction of $\Omega^{\mathcal{O}}_{\tau}$
to the left or right Hopf kernel of the projection $\Gamma$ of
$\mathcal{T}_v(H)$ onto $D(N_v)$. Let us take the right kernel. From
(\ref{tgarker}) we see that the elements of the right kernel are all
of the form $f\otimes\delta_e$, where $f$ is constant on left cosets
of $N_v$ in $H$; the right kernel is isomorphic to the algebra of
functions on the left $N_v$-cosets in $H$ The irreps $E_{hN}$ of
$\rker(\Gamma)$ are labeled by these cosets and given by
\begin{equation}
E_{hN}(f\otimes\delta_e)=f(h).
\end{equation}
It is easy to find the restriction of $\Omega^{\mathcal{O}}_{\tau}$ to
$\rker(\Gamma)$. Let $\phi^{i}_{\zeta}$ be the basis elements for
$V^{\mathcal{O}}_{\tau}$ as defined through formula
(\ref{traforepbas}), that is
$\phi^{i}_{\zeta}(y)=1_{x_{\zeta}N_{\mathcal{O}}}(y)
\tau(y^{-1}x_{\zeta})e_{i}^{\tau}$. Note that the $\zeta$ is are in
this case just elements of $H$ and that we have
$x_{\zeta}g_{\mathcal{O}} x_{\zeta}^{-1}=\zeta$. Also,
$N_{\mathcal{O}}$ is just the stabilizer of $g_{\mathcal{O}}$ in
$N$. Using this, we have
\begin{eqnarray}
(\Omega^{\mathcal{O}}_{\tau}(f\otimes\delta_e)\phi^{i}_{\zeta})(y)&=&
f(yg_{\mathcal{O}} y^{-1})\phi(y)
\nonumber \\
~&=& f(yg_{\mathcal{O}}y^{-1})1_{x_{\zeta}N_{\mathcal{O}}}(y)
\tau(y^{-1}x_{\zeta})e_{i}^{\tau} 
\nonumber \\
~&=& f(\zeta)\phi(y)=E_{\zeta N}(f)\phi(y).
\end{eqnarray}
So we see that each of the $\phi^{i}_{\zeta}$ spans a one dimensional
$\rker(\Gamma)$-submodule of $\Omega^{\mathcal{O}}_{\tau}$ isomorphic
to the module of $E_{\zeta N}$. This gives the decomposition of
$\Omega^{\mathcal{O}}_{\tau}$ into $\rker(\Gamma)$ modules: for each
$\zeta$ in the orbit $\mathcal{O}$, we have $d_{\tau}$ copies of
$E_{\zeta N}$. Of course, some of the cosets $\zeta N$ may coincide
and then $E_{\zeta N}$ will occur a multiple of $d_{\tau}$ times in
the decomposition. In particular, if $\Omega^{\mathcal{O}}_{\tau}$ is
not confined, then the orbit $\mathcal{O}$ is just a conjugacy class
of $N$ and we see that $\Omega^{\mathcal{O}}_{\tau}$ corresponds to
$|\mathcal{O}|d_{\tau}$ copies of the trivial
$\rker(\Gamma)$-representation $E_{N}$, a result which we showed in
general already in section \ref{wallsec}. Here, it is also easy to see
that none of the confined irreps of $\mathcal{T}_v$ has this
property. In other words, none of the non-confined irreps pull
strings, while all the confined ones do. The result we have got for
the labeling of the walls is what we should have expected; a
string is created by inserting a flux $g\not\in N_{v}$ into the
condensate. This string may be characterized by the fact that, if the
condensate state on one side of the string is given by $\phi(0)=v$,
then it must be given by $\alpha(g)v$ on the other side. But this
means that the fluxes $gn$, with $n\in N_{v}$, will all pull the same
string as the flux $g$, since $\alpha(gn)v=\alpha(g)v$. Hence, the
string may already be characterized by the coset $gN_{v}$. However,
the flux $g$ which pulls the string may be transformed into the fluxes
$ngn^{-1}$ by gauge transformations with elements $n\in N_v$. Hence
the walls should indeed be labeled by the set of cosets
$ngn^{-1}N_{v}$ which is just the set of cosets $\zeta N$ of the
elements $\zeta$ in the $N_v$-orbit of $g$.

\subsection{Examples of electric condensates}
\label{elexsec}

\subsubsection{Abelian $H$}
\label{abelbreaksec}

Suppose a particle in the irrep $\Pi^{e}_{\alpha}$ of $D(H)$ has
condensed in the state $v \in V^{e}_{\alpha}$. We have seen that the
residual symmetry algebra $\mathcal{T}_v(H)$ is the Hopf subalgebra of
$D(H)\cong \CC (H\times H) \cong F(H\times H)$ which consists of the
functions supported by $H\times N_v$ (cf. \ref{tvdef}). Because $H$ is
Abelian, the irrep $\alpha$ is one-dimensional and hence $N_v$ is just
the kernel $N_{\alpha}$ of $\alpha$. Thus, we have
$\mathcal{T}_v(H)\cong F(H\times N_{\alpha}) \cong \CC (H\times
N_{\alpha})$. Here, the action of $N_{\alpha}$ on $H$ is trivial,
since $H$ is Abelian and hence the irreps $\Omega^{h}_{\beta}$ of
$\mathcal{T}_v$ are labeled by an element $h$ of $H$ and an irrep
$\beta$ of $N_\alpha$. The decomposition of $D(H)$-irreps into
$\mathcal{T}_v(H)$-irreps is straightforward: we have
$\Pi^{h}_{\beta}\equiv \Omega^{h}_{\tilde{\beta}}$, where
$\tilde{\beta}$ is the restriction of $\beta$ to $N_{\alpha}$.

The irreps $\Omega^{h}_{\beta}$ of $\mathcal{T}_v$ which are not confined
are those for which $h \in N_{v}$ and they are of course in one to one
correspondence with the irreps of $D(N_v)$. The corresponding Hopf
projection $\Gamma: \mathcal{T}_{v}(H)\rightarrow D(N_v)$ is just
restriction of the functions in $\mathcal{T}_{v}(H)$ to $N_{v}$ in the
left argument. The left and right Hopf kernels of $\Gamma$ coincide
and they are both isomorphic to the space of functions on the quotient
$H/N_{v}$. the representations of this function space are just the
evaluation functionals on the classes $hN_{v}$ and as before, we
denote them $E_{hN_{v}}$. The restriction of an irrep
$\Omega^{h}_{\beta}$ of $\mathcal{T}$ to $\lker(\Gamma)$ is simply given
by $\Omega^{h}_{\beta}\equiv E_{hN_{v}}$.

To illustrate what happens a bit more explicitly, we will work out the
case of $H=\ZZ_n$. This group is generated by a single element which
we will call $r$ and it has $n$ irreps $\alpha_0,\ldots,\alpha_{n-1}$,
given by
\begin{equation}
\alpha_k:~ r^a \mapsto e^{2\pi i ka/n}. 
\end{equation}
The kernel of $\alpha_k$ consists of those $r^a$ for which $ak=0 ~{\rm
mod}~ n$. The minimal non-zero $a$ for which this holds is the
quotient $n/{\rm gcd} (n,k)=:x$. Hence, we have
$N_{\alpha}=\spn{r^x}\cong \ZZ_{{\rm gcd}(n,k)}$. The corresponding
residual symmetry algebra is 
\begin{equation}
\mathcal{T}_v(\ZZ_n)\cong F(\ZZ_n \times\ZZ_{{\rm gcd}(n,k)}) \cong
\CC\ZZ_n \times\ZZ_{{\rm gcd}(n,k)}.
\end{equation}
Thus we see that the electric symmetry can be broken in as many
different ways as $n$ has divisors (the magnetic symmetry is never
broken). The irreps $\Omega^{r^a}_{\alpha_{l}}$ of the residual
algebra are labeled by an element $r^{a}\in H$ and an irrep
$\alpha_{l}$ (with $0\le l <{\rm gcd}(n,k)$) of $\ZZ_{{\rm
gcd}(n,k)}$. The decomposition of the irreps $\Pi^{r^a}_{\alpha_{l}}$
(with $0\le l < n$) of $\ZZ_n$ is then given by
\begin{equation}
\Pi^{r^a}_{\alpha_l}\equiv \Omega^{r^a}_{\alpha_{l ~{\rm mod}~ {\rm
gcd}(n,k)}}.
\end{equation}
This follows from the fact that $\alpha_l(r^{px})=e^{2\pi i
lpx/n}=e^{2\pi i lp/{\rm gcd}(n,k)}$. 

The irreps $\Omega^{r^a}_{\alpha_{l}}$ which are not confined are those
for which $r^{a}\in N_{\alpha}$, i.e. $a=px$, with $0\le p< {\rm
gcd}(n,k)$. These correspond to the irreps $\Pi^{p}_{\alpha_l}$ of the
non-confined algebra $\mathcal{U}$, which is given by
\begin{equation}
\mathcal{U}_v(\ZZ_n)\cong D(N_{\alpha})\cong D(\ZZ_{{\rm gcd}(n,k)}).
\end{equation}
The Hopf map $\Gamma$ from $F(\ZZ_n \times\ZZ_{{\rm gcd}(n,k)})$ to
$\mathcal{U}$ is just restriction to $\ZZ_{{\rm gcd}(n,k)}$ in the
left argument. The Hopf kernel of this map consist of the functions
$f\otimes\delta_e$ in $F(\ZZ_n \times\ZZ_{{\rm gcd}(n,k)})$ for which
$f$ is constant on the cosets of $N_{\alpha}=\ZZ_{{\rm
gcd}(n,k)}$. Hence, we have 
\begin{equation}
\lker(\Gamma)\cong F(\ZZ_{x}).
\end{equation}
The representations of $\lker(\Gamma)$ may be denoted $E_{a}$ with $(0
\le a <x)$ and are given by $E_{a}(f\otimes\delta_{e})=f(r^a)$. The
restriction of $\Omega^{r^a}_{\alpha_{l}}$ to $\lker(\Gamma)$ is given
by
\begin{equation}
\Omega^{r^a}_{\alpha_{l}} \equiv E_{a~{\rm mod}~x}.
\end{equation}

\subsubsection{$H=D_{2m+1}$}

We will treat all the possible types of electric condensate in order.

\vspace{1mm}
\noindent {\bf 1.} First, we take a condensate $v \in V^{e}_{J_1}$.
We then have $N_v=\spn{r}\cong \ZZ_{2m+1}$ and hence
\begin{equation}
\mathcal{T}_v(D_{2m+1})\cong F(D_{2m+1})\sptp\CC\ZZ_{2m+1}.
\end{equation}
Here and in the sequel, the tilde on the tensor product sign indicates
that the factor on the right acts on the factor on the left through
conjugation. To find the irreps of $\mathcal{T}_v$, we first need to
find the orbits of the adjoint $\ZZ_{2m+1}$-action on $D_{2m+1}$ and
their stabilizers. One easily finds that the orbits are
$\{e\},\{r\},\{r^2\},\ldots,\{r^{2m}\}$ and $\{s,sr,sr^2,\ldots,
sr^{2m}\}$ Of these, all the orbits that contain a single element have
stabilizer $\spn{r}\cong \ZZ_{2m+1}$, while the remaining orbit has
the trivial stabilizer $\{e\}$. Thus, the irreps of $\mathcal{T}_v$
may be denoted $\Omega^{r^k}_{\beta_l}$ (with $0\le k,l < 2m+1$) and
$\Omega^s$. Here, we let $r^k$ and $s$ denote the orbits of $r^k$ and
$s$, in order not to overload the notation. We see that
$\mathcal{T}_v$ has $(2m+1)^2+1$ irreps, which are all
one-dimensional, except for $\Omega^s$, which is
$2m+1$-dimensional. It follows that the squares of the dimensions add
up to $2(2m+1)^2$, which equals the dimension of $\mathcal{T}_v$, as
it should. The decomposition of $D(D_{2m+1})$-irreps into
$\mathcal{T}_v$-irreps may be found directly or by means of the
orthogonality relations for the characters of $\mathcal{T}_v$. We have
\begin{equation}
\begin{array}{ll}
\Pi^{e}_{J_0}\equiv \Omega^{e}_{\beta_0} &
\Pi^{r^k}_{\beta_l}\equiv\Omega^{r^k}_{\beta_l}\oplus
\Omega^{r^{-k}}_{\beta_{-l}}\\ 
\Pi^{e}_{J_1}\equiv\Omega^{e}_{\beta_0}& 
\Pi^{s}_{\gamma_0}\equiv \Omega^{s}\\
\Pi^{e}_{\alpha_k}\equiv\Omega^{e}_{\beta_k}\oplus
\Omega^{e}_{\beta_{-k}} & \Pi^{s}_{\gamma_1}\equiv \Omega^{s}.
\end{array}
\end{equation}
Of the representations of $\mathcal{T}_v$, $\Omega^{s}$ is confined,
since $s \not \in N_{v}$. The others are not confined and are in one
to one correspondence with the irreps of the non-confined algebra
\begin{equation}
\mathcal{U}_v(D_{2m+1})\cong D(N_v)=D(\ZZ_{2m+1}). 
\end{equation}
The right Hopf kernel of the projection $\Gamma$ of
$\mathcal{T}_{v}(H)$ onto $D(N_v)$ is isomorphic to the algebra of
functions on the set of left $\spn{r}$-cosets. There are only two such
cosets, namely $R:=\spn{r}$ and $S:=s \spn{r}$ and hence two
corresponding one-dimensional representations $E_{R}$ and $E_{S}$ of
the right kernel. The decomposition of $\mathcal{T}_v$-irreps into
$\rker{\Gamma}$-irreps is given by
\begin{equation}
\begin{array}{ll}
\Omega^{r^k}_{\beta_l}\equiv E_{R} &
\Omega^{s} \equiv (2m+1) E_{S}.
\end{array}
\end{equation}

\vspace{1mm}
\noindent {\bf 2.}  Now we take a condensate $v$ in the module
$V^{e}_{\alpha_j}$. The stabilizer $N_v$ of $v$ consists by definition
of all the elements $g$ of $D(D_{2m+1})$ for which $v$ is an
eigenvector of $\alpha_j(g)$ with eigenvalue $1$. This includes in
particular all the elements of the kernel of $\alpha_j$. From the
character table of $D(D_{2m+1})$ (table \ref{doddchartab}), one may
read off that this kernel consists of those elements $r^a$ for which
$q^{ja}+q^{-ja}=2$, where $q=e^{2\pi i/(2m+1)}$, or in other words,
for which $\cos(2\pi ja/(2m+1))=1$. It follows that one has to have
$ja=0 ~{\rm mod}~ 2m+1$. The smallest non-zero $a$ for which this
holds is $(2m+1)/{\rm gcd}(2m+1,j)=:x$. Thus, one has
$N_{\alpha_{j}}=\spn{r^x}\cong \ZZ_{{\rm gcd}(2m+1,j)}$.

Of course, the stabilizer $N_v$ of $v$ may be larger than
$N_{\alpha_j}$, if $v$ is an eigenvector of $\alpha_j(g)$ for some $g
\not\in N_{\alpha_j}$. Thus, in order to find out what kinds of
stabilizers are possible, it is a good idea to have a look at the
eigenvalues of the matrices $\alpha_j(g)$. {}From the explicit
matrices in (\ref{doddmats}), we see that the eigenvalues of
$\alpha_j(r^p)$ are $q^{jp}$ and $q^{-jp}$, with $q=e^{2\pi
i/(2m+1)}$. It follows that, if one of the eigenvalues of $r^p$ equals
$1$, so does the other. Hence, the only elements of $\spn{r}$ whose
matrices have eigenvalues equal to one are those that are already
contained in the kernel of $\alpha_j$. The eigenvalues of each of the
matrices $\alpha_j(sr^p)$ are $1$ and $-1$. Thus, we have two
possibilities: either $v$ is not left invariant by any of the matrices
$\alpha_j(sr^p)$, in which case
$N_v=N_{\alpha_j}=\spn{r^x}\cong\ZZ_{{\rm gcd}(2m+1,j)}$, or $v$ is
left invariant by some of the $\alpha_j(sr^p)$. In this case, we may
without loss of generality choose $v$ to be the invariant vector of
$\alpha_j(s)$, since each of the $sr^p$ is a conjugate of $s$ in
$D_{2m+1}$ and hence the invariant vectors of the $sr^p$ are in the
same gauge orbit as the invariant vector of $s$. With this choice, one
sees easily that $N_v=\spn{r^x}\cup s\spn{r^x}\cong D_{{\rm
gcd}(2m+1,j)}$. We will now treat the two possibilities for $N_v$ in
order.

\noindent {\bf 2.a} When $N_v=\spn{r^x}\cong \ZZ_{{\rm gcd}(2m+1,j)}$, we have
\begin{equation}
\mathcal{T}_v(D_{2m+1})\cong F(D_{2m+1})\sptp\CC\ZZ_{{\rm gcd}(2m+1,j)}.
\end{equation}
The orbits of the $\spn{r^x}$-action on $D_{2m+1}$ are
$\{e\},\{r\},\ldots,\{r^{2m}\}$, with stabiliser $\spn{r^x}$, and
$s\spn{r^x}$, $sr\spn{r^x},\ldots, sr^{x-1}\spn{r^x}$, with stabilizer
$\{e\}$. This means that the irreps of $\mathcal{T}$ may be denoted as
$\Omega^{r^k}_{\beta_{l}}$ (with $0 \le k < 2m+1$, $0\le l < {\rm
gcd}(2m+1,j)$) and $\Omega^{sr^k}$ (with $0 \le k < x$). Here, we have
once again denoted orbits by representative elements. We see that
there are $(2m+1){\rm gcd}(2m+1,j)+(2m+1)/{\rm gcd}(2m+1,j)$
irreps. Of these, $(2m+1){\rm gcd}(2m+1,j)$ are one-dimensional and
the remaining $(2m+1)/{\rm gcd}(2m+1,j)$ (the $\Omega^{sr^p}$) are
${\rm gcd}(2m+1,j)$-dimensional, so that the squares of the dimensions
again add up to the dimension of $\mathcal{T}$, which is $2(2m+1){\rm
gcd}(2m+1,j)$. The decomposition of $D(D_{2m+1})$-irreps reads
\begin{equation}
\begin{array}{ll}
\Pi^{e}_{J_0}\equiv \Omega^{e}_{\beta_0} &
\Pi^{r^k}_{\beta_l}\equiv
\Omega^{r^k}_{\beta_{l}}\oplus 
\Omega^{r^{-k}}_{\beta_{-l}}\\
\Pi^{e}_{J_1}\equiv\Omega^{e}_{\beta_0} &
\Pi^{s}_{\gamma_0}\equiv \bigoplus_{0\le p <x}\Omega^{sr^p}\\
\Pi^{e}_{\alpha_l}\equiv 
\Omega^{e}_{\beta_{l}}\oplus 
\Omega^{e}_{\beta_{-l}}&
\Pi^{s}_{\gamma_1}\equiv \bigoplus_{0\le p <x}\Omega^{sr^p}
\end{array}
\end{equation}
where the labels $l$ and $-l$ should be read modulo $2m+1$ on the left
hand side and modulo \mbox{${\rm gcd}(2m+1,j)$} on the right hand
side. The non-confined irreps are those $\Omega^{r^k}_{\beta_{l}}$ for
which $r^k\in\spn{r^x}$ and they are in one correspondence with the
irreps of the non-confined algebra
\begin{equation}
\mathcal{U}_v(D_{2m+1}) \cong D(\ZZ_{{\rm gcd}(j,2m+1)}).
\end{equation}
The right and the left kernel of the Hopf map
$\Gamma:\mathcal{T}\rightarrow \mathcal{U}$ are equal and isomorphic
to the algebra of functions on the quotient group
$D_{2m+_1}/\spn{r^x}$. Since this quotient group is isomorphic to
$D_{x}$, we have 
\begin{equation}
\rker(\Gamma)\cong F(D_{x}).
\end{equation}
The representations of $\rker(\Gamma)$ are labeled by the elements
$R^{k},SR^{k}$ of $D_{x}$ and we denote them $E_{R^k},E_{SR^{k}}$. The
decomposition of $\mathcal{T}$-irreps into $\rker(\Gamma)$-irreps is
given by
\begin{equation}
\Omega^{r^k}_{\beta_{l}}\equiv E_{R^{k}} ~~~~ 
\Omega^{sr^k} \equiv x E_{SR^{k}}
\end{equation}
where, on the right hand side, $k$ should be taken modulo $x$. 

\noindent {\bf 2.b} When $N_v\cong D_{{\rm gcd}(2m+1,j)}$, we have
\begin{equation}
\mathcal{T}_v(D_{2m+1})\cong F(D_{2m+1})\sptp\CC D_{{\rm gcd}(2m+1,j)}.
\end{equation}
The $D_{{\rm gcd}(2m+1,j)}$-orbits in $D_{2m+1}$ are $\{e\}$,
$\{r,r^{-1}\}$, $\{r^2,r^{-2}\},\ldots,\{r^m,r^{-m}\}$, $s\spn{x}$ and
$sr\spn{x}\cup sr^{x-1}\spn{x},sr^{2}\spn{x}\cup
sr^{x-2}\spn{x}\ldots$. The stabilizer of $e$ is of course all of
$D_{{\rm gcd}(2m+1,j)}$, the stabilizer of $r^{k}$ is $\spn{r^x}\cong
\ZZ_{{\rm gcd}(2m+1,j)}$ and the stabilizer of $s$ is $\spn{s}\cong
\ZZ_2$. The stabilizer of the orbits $sr^{p}\spn{x}\cup
sr^{x-p}\spn{x}$ is just $\{e\}$ Hence, the irreps of $\mathcal{T}_v$
may be denoted $\Omega^{e}_{J_0}$, $\Omega^{e}_{J_1}$,
$\Omega^{e}_{\alpha_k}$ (with $1\le k \le \frac{1}{2}({\rm
gcd}(2m+1,j)-1) $), $\Omega^{r^{k}}_{\beta_{l}}$ (with $0<k\le m,\,
0\le l<{\rm gcd}(2m+1,j)$), $\Omega^{s}_{\gamma_0},
\Omega^{s}_{\gamma_1}$ and finally $\Omega^{sr^p}$ (with $1 \le
p<\frac{1}{2}(x-1)$). This yields $3+\frac{1}{2}(2m+1)({\rm
gcd}(2m+1,j)+1/{\rm gcd}(2m+1,j))$ irreps in total and one may check
that the squares of their dimensions sum correctly to the dimension of
$\mathcal{T}_v$, which is $4(2m+1){\rm gcd}(2m+1,j)$. The
decomposition of the $\Pi^{e}_{\alpha_{k}}$ into
$\mathcal{T}_v$-irreps is now
\begin{equation}
\Pi^{e}_{\alpha_k}\equiv \left\{
\begin{array}{lr}
\Omega^{e}_{J_0}\oplus \Omega^{e}_{J_1} &([k]=0)\\
\Omega^{e}_{\alpha_{[k]}}&([k]\le\frac{1}{2}({\rm gcd}(2m+1,j)-1)\\
\Omega^{e}_{\alpha_{{\rm gcd}(2m+1,j)-[k]}}&
([k]>\frac{1}{2}({\rm gcd}(2m+1,j)-1)\\
\end{array}\right.
\end{equation}
Here, $[k]$ denotes $k {\rm ~mod~}{\rm gcd}(2m+1,j)$.  The
decomposition of the other $D_{2m+1}$-irreps into $\mathcal{T}_v$-irreps is
given by
\begin{equation}
\begin{array}{ll}
\Pi^{e}_{J_0}\equiv \Omega^{e}_{J_0} &
\Pi^{s}_{\gamma_0}\equiv \Omega^{s_{\gamma_0}} \oplus
\bigoplus_{1\le p <\frac{1}{2}(x-1)}\Omega^{sr^p} \\
\Pi^{e}_{J_1}\equiv\Omega^{e}_{J_1} &
\Pi^{s}_{\gamma_1}\equiv \Omega^{s_{\gamma_1}} \oplus
\bigoplus_{1\le p <\frac{1}{2}(x-1)}\Omega^{sr^p} \\
\Pi^{r^k}_{\beta_l}\equiv \Omega^{r^k}_{\beta_l}
\end{array}
\end{equation}
The labels $l$ on the left should be read modulo $2m+1$, while those
on the right hand side should be read modulo \mbox{${\rm
gcd}(2m+1,j)$}. The non-confined irreps of $\mathcal{T}_{v}$ are
$\Omega^{e}_{J_0},\Omega^{e}_{J_1}$, the $\Omega^{e}_{\alpha_{k}}$,
$\Omega^{s}_{\gamma_0},\Omega^{s}_{\gamma_1}$ and those
$\Omega^{r^k}_{\beta_{l}}$ for which $r^k\in \spn{r^x}$. These irreps
correspond to the irreps of the non-confined algebra 
\begin{equation}
\mathcal{U}_v(D_{2m+1})\cong D(D_{{\rm gcd}(2m+1,j)}).
\end{equation}
The right kernel of the Hopf map $\Gamma:\mathcal{T}\rightarrow
\mathcal{U}$ is isomorphic to the algebra of functions on the space of
left cosets of $N_v\cong D_{{\rm gcd}(2m+1,j)}$ in $D_{2m+1}$. There
are $x$ distinct cosets, namely the cosets of $e,r,\ldots,r^{x-1}$. We
will denote these $E,R,\ldots,R^{x-1}$. The corresponding irreps of
$\rker(\Gamma)$ will again be denoted $E_{R^k}$. The restriction of
the irreps of $\mathcal{T}_v$ to $\lker(\Gamma)$ is given by
\begin{equation}
\begin{array}{lll}
\Omega^{e}_{J_0} \equiv E_{E}&
\Omega^{r^k}_{\beta_l}\equiv E_{R^{k}}&
\Omega^{sr^p}\equiv {\rm gcd}(2m+1,j) (E_{R^p}\oplus E_{R^{x-p}}) \\
\Omega^{e}_{J_1} \equiv E_{E}&
\Omega^{s_{\gamma_0}} \equiv {\rm gcd}(2m+1,j) E_{E} &~\\
\Omega^{e}_{\alpha_l} \equiv 2 E_{E} &
\Omega^{s_{\gamma_1}} \equiv {\rm gcd}(2m+1,j) E_{E}&~
\end{array}
\end{equation}
where the index $k$ should be read modulo $2m+1$ on the left hand side
and modulo $x$ on the right.  In the restriction of $\Omega^{sr^p}$, we
see our first example of a situation where the wall created by a
$\mathcal{T}$-particle carries a representation of $\rker(\Gamma)$
that contains two distinct irreps of $\rker(\Gamma)$, namely $E_{R^p}$
and $E_{R^{x-p}}$. The isotypical components of these irreps are sent
onto each other by gauge transformation with $s \in N_v$, since
$sR^{p}s^{-1}=R^{x-p}$.

\section{Gauge invariant magnetic condensates} 
\label{magcondsec}

\subsection{Symmetry breaking}
\label{magbreaksec}

There is precisely one gauge invariant state in every
magnetic representation $\Pi^{A}_{1}$. This state is represented by
the constant function
\begin{equation}
\phi: h \mapsto 1
\end{equation}
on $H$. To find the Hopf stabilizer of $\phi$, we need to find the
irreps $(\rho,g)$ of $D(H)^*$ which solve equation
(\ref{inveq}). Since $\phi$ is constant equal to one, this reduces to
\begin{equation}
\rho(g_{A}) = I.
\end{equation}
Hence, the unbroken symmetry algebra is the algebra generated by the
matrix elements of the representations $(\rho,g)$ for which $g_{A}$ is
contained in the kernel of $\rho$. Now define $K_A$ as the minimal
normal subgroup of $H$ that contains $g_A$ (and hence all of
$A$). Since the kernel of a representation is a normal subgroup, the
irreducible representations $\rho$ which have $g_A$ in their kernel
will be precisely the ones which contain all of $K_A$ in their
kernel. Such irreps are in one-to-one correspondence with the irreps
of $H/K_A$ \cite{jalie} and since the matrix elements of the irreps of
$H/K_A$ generate $F(H/K_A)$, the algebra generated by the matrix
elements of the irreps of $H$ which contain $g_A$ in their kernel is
precisely the algebra $\mathcal{T}_A$ of functions on $G$ which are
constant on the cosets of $K_A$. Hence, the unbroken symmetry algebra
in this case is the Hopf subalgebra $\mathcal{T}_A(H)$ of $D(H)$
defined by
\begin{equation}
\mathcal{T}_A(H):= \left\{ F \in D(H)|F(xk,y)=F(x,y) 
\forall k \in K_A \right\}.
\end{equation}
Clearly, $\mathcal{T}_A \cong F(H/K_A \times H)$ as a linear space and we see
that $\mathcal{T}_A$ is a transformation group algebra, with $H$ acting on $K_A$
by conjugation. This means we can once again make use of theorem
\ref{irrepth} to write down the irreps of $\mathcal{T}_A$. They are labeled by
an $H$-orbit $\mathcal{O}\subset H/K$ and an irrep $\tau$ of the
stabilizer $N_{\mathcal{O}}$ of a chosen element $g_{\mathcal{O}}\in
\mathcal{O}$. The irrep labeled by $\mathcal{O}$ and $\tau$ will be
denoted $\Omega^{\mathcal{O}}_{\tau}$.  It acts on the Hilbert space
$F_{\tau}(H,V_{\tau})$ in the usual way:
\begin{equation}
\label{TAaction}
\left(\Omega^{\mathcal{O}}_{\tau}(F)\phi\right)(x):=\int_{H}\,dz\,
F(x g_{\mathcal{O}} x^{-1},z)\, \phi(z^{-1}x).
\end{equation} 
The character $\psi^{\mathcal{O}}_{\tau}$ of
$\Omega^{\mathcal{O}}_{\tau}$ is given as a function on $H/K_A\times
H$ by (cf. (\ref{trafochar2}))
\begin{equation}
\psi^{\mathcal{O}}_{\tau}(\eta,h)=
1_{N_{\eta}}(h)1_{\mathcal{O}}(\eta)\psi_{\tau}(x_{\eta}^{-1}h x_{\eta}).
\end{equation} 
The decomposition of any $\mathcal{T}_A(H)$-module into irreps may be found by
calculating the inner products (defined in (\ref{trafoinprod}))
between the character of the module and the above characters of the
irreps. Of course, we can view any $D(H)$-module as a $\mathcal{T}_A(H)$-module
by restriction. The characters $\chi^{B}_{\beta}$ of the irreps
$\Pi^{B}_{\beta}$ of $D(H)$, viewed as $\mathcal{T}_A(H)$-modules are given by
\begin{equation}
\chi^{B}_{\beta}(gK_A,h)=
\sum_{k\in K_A} 1_{N_{gk}}(h)1_{B}(gk)\chi_{\beta}(x_{gk}^{-1}h x_{gk}).
\end{equation} 
We see that all the irreps $\Omega^{\mathcal{O}}_{\tau}$ in the
decomposition of $\Pi^{B}_{\beta}$ must be such that $B$ is a subset
of the set of elements of $H$ that constitute the $K_A$-classes in
$\mathcal{O}$. Clearly, there is only a single orbit $\mathcal{O}$ for
which this holds. The decomposition of a purely electric
representation $\Pi^{e}_{\beta}$ is very simple: such a
representation is irreducible and isomorphic to the purely electric
irrep $\Omega^{\mathcal{O}=K_A}_{\beta}$ (Note that $N_{K_A}=H$). On
the other hand, the decomposition of a purely magnetic representation
$\Pi^{B}_{1}$ may contain irreps $\Omega^{\mathcal{O}}_{\tau}$ which
are not purely magnetic (i.e. $\tau$ may be non-trivial).

\subsection{Confinement}

We will now find out which of the irreps $\Omega^{\mathcal{O}}_{\tau}$
of $\mathcal{T}_A$ are confined and which are not. The non-confined
irreps have to satisfy the requirements (\ref{modustarcond}). Since
$\mathcal{T}_A$ is isomorphic to a transformation group algebra, these
reduce to the conditions (\ref{tgacond1}) and (\ref{tgacond2}), with
$K=K_A$ and $N=H$. We have seen in section \ref{trafoconf},
proposition \ref{uprop}, that these requirements will be satisfied by
the set of irreps $\Omega^{\mathcal{O}}_{\tau}$ for which
$g_{\mathcal{O}}K=nK$ for some $n\in N$ and for which $\tau$ is
trivial on $K$. The first of these requirements is trivial here, since
$N=H$ and so this set consist of all $\Omega^{\mathcal{O}}_{\tau}$ for
which $\tau$ is trivial on $K_A$. These irreps correspond to the
irreps of the quotient $D(H/K_A)$ of $\mathcal{T}_{A}$. In the case at
hand, it turns out that this set of of solutions is actually complete
and hence the non-confined symmetry algebra $\mathcal{U}_{A}$ is just
the quantum double of the quotient group $H/K_A$. Let us demonstrate
this.

Equation (\ref{tgacond1}) is trivially satisfied by the matrix
elements of all the irreps of $\mathcal{T}_A$, since $\phi$ is the constant
function $1$. Thus, we are left with the requirement
(\ref{tgacond2}). Since the support of $\phi$ is all of $H$, this
becomes
\begin{equation}
(\forall x\in H)~~~\tau(xg_A x^{-1})=I
\end{equation}
or in other words
\begin{equation}
A \subset {\rm Ker}(\tau).
\end{equation}
The requirement that $A \subset {\rm Ker}(\tau)$ is equivalent to the
requirement that $K_A\subset {\rm Ker}(\tau)$, since $K_A$ is just the
subgroup of $H$ generated by the elements of $A$. Hence, the
non-confined irreps $\Omega^{\mathcal{O}}_{\tau}$ of $\mathcal{T}_A$
are exactly those for which $\tau$ is trivial on $K_A$, as we claimed.

The results we have obtained are quite satisfying when one thinks back
of the intuition that went into our method of finding the non-confined
irreps. We wanted the non-confined irreps to have trivial braiding
with the condensate. For a purely magnetic condensate, this means
roughly that the flux of the condensate should commute with the flux
of the non-confined irreps and should act trivially on the charges of
the non-confined irreps. The first of these conditions is
automatically met: the flux state of the condensate commutes with any
other flux state (the class sum is a central element of the group
algebra of $H$). Therefore, there is no requirement on
$\mathcal{O}$. The second condition is implemented by the demand that
$\tau$ is trivial on $K_A$, the group which is generated by the fluxes
in the class $A$. We also wanted to have well-defined fusion, spin and
braiding among the non-confined particles and these are now provided
by the Hopf structure, $R$-matrix and ribbon-element of $D(H/K_A)$.

Finally, let us say something about the characterization of strings
(or walls). From proposition \ref{kerprop}, we see that the Hopf
kernel of the projection $\Gamma:\mathcal{T}_{A}(H)\rightarrow
D(H/K_A)$ is just the set of elements $1\otimes f\in
\mathcal{T}_{A}(H)$ for which $f$ has support in $K_A$. In this case,
the left and right Hopf kernels coincide and hence the kernel is
itself a Hopf algebra. This Hopf algebra is clearly isomorphic to the
group algebra $\CC K_A$ (cf. corrollary \ref{kercorr}) and hence the
irreps of $\lker(\Gamma)$ correspond to the irreps of $K_A$. If $\rho$
is an irrep of $K_A$ then we also write $\rho$ for the corresponding
irrep of $\lker(\Gamma)$ and with this slight abuse of notation, we
may write
\begin{equation}
\rho(1\otimes \delta_{k})=\rho(k)
\end{equation}
for all $k\in K_A$. We will now calculate the decomposition of a
representation $\Omega^{\mathcal{O}}_{\tau}$ of $\mathcal{T}_{A}(H)$
into representations of $\lker(\Gamma)$ by means of the formula
(\ref{trfc}) for the character $\psi^{\mathcal{O}}_{\tau}$ of
$\Omega^{\mathcal{O}}_{\tau}$. For $g \in K_A$, we have
\begin{eqnarray}
(\psi^{\mathcal{O}}_{\tau}(1\otimes\delta_g)&=&
\int_{\mathcal{O}} d\zeta \int_{N_{\mathcal{O}}} dn \,
\delta_{g}(x_{\zeta}n x_{\zeta}^{-1})\chi_{\tau}(n)
\nonumber \\
~&=& 
\int_{\mathcal{O}} d\zeta \,\chi_{\tau}(x_{\zeta}^{-1} g x_{\zeta}).
\end{eqnarray}
{}From this, we see that the restriction of $\Omega^{\mathcal{O}}_{\tau}$
to $\lker(\Gamma)\cong\CC K_A$ contains exactly the irreps of $K_A$
that are contained in the restriction of $\beta$ to $K_A$, together
with the irreps obtained from these by composition with the
automorphisms of $K_A$ that are given by conjugation with the
$x_{\zeta}^{-1}$. As in the case of electric condensates, we see that
the non-confined irreps are exactly all those that have trivial
restriction to the Hopf kernel of $\Gamma$. 

\subsection{Examples of gauge invariant condensates}
\label{magexsec}

\subsubsection{Abelian $H$}

For Abelian $H$, every state in a purely magnetic representation
$\Pi^{A}$ is gauge invariant, so this section covers all purely
magnetic condensates for Abelian groups. Suppose we condense a state
in the purely magnetic representation labeled by the element $g_A$ of
$H$. Then we know that the residual symmetry algebra
$\mathcal{T}_A(H)$ is the Hopf subalgebra of $D(H)$ which consists of
the functions that are constant on cosets of $K_A$ in their left
argument. Here $K_A$ is the minimal normal subgroup of $H$ that
contains $g_A$, which, when $H$ is Abelian, is just the cyclic group
generated by $g_A$. As an algebra, $\mathcal{T}_A(H)$ is isomorphic to
the transformation group algebra $F(H/K_A\times H)$, where $H$ acts on
$H/K_A$ by conjugation. When $H$ is Abelian, the action of $H$ on
$H/K_A$ is thus trivial. The orbits are then just the elements of
$H/K_A=H/\spn{g_A}$ and the stabilizer of each orbit is all of
$H$. Thus, the irreps of $\mathcal{T}_A(H)$ may be denoted
$\Omega^{hK_A}_{\alpha}$, where $hK_A$ is an element of $H/K_A$ and
$\alpha$ is an irrep of $H$. The action of $\mathcal{T}_A$ in the
irrep $\Omega^{hK_A}_{\alpha}$ is given in formula
(\ref{TAaction}). The irreps of $D(H)$ may be easily decomposed into
irreps of $\mathcal{T}_A(H)$; we have $\Pi^{h}_{\alpha}\equiv
\Omega^{hK_A}_{\alpha}$. The non-confined irreps of $\mathcal{T}_A(H)$
are those $\Omega^{hK_A}_{\alpha}$ for which $\alpha$ is trivial on
$K_A$ These correspond to the irreps of the non-confined algebra
$\mathcal{U}\cong D(H/K_A)\cong D(H/\spn{g_A})$. The kernel of the
Hopf map $\Gamma:\mathcal{T}\rightarrow \mathcal{U}$ is isomorphic to
$\CC K_A$ and hence its irreps are just the irreps of $K_A$. Since
$K_A=\spn{g_A}$, it follows that the number of these irreps equals the
order of the element $g_A$. We may indeed give the irreps explicitly;
denoting them as $\rho_k$ (with $0\le k< {\rm ord}(g_A)$), we have
$\rho_k((g_A)^p)=\exp(2\pi i kp/{\rm ord}(g_A))$ The restriction of
the irreps of $\mathcal{T}_A$ to $\lker(\Gamma)$ is also easily
found. We have $\Omega^{hK_A}_{\alpha}\equiv \alpha|_{K_{A}}$. In
other words, the wall corresponding to $\Omega^{hK_A}_{\alpha}$ can be
labeled by the phase factor $\alpha(g_A)$.

We once again explicitly work out the case of $H=\ZZ_n$. As in section
\ref{abelbreaksec}, we will denote our preferred generator for $\ZZ_n$
by $r$ and we write $\alpha_{0},\ldots,\alpha_{n-1}$ for the irreps of
$\ZZ_n$. Now suppose we condense the magnetic flux $g_A=r^k$. Then we
have $K_A=\spn{r^k}=\spn{r^{{\rm gcd}(k,n)}}\cong \ZZ_{x}$, where
$x=n/{\rm gcd}(k,n)$. As a consequence, we have $H/K_A\cong
\ZZ_{n}/\ZZ_x \cong \ZZ_{{\rm gcd}(k,n)}$. Thus, 
\begin{equation}
\mathcal{T}_{[r^k]}(\ZZ_n)\cong F(\ZZ_{{\rm gcd}(k,n)}\times \ZZ_n) \cong
\CC(\ZZ_{{\rm gcd}(k,n)}\times \ZZ_n)
\end{equation}
and we see that there is one type of broken symmetry for each divisor
of $n$.  The irreps $\Omega^{r^a K_A}_{\alpha_{l}}$ of $\mathcal{T}$
may be labeled by an element $r^{a}$ of $\ZZ_{\rm gcd}(k,n)$ and an
irrep $\alpha_{l}$ of $\ZZ_{n}$. The decomposition of the irreps
$\Pi^{r^a}_{\alpha_{l}}$ (with $0\le a < n$) of $\ZZ_n$ is then given
by
\begin{equation}
\Pi^{qr^a}_{\alpha_l}\equiv 
\Omega^{r^{a~{\rm mod}~ {\rm gcd}(n,k)}K_A}_{\alpha_l} 
\end{equation}
Of the $\Omega^{r^a K_A}_{\alpha_{l}}$, the ones that are not confined
are those for which $\alpha_l(r^k)=\exp(2\pi i kl/n)=1$, or
equivalently $kl=0 {\rm ~mod~} n$. These are exactly those for which
$l$ is a multiple of $x$ and we see that the non-confined irreps of
$\mathcal{T}_v$ correspond to the irreps of
\begin{equation}
\mathcal{U}_{[r^k]}(\ZZ_n)\cong D(H/\spn{r^k})\cong D(\ZZ_{{\rm
gcd}(k,n)}).
\end{equation}
The kernel of the Hopf map $\Gamma:\mathcal{T}\rightarrow \mathcal{U}$
is isomorphic to $\CC\ZZ_x$ and has representations
$\tilde{\alpha}_{j}$ (with $(0\le j< x)$) defined in the usual way,
with $r^{{\rm gcd}(k,n)}$ as the preferred generator. That is, we take
$\tilde{\alpha}_{j}(r^{{\rm gcd}(k,n)})=\exp(2\pi i j/x)$. The
restriction of $\mathcal{T}$-irreps to $\rker{\Gamma}$ is given by
\begin{equation}
\Omega^{r^{a}K_A}_{\alpha_l} \equiv \tilde{\alpha}_{l {\rm ~mod~}x}.
\end{equation}
One should notice the duality between the situation described here and
that for symmetry breaking by electric condensates described in
section \ref{abelbreaksec}.

\subsubsection{$H=D_{2m+1}$}
\label{doddmagbreaksec}

\vspace{1mm}\noindent {\bf 1.}  First, we take the condensate state in
the module $V^{r^k}_{\beta_0}$. To find the residual symmetry algebra
$\mathcal{T}_{[r^k]}$, we first need to find the minimal normal
subgroup $K_{[r^k]}$ of $D_{2m+1}$ that contains $r^k$. This is just
the subgroup generated by the elements of the conjugacy class of
$r^k$, which are $r^k$ and $r^{-k}$. In other words, we have
$K_{[r^k]}=\spn{r^k}=\spn{r^{{\rm gcd}(k,2m+1)}}\cong \ZZ_{x}$, where
$x=(2m+1)/{\rm gcd}(k,2m+1)$. One checks easily that
$D_{2m+1}/\spn{r^k} \cong D_{{\rm gcd}(k,2m+1)}$, where this $D_{{\rm
gcd}(k,2m+1)}$ is generated in the usual way by the rotation
$R=r\spn{r^k}$ and the reflection $S=s\spn{r^k}$. We will also use the
notation $E$ for the class $e\spn{r^k}$. The residual algebra is now
\begin{equation}
\mathcal{T}_{[r^k]}(D_{2m+1})\cong 
F(D_{{\rm gcd}(k,2m+1)})\sptp \CC D_{2m+1}.
\end{equation}
The orbits of the $D_{2m+1}$ action are exactly the conjugacy classes
of $D_{{\rm gcd}(k,2m+1)}$, i.e. $\{E\}$, $\{R^p,R^{-p}\}$ (with $0\le
p <({\rm gcd}(k,2m+1)-1)/2$) and $\{S,SR,\ldots,SR^{({\rm
gcd}(k,2m+1)-1)}\}$. The stabilizers of these orbits are
$N_{E}=D_{2m+1}$, $N_{R^p}=\spn{r}\cong \ZZ_{2m+1}$ and $N_{S}=E\cup
sE\cong D_{x}$. Thus the irreps of $\mathcal{T}_{[r^k]}$ may be
written as $\Omega^{E}_{J_0}$, $\Omega^{E}_{J_1}$,
$\Omega^{E}_{\alpha_j}$ (with $1 \le j \le m$),
$\Omega^{R^p}_{\beta_l}$ (with $1 \le p\le \frac{1}{2}({\rm
gcd}(k,2m+1)-1)$, $0\le l<2m+1$), $\Pi^{S}_{J_0}$, $\Pi^{S}_{J_1}$ and
$\Pi^{S}_{\alpha_j}$ (with $1\le j \le \frac{1}{2}(x-1)$). This yields
$3+\frac{1}{2}(2m+1)({\rm gcd}(k,2m+1)+\frac{1}{{\rm gcd}(k,2m+1)})$
irreps in total and one may check that the squares of their dimensions
add up to the dimension of $\mathcal{T}_{[r^k]}$, which is $4{\rm
gcd}(k,2m+1)(2m+1)$.  The decomposition of $D(D_{2m+1})$-irreps is as
follows:
\begin{equation}
\begin{array}{ll}
\Pi^{e}_{J_0}\equiv \Omega^{E}_{J_0} &
\Pi^{r^p}_{\beta_l}\equiv
\Omega^{R^p}_{\beta_{l}}\\
\Pi^{e}_{J_1}\equiv \Omega^{E}_{J_1} &
\Pi^{s}_{\gamma_0}\equiv \Omega^{S}_{J_0} \oplus 
\bigoplus_{j} \Omega^{S}_{\alpha_{j}}\\
\Pi^{e}_{\alpha_l}\equiv \Omega^{E}_{\alpha_{l}}&
\Pi^{s}_{\gamma_1}\equiv \Omega^{S}_{J_1} \oplus 
\bigoplus_{j} \Omega^{S}_{\alpha_{j}}
\end{array}
\end{equation}
In the decomposition of $\Pi^{s}_{\gamma_0}$, we see that a chargeless
flux may be turned into a charged flux upon formation of a magnetic
condensate.

The irreps $\Omega^{\mathcal{O}}_{\tau}$ of $\mathcal{T}_{[r^k]}$ that are not
confined are exactly those for which $\tau$ is trivial on
$K_{[r^k]}=\spn{r^k}$. These are $\Omega^{E}_{J_0},\Omega^{E}_{J_1}$, the
$\Omega^{E}_{\alpha_{l}}$ and $\Omega^{R^p}_{\alpha_{l}}$ for which $l$ is a
multiple of $x$, $\Omega^{S}_{J_0}$ and $\Omega^{S}_{J_1}$. These correspond
to the irreps of the non-confined algebra 
\begin{equation}
\mathcal{U}_{[r^k]}(D_{2m+1})\cong D(D_{2m+1}/\spn{r^k})\cong D(D_{{\rm
gcd}(k,2m+1)}).
\end{equation}
The kernel of the Hopf map $\Gamma:\mathcal{T}\rightarrow \mathcal{U}$
is isomorphic to $\CC\spn{r^k}=\CC\ZZ_x$. We will denote its
representations by $\rho_l$ (with $0\le l <x$). They are defined in
the usual way, with $r^{{\rm gcd}(k,2m+1)}$ taken as the preferred
generator. The restriction of the irreps of $\mathcal{T}_{[r^k]}$ to
$\lker(\Gamma)$ is given by
\begin{equation}
\begin{array}{lll}
\Omega^{E}_{J_0}\equiv \rho_0&
\Omega^{E}_{\alpha_{l}}\equiv \rho_{l}\oplus\rho_{-l}&
\Omega^{S}_{J_0}\equiv {\rm gcd}(k,2m+1)\rho_0 \\
\Omega^{E}_{J_1}\equiv \rho_0 &
\Omega^{r^p}_{\beta_l}\equiv \rho_{l}\oplus\rho_{-l}&
\Omega^{S}_{J_1}\equiv {\rm gcd}(k,2m+1)\rho_0 \\
~&~&\Omega^{S}_{\alpha_{j}}\equiv {\rm gcd}(k,2m+1)(\rho_{j}\oplus\rho_{-j})
\end{array}
\end{equation}
Here, the indices on the $\rho$'s should be read modulo $x$.

\vspace{1mm}\noindent {\bf 2.}  Now we take the condensate state in
the module $V^{s}_{\gamma_0}$. Since the minimal normal subgroup of
$D_{2m+1}$ that contains $s$ is $D_{2m+1}$ itself, this condensate
breaks the magnetic part of $D(D_{2m+1})$ completely and we are left
with just the electric group $D_{2m+1}$, that is, we have
\begin{equation}
\mathcal{T}_{[s]}(D_{2m+1})\cong \CC D_{2m+1}.
\end{equation}
Thus, the irreps of $\mathcal{T}_{[s]}$ are just the irreps $J_0$,
$J_1$ and $\alpha_1,\ldots,\alpha_{m}$ of $D_{2m+1}$. The
decomposition of $D_{2m+1}$-irreps into these gauge group irreps is
\begin{equation}
\begin{array}{lll}
\begin{array}{l}
\Pi^{e}_{J_0}\equiv J_0\\
\Pi^{e}_{J_1}\equiv J_1 
\end{array} &
\Pi^{r^k}_{\beta_l}\equiv \left\{
\begin{array}{ll} 
J_0\oplus J_1  &(l=0) \\
\alpha_l       &(1 \le l \le m)\\
\alpha_{2m+1-l}&(m+1 \le l < 2m+1)
\end{array}\right.&
\begin{array}{l}
\Pi^{s}_{\gamma_0}\equiv J_0 \oplus \bigoplus_{j} \alpha_{j} \\
\Pi^{s}_{\gamma_1}\equiv J_1 \oplus \bigoplus_{j} \alpha_{j}
\end{array}
\end{array}
\end{equation}
Again, we see that pure fluxes may be turned into particles which
carry a charge with respect to the residual symmetry.

Of the irreps of $\mathcal{T}_{[s]}$, only the trivial representation $J_0$ is
not confined. In other words, all non-confined excitations over this
condensate are ``color'' singlets. This means that
\begin{equation}
\mathcal{U}_{[s]}(D_{2m+1})\cong \CC\{e\}.
\end{equation}
The Hopf kernel of the associated map
$\Gamma:\mathcal{T}_{[s]}\rightarrow \mathcal{U}_{[s]}$ is all of
$\mathcal{T}_{[s]}$. Hence the ``restriction'' to $\lker(\Gamma)$ is
trivial; the walls are just labeled by the irreps of
$\mathcal{T}_{[s]}$.

\section{Condensates of pure magnetic flux}
\label{purecondsec}

\subsection{Symmetry breaking}
\label{magelbreaksec}

We will now study symmetry breaking by a state with pure flux
$yg_Ay^{-1}$ in the conjugacy class $A\subset H$. The vector $\phi\in
\Pi^{A}_{1}$ that corresponds to this state is given by
$\phi(x)=1_{yN_A}(x)$. According to proposition \ref{tprop}, the
residual symmetry algebra $\mathcal{T}_{yg_Ay^{-1}}(H)$ is spanned by
the matrix elements of the set of irreps $(\rho,g)$ of $D(H)^*$ which
have the property that $\phi$ is an eigenstate of the action of
$g^{-1}$ with eigenvalue equal to
$\frac{\chi_{\rho}(g_A)}{d_{\rho}}$. In the case at hand, it is clear
that the only eigenvalue of the action of any element of $H$ that may
occur is the value $1$. It follows that $\rho$ must be such that
$\chi_{\rho}(g_A)=d_{\rho}$ and hence such that $g_A$ lies in the
kernel of $\rho$. Given such a $\rho$, we can find the corresponding
elements $g$ by solving the equation $\phi(gx)=\phi(x)$. In this
case,we have
\begin{equation}
1_{yN_A}(x)=1_{yN_A}(gx)=1_{g^{-1}yN_A}(x).
\end{equation}
Now the functions $1_{yN_A}$ and $1_{g^{-1}yN_A}$ are equal exactly if
$g^{-1}\in yN_{A}y^{-1}$, or equivalently, $g\in
yN_{A}y^{-1}=N_{yg_Ay^{-1}}$. Thus, the admissible irreps $(\rho,g)$
are those for which $g_A$ lies in the kernel of $\rho$ and $g$
commutes with the condensed flux $yg_Ay^{-1}$. Following the same
arguments as in section \ref{magbreaksec}, we see that the allowed
$\rho$ span exactly the space of functions on $H$ that are constant on
the cosets of the minimal normal subgroup $K_A$ of $H$ that contains
the class $A$. Thus the residual symmetry algebra is the Hopf
subalgebra $\mathcal{T}_{yg_Ay^{-1}}(H)$ of $D(H)$ defined by
\begin{equation}
\mathcal{T}_{yg_Ay^{-1}}(H):= \left\{ F \in D(H)|F(xk,y)=F(x,y) \forall k \in
K_A ~{\rm and}~ {\rm supp}(F) \subset H\times N_{yg_A y^{-1}}\right\}.
\end{equation}
Clearly, $\mathcal{T}_{yg_Ay^{-1}} \cong F(H/K_A \times N_{yg_A y^{-1}})$ as a
vector space and we see that $\mathcal{T}_{yg_Ay^{-1}}$ is a transformation
group algebra, with $N_{yg_Ay^{-1}}$ acting on $K_A$ through
conjugation. Thus we may again use theorem \ref{irrepth} to write down
the irreps of $\mathcal{T}_{yg_Ay^{-1}}$. They are labeled by an
$N_{yg_Ay^{-1}}$-orbit $\mathcal{O}\subset H/K_A$ and an irrep $\tau$ of
the stabilizer $N_{\mathcal{O}}$ of a chosen element
$g_{\mathcal{O}}\in \mathcal{O}$. The irrep labeled by $\mathcal{O}$
and $\tau$ will be denoted $\Omega^{\mathcal{O}}_{\tau}$.  It acts on
the Hilbert space $F_{\tau}(N_{yg_Ay^{-1}},V_{\tau})$ in the usual
way:
\begin{equation}
\label{Tgaction}
\left(\Omega^{\mathcal{O}}_{\tau}(F)\phi\right)(x):=\int_{N_{yg_Ay^{-1}}}\,dz\,
F(x g_{\mathcal{O}} x^{-1},z)\, \phi(z^{-1}x).
\end{equation} 
The character $\psi^{\mathcal{O}}_{\tau}$ of
$\Omega^{\mathcal{O}}_{\tau}$ is given as a function on $H/K_A\times
N_{yg_Ay^{-1}}$ by (cf. (\ref{trafochar2}))
\begin{equation}
\psi^{\mathcal{O}}_{\tau}(\eta,n)=
1_{N_{\eta}}(h)1_{\mathcal{O}}(\eta)\psi_{\tau}(x_{\eta}^{-1}h x_{\eta}).
\end{equation} 
The characters $\chi^{B}_{\beta}$ of the irreps $\Pi^{B}_{\beta}$ of
$D(H)$, viewed as $\mathcal{T}_{yg_Ay^{-1}}(H)$-modules are given by
\begin{equation}
\chi^{B}_{\beta}(gK_A,n)=
\sum_{k\in K_A} 1_{N_{gk}}(n)1_{A}(gk)\chi_{\beta}(x_{gk}^{-1}n x_{gk}),
\end{equation} 
where $n\in N_{yg_Ay^{-1}}$.

\subsection{Confinement}

We want to find out which of the irreps $\Omega^{B}_{\beta}$ of
$\mathcal{T}_{yg_Ay^{-1}}$ are confined and which are not. To keep
things simple, we take the condensed flux $yg_Ay^{-1}$ to be just
$g_A$. This can be done without any real loss of generality, since
$g_A$ was chosen arbitrarily in $A$. Again, the non-confined irreps
have to satisfy the requirements (\ref{modustarcond}) and again, these
reduce to (\ref{tgacond1}) and (\ref{tgacond2}) (with $K=K_A$ and
$N=N_{g_A}=N_A$), since $\mathcal{T}_{g_A}$ is isomorphic to a
transformation group algebra. In the case at hand, where
$\phi=1_{N_A}$, (\ref{tgacond1}) reduces to the requirement that
\begin{equation}
(\forall x \in N_A,\forall \eta \in B)~~
|x_{\eta}g_{\mathcal{O}}^{-1}x_{\eta}^{-1}x K_A\cap N_A|=
|x K_A\cap N_A|.
\end{equation}
By definition of $x_{\eta}$ and $g_{\mathcal{O}}$, we have
$x_{\eta}g_{\mathcal{O}}^{-1}x_{\eta}K_A=\eta$. using this and multiplying the
sets in the above equation with $x^{-1}$ from the right, we see that
it reduces to 
\begin{equation}
\label{purecond1}
(\forall \eta \in B)~~
|\eta K_A \cap N_A|=|K_A\cap N_A|.
\end{equation}
We know that $g_A\in K_A\cap N_A$ and thus that, if the above
requirement is to hold, $\eta K_A\cap N_A$ must be non-empty. But this
implies that $\eta=nK_A$ for some $n\in N_A$. On the other hand, if
this is the case, then the above equation is always satisfied. Hence,
the orbits $B$ which are not confined are those whose elements can be
written in the form $nK_A$ for some $n\in N_A$.

The condition (\ref{tgacond2}) becomes
\begin{equation}
(\forall x \in N_A,\forall \eta \in B)~~
\beta(x_{\eta}^{-1}xg_{A}x^{-1}x_{\eta})=I.
\end{equation}
Since $x\in N_A$ and $x_{\eta}\in N_A$ for all $\eta$, this reduces
further to yield the condition
\begin{equation}
\label{purecond2}
\beta(g_{A})=I
\end{equation}
on $\beta$. Basically, this says that $\beta$ must be trivial on the
minimal normal subgroup of $N_{B}$ that contains $g_A$. 

Let us compare the solutions that we have found to the set of
solutions that we had found already in proposition \ref{uprop} in
section \ref{trafoconf}. The latter set consists of all
$\Omega^{\mathcal{O}}_{\tau}$ for which the orbit $\mathcal{O}$ is
made up of cosets of the form $nK_A$ (with $n\in N_A$) and for which
$\tau$ is trivial on $K_A\cap N_A$. Thus, we see that we have not
found any extra orbits $\mathcal{O}$, but, depending on $H$, $A$ and
$N_{\mathcal{O}}$, we may have found extra irreps $\tau$ of
$N_{\mathcal{O}}$, since the minimal normal subgroup of
$N_{\mathcal{O}}$ that contains $g_A$ can be smaller than $K_A\cap
N_A$.

Thus we come back to a point that we touched upon already in section
\ref{genconfsec}, namely the fact that we are in doubt whether it is
always possible to give the full set of solutions to
(\ref{modustarcond}) a well-defined spin and and a well-defined
braiding. We do know that braiding and spin are well-defined for the
set of solutions that we had already found in section \ref{trafoconf},
since these are in one to one correspondence with the irreps of the
quantum double of $N_{A}/(K_A\cap N_{A})$. Therefore, we expect that
the non-confined symmetry algebra for the condensates treated in this
section should be $D(N_{A}/(K_A\cap N_{A}))$.

If the unconfined algebra is $D(N_{A}/(K_A\cap N_{A}))$, then the
walls that are created by $\mathcal{T}_{g_A}$-excitations can be
classified by the left or right Hopf kernel of the map
$\Gamma:\mathcal{T}_{g_A}\rightarrow D(N_{A}/(K_A\cap N_{A}))$. We
will take the right kernel, as given in proposition
\ref{kerprop}. Corollary \ref{kercorr} tells us that this Hopf kernel
is isomorphic as an algebra to the tensor product
$F((H/K_A)/\overline{N_A})\otimes\CC(N_A\cap K_A)$, where
$\overline{N_A}$ is the subgroup of $H/K_A$ which consists of elements
of the form $nK_A$, with $n\in N$. In fact, $\rker(\Gamma)$ is spanned
by the elements of $\mathcal{T}$ which are of the form $1_{hK_A
N_A}\otimes\delta_{g}$ with $g \in N_A\cap K_A$. The irreps of
$\rker(\Gamma)$ are tensor products of an irrep $E_{[\zeta]}$ of
$F((H/K_A)/\overline{N_A})$ and an irrep $\rho_l$ of $N_A\cap K_A$. We
will denote them $E_{[\zeta]}\otimes \rho_l$. Here, $[\zeta]$ is
notation for the $\overline{N_A}$-coset of $\zeta$ in $H/K_{A}$. We
have
\begin{equation}
E_{[\zeta]}\otimes \rho_l (1_{hK_A N_A}\otimes g)=
\delta_{[\zeta],[hK_A]}\rho_l(g).
\end{equation}
The decomposition of the $\mathcal{T}_{g_A}$-irrep $\Omega^{\mathcal{O}}_{\tau}$
into $\rker(\Gamma)$ irreps may be found using the formula (\ref{trfc})
for the character $\psi^{\mathcal{O}}_{\tau}$. We have 
\begin{eqnarray}
\psi^{\mathcal{O}}_{\tau}(1_{hK_A N_A}\otimes\delta_{g})&=&
\int_{\mathcal{O}} d\zeta \int_{K_A\cap N_A} dn \,
1_{hK_A N_A}(x_{\zeta}\xi_{\mathcal{O}}x_{\zeta}^{-1})
\delta_{g}(x_{\zeta}n x_{\zeta}^{-1})\chi_{\alpha}(n) \nonumber \\
~&=&
\int_{\mathcal{O}} d\zeta \,
1_{hK_A N_A}(x_{\zeta}\xi_{\mathcal{O}}x_{\zeta}^{-1})
\chi_{\alpha}(x_{\zeta}^{-1} g x_{\zeta}) \nonumber \\
~&=&
\int_{\mathcal{O}} d\zeta \,
\delta_{[\zeta],[hK_A]}\chi_{\alpha}(x_{\zeta}^{-1} g x_{\zeta}). 
\end{eqnarray} 
{}From this, we read off that $\Omega^{\mathcal{O}}_{\tau}$ is the sum over
$\zeta \in \mathcal{O}$ of those $E_{[\zeta]}\otimes \rho_l$ for which
$\rho_l$ is related to one of the $N_A\cap K_A$-irreps contained in
$\tau$ by conjugation with $x_{\zeta}^{-1}$. Of course, there may be
multiplicities in the decomposition, for example if the coset
$[\zeta]$ is the same for several $\zeta \in \mathcal{O}$. Also, note
that the non-confined irreps of $\mathcal{T}_{g_A}$ all correspond to the
trivial irrep $E_{[K_A]}\otimes 1$, as they should.

\subsection{Examples of pure flux condensates}
\label{purexsec}

Pure flux condensates whose flux is central in $H$ are gauge invariant
and examples may be found in section \ref{magexsec}. Here we treat the
case where the flux of the condensate is non-central, so that not only
the magnetic part of the double, but also the electric group is
broken.

\subsubsection{$H=D_{2m+1}$}

\vspace{1mm}\noindent {\bf 1.}  Suppose the condensed flux is $r^k \in
D_{2m+1}$. In that case, the residual symmetry algebra is the
transformation group algebra $F(D_{2m+1}/K_{r^k})\sptp \CC N_{r^k}$,
where $K_{r^k}$ is the minimal normal subgroup that contains $r^k$ and
$N_{r^k}=\spn{r}$ is the centralizer of $r^k$ in $D_{2m+1}$. From
section \ref{doddmagbreaksec}, we know that
$K_{r^k}=\spn{r^k}=\spn{r^{{\rm gcd}(k,2m+1)}}\cong \ZZ_{x}$, where
$x=(2m+1)/{\rm gcd}(k,2m+1)$. We also recall that $D_{2m+1}/\ZZ_x
\cong D_{{\rm gcd}(k,2m+1)}$. Hence,
\begin{equation}
\mathcal{T}_{r^k}(D_{2m+1})\cong F(D_{{\rm gcd}(k,2m+1)})\sptp \CC\ZZ_{2m+1}.
\end{equation}
The $D_{{\rm gcd}(k,2m+1)}$ is generated $R=r\ZZ_x$ and $S=s\ZZ_x$ and
we will write $E$ for its unit element $e\ZZ_x$. The normalizer
$N_{r^k}\subset D_{2m+1}$ is just the group of rotations:
$N_{r^k}=\spn{r}\cong \ZZ_{2m+1}$. The orbits of the action of
$N_{r^k}$ on $K_{r^k}$ are $\{E\}$, $\{R\}$, $\{R^2\},\ldots,\{R^{{\rm
gcd}(k,2m+1)-1}\}$ and $\{S,SR,\ldots,SR^{{\rm gcd}(k,2m+1)-1}\}$. The
stabilizer of the orbits with one element is of course
$N_{r^k}\cong\ZZ_{2m+1}$, while the orbit of $S$ has stabilizer
$K_{r^k} \cong \ZZ_x$. It follows that the representations of
$\mathcal{T}_{r^k}$ may be written $\Omega^{R^p}_{\beta_{l}}$ (with
$0\le p <{\rm gcd}(k,2m+1)$, $0\le l <2m+1$) and
$\Omega^{S}_{\beta_{l}}$ (with $0\le l <x$). We see that
$\mathcal{T}_{r^k}$ has $(2m+1){\rm gcd}(k,2m+1)$ irreps of dimension
$1$ and $x$ irreps of dimension ${\rm gcd}(k,2m+1)$. The squares of
the dimensions add to the dimension of $\mathcal{T}_{r^k}$, which is
$2(2m+1)^2/{\rm gcd}(k,2m+1)$. The decomposition of
$D(D_{2m+1})$-irreps into $\mathcal{T}_{r^k}$-irreps is as follows:
\begin{equation}
\begin{array}{ll}
\Pi^{e}_{J_0}\equiv \Omega^{E}_{\beta_0} &
\Pi^{r^k}_{\beta_l}\equiv \Omega^{R^k}_{\beta_l}\oplus \Omega^{R^{-k}}_{\beta_{-l}}\\
\Pi^{e}_{J_1}\equiv \Omega^{E}_{\beta_0} &
\Pi^{s}_{\gamma_0}\equiv \bigoplus_{l} \Omega^{S}_{\beta_{l}}\\
\Pi^{e}_{\alpha_l}\equiv \Omega^{E}_{\beta_{l}}\oplus \Omega^{E}_{\beta_{-l}}&
\Pi^{s}_{\gamma_1}\equiv \bigoplus_{l} \Omega^{S}_{\beta_{l}}\\
\end{array}
\end{equation}
The irreps of $\mathcal{T}_{r^k}$ which are not confined are the
$\Omega^{R^k}_{\beta_l}$ for which $\beta_{l}(r^k)=1$, or in other words,
those for which $l$ is a multiple of $x$. It follows that the
unconfined representations are automatically in one to one
correspondence with the irreps of 
\begin{equation}
\mathcal{U}_{r^k}(D_{2m+1})\cong D(N_{r^k}/(K_{r^k}\cap
N_{r^k}))\cong D(\ZZ_{{\rm gcd}(k,2m+1)}). 
\end{equation}
The right Hopf kernel of the Hopf map $\Gamma:
\mathcal{T}_{r^k}\rightarrow D(\ZZ_{{\rm gcd}(k,2m+1)})$ is isomorphic
to $F(\ZZ_2)\otimes \CC\ZZ_{x}$ and we may denote its representations
as $E_{[E]}\otimes \rho_{l}$ and $E_{[S]}\otimes \rho_{l}$ (with $0\le
l<x$). Here, $[E]$ and $[S]$ denote the $\overline{N_{r^k}}$-cosets of
the $K_{r^k}$-cosets $E$ and $S$ and $\rho_l$ denotes the $l^{\rm th}$
representation of $\ZZ_x$, defined in the usual way, with $r^{{\rm
gcd}(k,2m+1)}$ taken as the preferred generator of $\ZZ_x$. The
restriction of the irreps of $\mathcal{T}_{r^k}$ to $\rker{\Gamma}$ is
given by
\begin{equation}
\Omega^{R^k}_{\beta_l}\equiv E_{[E]}\otimes \rho_{l {\rm ~mod~}x} ~~~~
\Omega^{S}_{\beta_{l}}\equiv {\rm gcd}(k,2m+1)
E_{[S]}\otimes \rho_{l {\rm ~mod~}x}
\end{equation}

\vspace{1mm}\noindent {\bf 2.}
Now suppose the condensate has flux $s \in D_{2m+1}$. The
minimal normal subgroup of $D_{2m+1}$ that contains $s$ is $D_{2m+1}$
itself and the normalizer $N_s$ of $s$ is just $\{e,s\}\cong
\ZZ_2$. Hence, this condensate leaves us with the symmetry algebra
\begin{equation}
\mathcal{T}_{s}(D_{2m+1})\cong F(\ZZ_2)\cong \CC\ZZ_2.
\end{equation}
The irreps of this $\ZZ_2$ may be labeled $\Omega_{J_0}$ and
$\Omega_{J_1}$ and the decomposition of $D(D_{2m+1})$-irreps is then
given by
\begin{equation}
\begin{array}{ll}
\Pi^{e}_{J_0}\equiv \Omega_{J_0} &
\Pi^{r^k}_{\beta_l}\equiv \Omega_{J_0}\oplus \Omega_{J_1}\\
\Pi^{e}_{J_1}\equiv \Omega_{J_1} &
\Pi^{s}_{\gamma_0}\equiv (m+1) \Omega_{J_0}\oplus m \Omega_{J_1}\\
\Pi^{e}_{\alpha_l}\equiv \Omega_{J_0}\oplus \Omega_{J_1}&
\Pi^{s}_{\gamma_1}\equiv  m \Omega_{J_0}\oplus (m+1) \Omega_{J_1}\\
\end{array}
\end{equation}
Since $J_1(s)=-1\neq 1$, it follows that $\Omega_{J_1}$ is
confined, so that the only non-confined irrep of $\mathcal{T}_{s}$ is
the ``color singlet'' $\Omega_{J_0}$. Hence
\begin{equation}
\mathcal{U}_{s}(D_{2m+1})\cong \CC\{e\}
\end{equation}
and the corresponding Hopf kernel equals $\mathcal{T}_{s}$.

\section{Dyonic condensates}
\label{dyonsec}

Attempts to study dyonic condensates in the same generality as
electric or magnetic condensates meet with some problems of a
technical nature. For example, the residual algebra after symmetry
breaking does not have to be a transformation group algebra of the
kind we discussed in section \ref{trafoconf} (see the second part of
section \ref{abdysec} for an example). Therefore, we will only treat
some examples with specific groups and condensate vectors here, in
order to give an idea of what one may expect. In the process, we also
complete our treatment of condensates in theories where the gauge
group is an odd dihedral group.

\subsection{$H=\ZZ_n$}
\label{abdysec}

First of all, let us check which condensates satisfy the requirements
of trivial spin and self-braiding that we gave in section
\ref{stabsec}. As before, we denote our favorite generator of $\ZZ_n$
as $r$ and we denote the representations of this group, defined in the
usual way, as $\alpha_{l}$ (with $0\le l<n$). The representations of
$D(\ZZ_n)$ may then be written $\Pi^{r^{k}}_{\alpha_{l}}$. The spin
factor $s^{r^{k}}_{\alpha_{l}}$ of $\Pi^{r^{k}}_{\alpha_{l}}$ is just
$\exp(-2\pi i kl/n)$ and so the requirement of trivial spin selects
those $\Pi^{r^{k}}_{\alpha_{l}}$ for which we have
\begin{equation}
kl=0 {\rm ~mod~}n.
\end{equation}
Thus, given $k$, the allowed $l$ are those which are $0$ modulo
$n/{\rm gcd}(k,n)$ and given $l$, the allowed $k$ are those which are
$0$ modulo $n/{\rm gcd}(l,n)$. From this, we see immediately that, if
$n$ is a prime, there will be no allowed dyonic condensates (either
$l$ or $k$ has to be zero). We will thus assume from now on that $n$
is composite. For Abelian groups, the requirement of trivial
self-braiding is automatically satisfied for states with trivial spin,
so the $\Pi^{r^{k}}_{\alpha_{l}}$ with $kl=0 {\rm ~mod~}n$ all give
good condensates.

To find the residual algebra $\mathcal{T}^{k}_{l}$ for a
$\Pi^{r^{k}}_{\alpha_{l}}$-condensate, we have to find the
representations of $D(\ZZ_n)^{*}$ that satisfy equation
(\ref{invareq}). Since $D(\ZZ_n)^{*}\cong \CC\ZZ_n\otimes F(\ZZ_n)$,
its representations may be labeled by an irrep $\alpha_{q}$ of $\ZZ_n$
and an element $r^{p}$ of $\ZZ_n$. The equation (\ref{invareq}) then
selects those $(\alpha_p,r^q)$ for which
$\alpha_{p}(r^k)\alpha_{l}(r^q)=1$, or more explicitly, those
$(\alpha_p,r^q)$ for which $\exp(2\pi i (kp+lq)/n)=1$. This means
that, to find the allowed $p$ and $q$, we have to solve the equation 
\begin{equation}
\label{abdyreseq}
kp+lq = 0 {\rm ~mod~}n.
\end{equation}
Rather than looking at the general solution of this equation for all
$k$ and $l$, we will examine two illustrative special cases:

\vspace{1mm}\noindent {\bf 1.}
First, let us take $(n,k,l)$ such that $n=kl$ and ${\rm
gcd}(k,l)=1$. In this situation, we can easily find the solution to
equation (\ref{abdyreseq}). Since $kp+lq$ is a multiple of $n$, say
$mn$, we may solve for $p$ to get 
\begin{equation}
p = \frac{1}{k}(mn-lq) = lm - \frac{l}{k}q,
\end{equation}
using $n=kl$ in the second equality. Since $p$ has to be integer, it
follows that $\frac{l}{k}q$ must be an integer. Since ${\rm
gcd}(k,l)=1$, the fraction $\frac{l}{k}$ is irreducible and hence
$\frac{l}{k}q$ can only be integer if $q$ is a multiple of $k$. But
then it follows from the equation above that $p$ is a multiple of
$l$. On the other hand, it is clear from $n=kl$ that any $(p,q)$ for
which $p$ is an $l$-fold and $q$ is a $k$-fold will solve
(\ref{abdyreseq}). Thus the residual algebra $\mathcal{T}^{k}_{l}$ is
spanned by the (matrix elements of) the representations
$(\alpha_{p},r^{q})$ for which $p=0 {\rm ~mod~}l$ and $q=0 {\rm
~mod~}k$. Now since $n=kl$, the irreps $\alpha_{p}$ of $\ZZ_n$ with
$p=0 {\rm ~mod~}l$ correspond exactly to the irreps of the quotient
group $\ZZ_n/\spn{r^k}\cong\ZZ_k$. Hence,
\begin{equation}
\mathcal{T}^{k}_{l}\cong F(\ZZ_n/\spn{r^k})\sptp \CC\spn{r^{k}}
\cong F(\ZZ_k)\sptp \CC\ZZ_l.
\end{equation}
We see that $\mathcal{T}^{k}_{l}$ is a transformation group algebra of
the kind treated in section \ref{trafoconf}, where both the normal
subgroup $K$ and the subgroup $N$ of these sections equal
$\spn{r^k}\cong\ZZ_l$ in this case. The representations of
$\mathcal{T}^{k}_{l}$ may thus be denoted $\Omega^{r}_{s}$, with $0\le
r<k$ and $0\le s<l$ and the restriction of the irreps of $D(\ZZ_n)$ to
$\mathcal{T}^{k}_{l}$ is given by
\begin{equation}
\Pi^{r^a}_{\alpha_b}\equiv \Omega^{a {\rm ~mod~}k}_{b {\rm ~mod~}l}.
\end{equation}
Using the theory of section \ref{trafoconf}, one may see that all the
$\Omega^{r}_{s}$ with $(r,s)\neq (0,0)$ confined. The unconfined
algebra $\mathcal{U}^{k}_{l}$ is thus the group algebra of the trivial
group and the Hopf kernel of the Hopf map
$\Gamma:\mathcal{T}^{k}_{l}\rightarrow \mathcal{U}^{k}_{l}$ is all of
$\mathcal{T}^{k}_{l}$, implying that walls and
$\mathcal{T}^{k}_{l}$-particles are in one-to-one correspondence.

\vspace{1mm}\noindent {\bf 2.}  Now consider the case where $l=-k {\rm
~mod~}n$. Equation (\ref{abdyreseq}) then becomes
\begin{equation}
k(p-q) = 0 {\rm ~mod~}n
\end{equation}
so that the allowed $(p,q)$ are those for which $p=q {\rm ~mod~}n/{\rm
gcd}(k,n)$. It follows that
\begin{equation}
\mathcal{T}^{k}_{l}\cong \CC(\ZZ_n\times\ZZ_{{\rm gcd}(k,n)}),
\end{equation}
where $(\alpha_1,r)$ generates the $\ZZ_n$ and where either
$(\alpha_{n/{\rm gcd}(k,n)},e)$ or $(\alpha_0,r^{n/{\rm gcd}(k,n)})$
can be taken as the generator for the $\ZZ_{{\rm gcd}(k,n)}$. We will
take the latter possibility. One should notice that, in contrast to
everything we have seen up to now, the full residual algebra is not
generated by the residual magnetic and the residual electric symmetry
algebra. The residual electric and magnetic algebra are generated by
$(\alpha_0,r^{n/{\rm gcd}(k,n)})$ and $(\alpha_{n/{\rm gcd}(k,n)},e)$
respectively and are both isomorphic to $\CC\ZZ_{{\rm gcd}(k,n)}\cong
F(\ZZ_{{\rm gcd}(k,n)})$. The total residual algebra is
$\CC(\ZZ_n\times\ZZ_{{\rm gcd}(k,n)})$ and contains for example the
element $(\alpha_1,r)$, which cannot be generated from the elements of
the residual electric and magnetic algebras. Clearly, the residual
algebra is not a transformation group algebra of the kind treated in
section \ref{trafoconf}.  This phenomenon is not limited to Abelian
$H$, but can also occur for non-Abelian $H$. In fact, one may check
that it does so already for some condensates in a $D(D_4)$-theory.

The representations of $\mathcal{T}^{k}_{l}$ may be written
$\chi_{a,b}$, with $0\le a<n$, $0\le b<{\rm gcd}(k,n)$. They are
defined in the usual way, through
\begin{eqnarray}
\chi_{a,b}(\alpha_1,r)&=&e^{2\pi i a/n} \nonumber \\
\chi_{a,b}(\alpha_0,r^{n/{\rm gcd}(k,n)})&=&e^{2\pi i b/{\rm
gcd}(n,k)}.
\end{eqnarray}
On the magnetic part of $\mathcal{T}^{k}_{l}$, $\chi_{a,b}$ is given by
$\chi_{a,b}(\alpha_{n/{\rm gcd}(k,n)},e)=e^{2\pi i (a-b)/{\rm
gcd}(n,k)}$, as follows from the definition above. The restriction of
the irreps of $D(\ZZ_n)$ to $T$ is given by
\begin{equation}
\Pi^{r^p}_{\alpha_q}\equiv \chi_{p+q,\,q},
\end{equation}
where the second $q$ on the right hand side should be read modulo
${\rm gcd}(k,n)$. 

Since we cannot apply the theory of section \ref{trafoconf} here, we
have to refer back to the requirements (\ref{modustarcond}) in section
\ref{genconfsec} in order to determine which of the representations of
$\mathcal{T}^{k}_{l}$ are confined and which are not. After some algebra, the
first of these requirements, applied to $f=\chi_{a,b}$, reduces to
\begin{equation}
e^{-2\pi i (a-b)k/n}=1,
\end{equation}
{}from which it follows that 
\begin{equation}
a=b {\rm ~mod~}n/{\rm gcd}(n,k).
\end{equation}  
Note that $n/{\rm gcd}(n,k)$ is a divisor of both $n$ and ${\rm
gcd}(n,k)$, since $k^2=0 {\rm ~mod~}n$. As a consequence, the above
equation retains its usual meaning, despite the fact that $a$ is only
defined modulo $n$ and $b$ is only defined modulo ${\rm gcd}(n,k)$.

\begin{figure}[h,t,b]
\psfig{figure=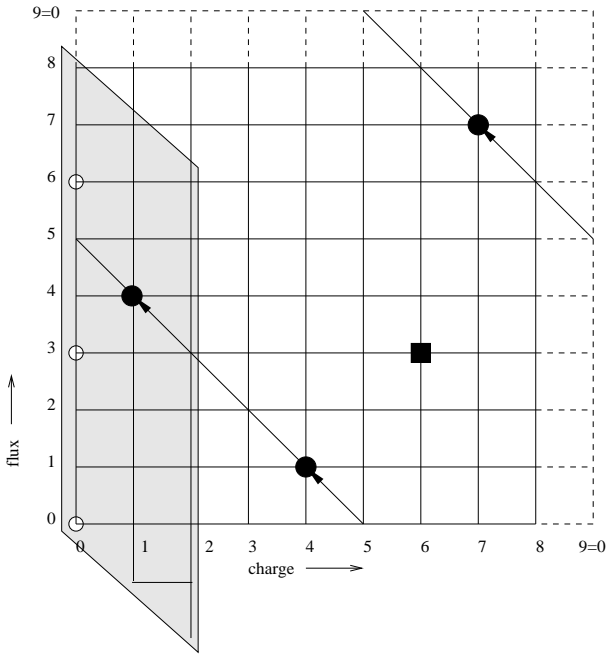,width=8cm}
\caption[something]{\footnotesize Flux-charge lattice for a $D(\ZZ_9)$
theory. We assume that a condensate of particles with flux $r^3 \equiv
3$ and charge $\alpha_{-3}\equiv-3\equiv 6$ forms. The condensed irrep
is indicated as a square dot. The residual symmetry algebra
$\mathcal{T}^{3}_{-3}$ is a group algebra $\CC(\ZZ_9\times\ZZ_3)$. Two
$D(H)$-irreps in the lattice are equivalent as
$\mathcal{T}^{3}_{-3}$-irreps if one can be reached from the other
through translations by the ``condensate vector'' $(-3,3)$. This way,
$D(H)$-irreps are identified in trios. The trio in the picture
corresponds to the $\mathcal{T}$-irrep $\chi_{5,1}$. The shaded region
contains one representative from each trio and is thus a diagram of
all $\mathcal{T}$-irreps. The small white circles indicate the three
unconfined irreps of $\mathcal{T}$, which correspond to the irreps of
$\mathcal{U}\cong\CC\ZZ_3$.}
\label{dycondfig}
\end{figure}

The second requirement in (\ref{modustarcond}), applied
to $f=\chi_{a,b}$, becomes
\begin{equation}
e^{2\pi ibk/n}=1,
\end{equation}
so that we have  
\begin{equation}
b=0 {\rm ~mod~}n/{\rm gcd}(n,k).
\end{equation}
Hence, the non-confined representations of $\mathcal{T}^{k}_{l}$ are
just those $\chi_{a,b}$ for which both $a$ and $b$ are multiples of
$n/{\rm gcd}(n,k)$. This leaves ${\rm gcd}(n,k)$ possibilities for $a$
and $({\rm gcd}(n,k))^2/n$ possibilities for $b$, so that the
non-confined algebra $\mathcal{U}^{k}_{l}$ is given by
\begin{equation}
\mathcal{U}^{k}_{l}\cong \CC(\ZZ_{{\rm gcd}(n,k)}\times
\ZZ_{({\rm gcd}(n,k))^2/n}).
\end{equation}
As an example, consider the case of $D(\ZZ_9)$, with a condensate
given by $k=-l=3$. The only non-confined irreps of
$\mathcal{T}^{3}_{-3}$ are then $\chi_{0,0},\chi_{3,0}$ and
$\chi_{6,0}$ and the unconfined algebra is $\CC\ZZ_3$. We give a
graphical representation of our results for this case in figure
\ref{dycondfig}.

\subsection{$H=D_{2m+1}$}

In this section, we complete our treatment of condensates in the odd
dihedral gauge theories. First, we find out which states in dyonic
representations of $D(D_{2m+1})$ satisfy the conditions of trivial
spin and trivial self-braiding. From table \ref{ddoddspintab}, we read
off that the only dyonic irreps of $D(D_{2m+1})$ which have trivial
spin are those $\Pi^{r^k}_{\beta_{l}}$ for which $\exp(2\pi i
kl/(2m+1))=1$, or in other words, for which $kl=0 {\rm
~mod~2m+1}$. It follows that there are no admissible dyonic
condensates when $2m+1$ is prime. If $2m+1$ is not a prime, then there
will be dyons with trivial spin and one may check easily that any
state in the module of one of the $\Pi^{r^k}_{\beta_{l}}$ with $kl=0 {\rm
~mod~2m+1}$ also has trivial self-braiding. Therefore, all states in
the modules of these dyonic irreps may in principle be condensed.
 
Now suppose that we have condensed a state $\phi$ in the module of
$\Pi^{r^k}_{\beta_{l}}$. To find the residual symmetry algebra, we
have to solve equation (\ref{inveq}). The representations $(\rho,g)$
of $D(D_{2m+1})^{*}$ which satisfy this equation are those for which
$\frac{\chi_{\rho}(r^k)}{d_{\rho}}$ is a root of unity and $\phi$ is
an eigenvector of $g^{-1}$ with eigenvalue equal to this root of
unity. Thus, let us first find all irreps $\rho$ of $D_{2m+1}$ for
which $\frac{\chi_{\rho}(r^k)}{d_{\rho}}$ is a root of unity. From
table \ref{doddchartab} one may read off that these are $J_0, J_1$ and
those $\alpha_{j}$ for which $2\cos(2\pi jk/(2m+1))=2$, or
equivalently $jk=0 {\rm ~mod~}2m+1$. This leaves exactly those $j$
which are multiples of $x_k:=(2m+1)/{\rm gcd}(k,2m+1)$. In all these
cases, $\frac{\chi_{\rho}(r^k)}{d_{\rho}}$ actually equals $1$, or
equivalently, $\rho$ is trivial on $r^k$. It follows that the allowed
$\rho$ correspond to the irreps of the quotient group
$D_{2m+1}/\spn{r^k}$, which is isomorphic to $D_{{\rm gcd}(k,2m+1)}$.
The residual symmetry algebra will now be spanned by the $(\rho,g)$
with $\rho$ in the set we have just found and $g$ an element of the
subgroup of $D_{2m+1}$ that leaves $\phi$ invariant.  This subgroup
will depend on $\phi$.  Therefore, let us write $\phi$ more explicitly
as $a\phi_{r^k}+b\phi_{r^{-k}}$. Here $\phi_{r^k}$ and $\phi_{r^{-k}}$
are just the two basis functions for $V^{r^k}_{\beta_l}$ as defined
through (\ref{traforepbas}). We have dropped the index $i$ in
$(\ref{traforepbas})$, since the module of $\beta_l$ is
one-dimensional. Using the formula (\ref{trafomatelts}) for the matrix
elements of $\Pi^{r^k}_{\beta_l}$ with respect to this basis, we can
now write the action of the elements of $D_{2m+1}$ on $\phi$
explicitly as
\begin{eqnarray}
\Pi^{r^k}_{\beta_l}(1\otimes\delta_{r^p})(a\phi_{r^k}+b\phi_{r^{-k}})&=&
e^{\frac{2\pi i lp}{2m+1}}a\phi_{r^k}+
e^{\frac{-2\pi i lp}{2m+1}}b\phi_{r^{-k}}
\nonumber \\
\Pi^{r^k}_{\beta_l}(1\otimes sr^p)(a\phi_{r^k}+b\phi_{r^{-k}})&=&
e^{\frac{-2\pi i lp}{2m+1}}b\phi_{r^k}+
e^{\frac{2\pi i lp}{2m+1}}a\phi_{r^{-k}}.
\end{eqnarray}
{}From the first of these equations, we see that, independently of the
choice of $(a,b)$, $r^p$ will leave $\phi$ invariant only if
$\exp(2\pi i lp/(2m+1))=1$. In other words, $p$ has to be a multiple
of $x_l:=(2m+1)/{\rm gcd}(2m+1,l)$. From the second equation above, we
see that $sr^p$ will leave $\phi$ invariant only if $b=\exp(2\pi i
lp/(2m+1))a$. If no such relation between $a$ and $b$ exists, then
none of the elements $sr^p\in D_{2m+1}$ will leave $\phi$ invariant
and the subgroup of $D_{2m+1}$ that does leave $\phi$ invariant is
just the $\ZZ_{{\rm gcd}(2m+1,l)}$ generated by $r^{x_l}$. If we do
have $b=\exp(2\pi i lp/(2m+1))a$ for some $p$, then the required
subgroup of $D_{2m+1}$ is the $D_{{\rm gcd}(2m+1,l)}$ generated by
$r^{x_l}$ and $sr^p$. All these $D_{{\rm gcd}(2m+1,l)}$ subgroups
actually represent the same physics, since they are conjugates in
$D_{2m+1}$ (or equivalently, the corresponding condensates are all
related by gauge transformations). We have thus found two distinct
possibilities for the residual symmetry algebra $\mathcal{T}$, which
we will call $\mathcal{T}^{k}_{l}$ and
$\bar{\mathcal{T}}^{k}_{l}$. These algebras are given by
\begin{eqnarray}
\mathcal{T}^{k}_{l}&\cong&F(D_{2m+1}/\spn{r^k})\sptp\CC\spn{r^{x_l}}
\cong F(D_{{\rm gcd}(2m+1,k)})\sptp\CC\ZZ_{{\rm gcd}(2m+1,l)} \nonumber \\
\bar{\mathcal{T}}^{k}_{l}&\cong&F(D_{2m+1}/\spn{r^k})\sptp\CC\spn{r^{x_l},s}
\cong F(D_{{\rm gcd}(2m+1,k)})\sptp\CC D_{{\rm gcd}(2m+1,l)}.
\end{eqnarray}
Both $\mathcal{T}^{k}_{l}$ and $\bar{\mathcal{T}}^{k}_{l}$ are thus
transformation group algebras of the for $F(H/K)\sptp\CC N$ (for
$\mathcal{T}^{k}_{l}$, we have $K=\spn{r^k}$ and $N=\spn{r^{x_l}}$,
whereas for $\bar{\mathcal{T}}^{k}_{l}$, we have $K=\spn{r^k}$ and
$N=\spn{r^{x_l},s}$). Because of this, the decomposition of
$D(D_{2m+1})$-irreps into $\mathcal{T}$-irreps proceeds in the same
way as for the electric and magnetic cases. Also, the theory of
section \ref{trafoconf} may be applied to treat confinement. One finds
that the unconfined algebras $\mathcal{U}^{k}_{l}$ and
$\bar{\mathcal{U}}^{k}_{l}$ are isomorphic to the quantum doubles of
the groups $N/(K\cap N)$. Now it turns out that we have $K\subset
\spn{r^{x_l}}$ and hence $K\subset N$ in both cases. To see this,
remember that we have $kl=0 {\rm ~mod~2m+1}$ and thus ${\rm
gcd}(k,2m+1){\rm gcd}(l,2m+1)=0 {\rm ~mod~2m+1}$. Hence,
\begin{equation}
{\rm gcd}(k,2m+1){\rm gcd}(l,2m+1)=q\, (2m+1)
\end{equation}
for some integer $q$ and it follows that ${\rm gcd}(k,2m+1)=q x_l$ and
$\spn{r^k}\subset \spn{r^{x_l}}$ (note that $\spn{r^k}=\spn{r^{{\rm
gcd}(k,2m+1)}})$. The integer $q$ has the property that it divides
both ${\rm gcd}(k,2m+1)$ and ${\rm gcd}(l,2m+1)$. Using this, one now
sees easily that 
\begin{eqnarray}
\mathcal{U}^{k}_{l}&\cong& D(\ZZ_{q}) \nonumber \\
\bar{\mathcal{U}}^{k}_{l}&\cong& D(D_{q}).
\end{eqnarray}
When $q=1$, this means that there is full confinement of
$\mathcal{T}^{k}_{l}$-irreps, while on the other hand, there are still
four unconfined $\bar{\mathcal{T}}^{k}_{l}$-irreps, since
$\bar{\mathcal{U}}^{k}_{l}\cong D(\ZZ_2)$. 

The Hopf kernels of the maps $\Gamma:\mathcal{T}^{k}_{l}\rightarrow
\mathcal{U}^{k}_{l}$ and
$\bar{\Gamma}:\bar{\mathcal{T}}^{k}_{l}\rightarrow
\bar{\mathcal{U}}^{k}_{l}$ can also be determined, following the
treatment in section \ref{trafoconf}. We find that
\begin{eqnarray}
\rker(\Gamma)&\cong&F(D_{x_{l}})\otimes\CC\ZZ_{x_k} \nonumber \\
\rker(\bar{\Gamma})&\cong&
F(D_{{\rm gcd}(k,2m+1)}/D_{{\rm gcd}(k,2m+1)/x_l})\otimes\CC\ZZ_{x_k}.
\end{eqnarray}

\section{Summary and Outlook}
\label{conclusec}

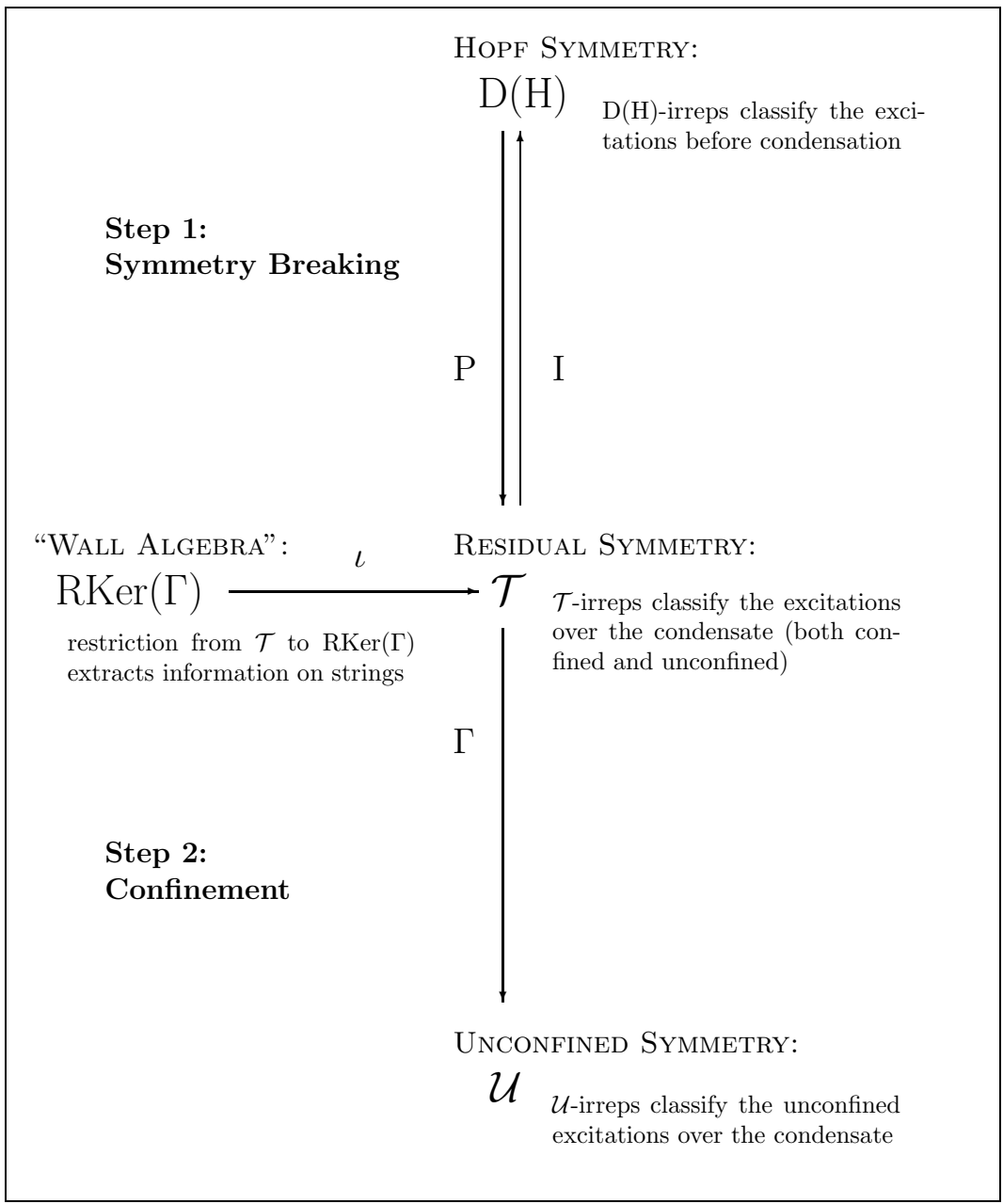
\begin{figure}[h,b,t]
\centerline{
\begin{picture}(400,480)(0,120)
\put(0,600){\line(1,0){400}}
\put(0,600){\line(0,-1){480}}
\put(0,120){\line(1,0){400}}
\put(400,600){\line(0,-1){480}}
\put(180,580){\sc Hopf Symmetry:}
\put(190,560){\Large D(H)}
\put(240,550){\parbox{130pt}{\footnotesize D(H)-irreps classify the
excitations before condensation}}
\put(200,550){\vector(0,-1){150}}
\put(180,450){\large P}
\put(40,500){\parbox{130pt}{\bf Step 1:\\ Symmetry Breaking}}
\put(207,400){\vector(0,1){150}}
\put(220,450){\large I}
\put(180,380){\sc Residual Symmetry:}
\put(195,360){\Large $\mathcal{T}$}
\put(220,345){\parbox{140pt}{\footnotesize $\mathcal{T}$-irreps classify the
excitations over the condensate (both confined and unconfined)}}
\put(200,350){\vector(0,-1){150}}
\put(180,300){\large \mbox{$\Gamma$}}
\put(40,250){\parbox{130pt}{\bf Step 2:\\ Confinement}}
\put(180,180){\sc Unconfined Symmetry:}
\put(195,160){\Large $\mathcal{U}$}
\put(220,150){\parbox{140pt}{\footnotesize $\mathcal{U}$-irreps
classify the unconfined excitations over the condensate}}
\put(90,365){\vector(1,0){100}}
\put(140,375){\large \mbox{$\iota$}}
\put(10,380){\sc ``Wall Algebra'':}
\put(20,360){\Large \mbox{$\rker(\Gamma)$}}
\put(25,335){\parbox{140pt}{\footnotesize restriction from $\mathcal{T}$ to
\mbox{$\rker(\Gamma)$} extracts information on strings}}
\end{picture}}
\caption[something]{Schematic picture of the structures that play a
role in this paper}
\label{finalpic}
\end{figure}

The general picture that emerges from this paper can be seen in figure
\ref{finalpic}. In words, it is as follows. The formation of a
condensate induces symmetry breaking from $D(H)$ to the Hopf
subalgebra $\mathcal{T}\subset D(H)$ which is the Hopf stabilizer of
the condensate state. The ensuing confinement is described by a Hopf
projection $\Gamma$ of $\mathcal{T}$ onto an ``unconfined'' symmetry
algebra $\mathcal{U}$, whose irreps label the free charges over the
condensate. Walls or strings in the condensate are labeled by the
restrictions of $\mathcal{T}$-irreps to the right Hopf kernel of
$\Gamma$. In the diagram, $I$ denotes the (Hopf) inclusion of
$\mathcal{T}$ into $D(H)$, $\iota$ denotes the inclusion of
$\lker(\Gamma)$ into $\mathcal{T}$ and $P$ denotes the orthogonal
projection of $D(H)$ onto $\mathcal{T}$, which we use in our
definition of $\mathcal{U}$.  To the information in the diagram, we
should add that all ``baryonic'' excitations on the condensate can be
constructed by fusing together a number of confined particles, labeled
by $\mathcal{T}$-irreps, in such a way that the overall fusion product
has a non-confined charge, labeled by a $\mathcal{U}$-irrep.
\pagebreak

Note that the role that the unconfined algebra $\mathcal{U}$ plays in
the $D(H)$-theory is quite comparable to the role that $D(H)$ plays in
the gauge theory with continuous gauge group $G$ of which our discrete
gauge theory is a Higgsed version. Just like $D(H)$ classifies the
free excitations over the Higgs condensate in the continuous gauge
theory, $\mathcal{U}$ classifies the free excitations over the
condensate in the $D(H)$-theory.  In fact, the different unconfined
algebras we have found for specific condensates are typically
themselves quantum doubles of a group related to $H$. For example:
\begin{itemize}
\item
For purely electric condensates, we have found that $\mathcal{U}$ is
the quantum double of the stabilizer $N$ of the condensate in
$H$. This is just what we expected, since the only effect of
condensing one of the electric particles of the $D(H)$-theory is to
modify the electric condensate of the $G$-theory in such a way that
the residual gauge group is now $N$ rather than $H$.
\item
For gauge invariant magnetic condensates, we have found that $\mathcal{U}$ is
the quantum double of the quotient group $H/K$, where $K$ is the group
generated by the elements of conjugacy class that labels the
condensate. This is also in accordance with the intuition, since the
division by $K$ can be seen as a consequence of the fact that, after
condensation, the flux of any particle can only be determined up to
the condensed flux. 
\end{itemize}

In a sense, we can describe the condensed phases of the $D(H)$-theory
even better than the $D(H)$-theory itself describes the Higgs phase of
the $G$-theory, since the algebra $\mathcal{T}$ that we obtain after symmetry
breaking gives us information on the possible substructures of the
free excitations over the condensate.

Nevertheless, there is still much work to be done. First of all, from
the requirements (\ref{modustarcond}) that we found in section
\ref{confsec}, it is not clear that the set of irreps of $\mathcal{U}$ will
always have a well-defined braiding. Although this does happen in the
examples with electric and gauge invariant magnetic condensates (where
$\mathcal{U}$ is even quasitriangular), we do not expect that the equations
(\ref{modustarcond}) will guarantee this in general. Therefore, we
expect that supplementary conditions will be necessary for a
completely satisfactory definition of $\mathcal{U}$. Secondly, it would be good
to have some ``independent'' theoretical confirmation of the results
in this paper. One could for example try to find the phases that we
are predicting in numerical calculations on a lattice. It is also
important to generalize the techniques for the breaking of Hopf
algebra symmetries that we have developed in this paper, both to the
case where the symmetry algebra is infinite dimensional and to the
case where it is no longer a Hopf algebra, but only a quasi-Hopf
algebra or even a weak quasi-Hopf algebra or Hopf algebroid. This
would extend the applicability of our symmetry breaking scheme
enormously. For example, physical systems which have an effective
description in terms of a Chern-Simons theory, such as fractional
quantum Hall states, would then come within the reach of our
methods. Finally, it would of course be extremely interesting if the
treatment of symmetry breaking and coninement that we give here could
be extended to gauge theories in $3+1$ (or higher) dimensions. One
might begin to think of such an extension starting from the ideas
presented in \cite{hooftflux}.
\vspace{1cm}

\vbox{
\noindent{\bf Acknowledgments}

\noindent
We would like to thank Yorck Sommerh\"auser for many useful comments
on Hopf subalgebras, Hopf quotients and Hopf kernels and also for
pointing out references \cite{nikshych} and \cite{nichrich}. We would
also like to thank Tom Koornwinder for useful discussions. BJS
acknowledges financial support through an Advanced Research Fellowship
of the Engineering and Physical Sciences Research Council. }

%\bibliographystyle{unsrt}
%\bibliography{symbreco}

\end{document}